\documentclass[12pt,preprint,twocolappendix]{aastex6}

\def\apjl{ApJL}
\def\apjs{ApJS}

\def\aap{A\&A}

\def\masyr{mas\,yr$^{-1}$}

\begin{document}


\title{Radio Jet Proper Motion Analysis of Nine Distant Quasars above Redshift 3.5}


\author{Yingkang Zhang\altaffilmark{1}, Tao An \altaffilmark{1,2}, S\'andor Frey\altaffilmark{3,4,5},
Krisztina \'Eva Gab\'anyi\altaffilmark{3,4,6,7},
Yulia Sotnikova\altaffilmark{8}
}

\altaffiltext{1}{SKA Regional Centre Joint Lab, Shanghai Astronomical Observatory, CAS, Nandan Road 80, Shanghai, 200030, China; e-mail: antao@shao.ac.cn}
\altaffiltext{2}{SKA Regional Centre Joint Lab, Peng Cheng Lab, Shenzhen, 518066, China}
\altaffiltext{3}{Konkoly Observatory, ELKH Research Centre for Astronomy and Earth Sciences, Konkoly Thege Mikl\'os \'ut 15-17, H-1121 Budapest, Hungary}
\altaffiltext{4}{CSFK, MTA Centre of Excellence, Konkoly Thege Mikl\'os \'ut 15-17, H-1121 Budapest, Hungary}
\altaffiltext{5}{Institute of Physics, ELTE E\"otv\"os Lor\'and University, P\'azm\'any P\'eter s\'et\'any 1/A, H-1117 Budapest, Hungary}
\altaffiltext{6}{Department of Astronomy, Institute of Geography and Earth Sciences, ELTE E\"otv\"os Lor\'and University, P\'azm\'any P\'eter s\'et\'any 1/A, H-1117 Budapest, Hungary}
\altaffiltext{7}{ELKH-ELTE Extragalactic Astrophysics Research Group, ELTE E\"otv\"os Lor\'and University, P\'azm\'any P\'eter s\'et\'any 1/A, H-1117 Budapest, Hungary}
\altaffiltext{8}{Special Astrophysical Observatory of RAS, Nizhny Arkhyz, 369167, Russia}
\begin{abstract}
Up to now, jet kinematic studies of radio quasars have barely reached beyond the redshift range at $z>3.5$.
This significantly limits our knowledge of high-redshift jets, which can provide key information for understanding the jet nature and the growth of the black holes in the early Universe.
In this paper, we selected 9 radio-loud quasars at $z>3.5$ which display milliarcsec-scale jet morphology. We provided evidence on the source nature by presenting high-resolution very long baseline interferometry (VLBI) images of the sample at 8.4~GHz frequency and making spectral index maps. We also consider Gaia optical positions that are available for 7 out of the 9 quasars, for a better identification of the jet components within the radio structures.
We find that 6 sources can be classified as core--jet blazars. The remaining 3 objects are more likely young jetted radio sources, compact symmetric objects.
By including multi-epoch archival VLBI data, we also obtained jet component proper motions of the sample and estimated the jet kinematic and geometric parameters (Doppler factor, Lorentz factor, viewing angle). Our results show that at $z>3.5$, the jet apparent transverse speeds do not exceed 20 times the speed of light ($c$). This is consistent with earlier high-redshift quasar measurements in the literature and the tendency derived from low-redshift blazars that fast jet speeds ($>40\,c$) only occur at low redshifts.
The results from this paper contribute to the understanding of the cosmological evolution of radio AGN.
\end{abstract}




\section{Introduction} \label{sec:intro}

Active galactic nuclei (AGNs) are the most powerful persistent objects in the Universe, luminous across the whole range of the electromagnetic spectrum. They can be found from the local Universe up to very high redshifts. AGNs have been observed throughout almost the entire history of the Universe, providing important information for studying the co-evolution of supermassive black holes (SMBHs) and their host galaxies. The most distant quasars discovered so far are ULAS J134208.10+092838.61 at redshift $z=7.54$ \citep{2018Natur.553..473B} and J0313$-$1806 at $z=7.64$ \citep{2021ApJ...907L...1W}. Representing the end era of the cosmic reionization, they offer important clues to explore this episode when the neutral hydrogen gas in the Universe became ionized. More than 300 quasars with $z>5.7$ are known to date, most of which were discovered in optical observations \citep{2015ApJ...804..118B, 2016ApJ...833..222J,2019ApJ...873...35S}. In recent years, in addition to traditional near-infrared spectroscopic observations, X-ray and mm-wavelength radio observations 
have also contributed to the discovery of quasars with $z>6$ \citep[e.g.,][]{2005AA...440L..51M,2020MNRAS.497.1842M}. Among these high-redshift AGNs, SMBHs with masses exceeding $10^9 M_\odot$ are found. Although there is an observational bias that the most massive and highly accreting black holes are the easiest to observe due to sensitivity limitations, 
at least the existence of these SMBHs indicates that they have completed their rapid growth in less than one tenth of the current age of the Universe. The formation and rapid growth of the first SMBHs is one of the mysteries of current cosmology and AGN astrophysics.

Relativistic jets are thought to play an important role in promoting rapid accretion of early AGNs. If a significant fraction of the gravitational energy can be channeled into forming and maintaining relativistic jets, then their black holes can remain at high-rate accretion for a long time \citep{2008MNRAS.386..989J,2013MNRAS.432.2818G}. 
In this sense, the dynamical properties and the lifetime of the high-redshift jets are key to testing this scenario.
Although the number of high-$z$ AGNs discovered has been increasing over the years \citep{2016ApJ...833..222J,2019ApJ...884...30W}, their radio counterparts are less intensely studied. An important reason is that only $\sim10\%$ of optically detected quasars have active radio counterparts. The ratio of radio-loud and radio-quiet AGNs seems to be consistent at high redshifts and in the local Universe \citep{2015ApJ...804..118B}, but it has also been suggested that the fraction of radio-loud AGNs tends to evolve with redshift \citep{2022MNRAS.511.5436D}. 
Producing relativistic jets is not only one of the ways of AGN energy release \citep{2019ARAA..57..467B}, but also has an important impact on the feedback to the environment in the host galaxy \citep{2012ARAA..50..455F}. To date, the role of jets in the formation and evolution of first SMBHs remains an open question.

The extreme subclass of AGNs are blazars, remarkable for their high bulk Lorentz factors in the jet plasma, the small viewing angles of the jets with respect to the line of sight, and thus the large Doppler-boosting factors. Blazars make up nearly half of the radio-loud AGNs with flux density $S_{\rm  1.4GHz}>100$~mJy in the high-redshift population \citep{2021MNRAS.508.2798S}. The jets of high-redshift blazars are typically short, possibly due to the projection effect caused by the small viewing angle. Perhaps also the environment of the early galaxies is not favorable for the development of large-scale jets. In the latter case, the relativistic electrons in the extended jets are heavily depleted via inverse-Compton scattering by the enhanced number density of cosmic microwave background (CMB) photons in the high-redshift Universe \citep{2014MNRAS.438.2694G}. Also, the interaction between the jet and the dense interstellar medium may cause a large amount of the jet's mechanical energy to be consumed and the growth of the jet to be stalled \citep{2022MNRAS.511.4572A}. Typical radio images of the distant jetted AGNs are characterized by compact flat-spectrum cores or core--jet structures. Very long baseline interferometry (VLBI), with its unique high-resolution capability, provides a direct way to probe the pc-scale compact structure and jet properties of high-$z$ AGNs. 

To date, there are only six radio-loud quasars imaged with VLBI at extremely high redshifts ($z>6$). Their VLBI morphologies indicate different types of radio sources. They do not show a strong tendency towards highly-beamed sources (i.e. blazars) that would be expected from a selection effect caused by relativistic beaming. Here we list the radio-loud quasars at $z>5$ with VLBI imaging available in the literature (Table~\ref{tab:highz-list}). As one can see, most of the objects are marked as a compact quasar. These sources show a compact structure or minor extension on milliarcsec (mas) angular scale (down to a few tens of pc in linear scale). They typically do not have very high brightness temperatures or show flat radio spectrum. Their VLBI studies suggest that these compact high-redshift quasars could be young jetted AGNs in the early Universe (see references in Table~\ref{tab:highz-list}). The observing wavelengths of current VLBI facilities are typically at a few cm, and the corresponding rest-frame frequencies of high-redshift quasars are much higher than those of their local counterparts studied at the same observing frequency. The emitted frequency is $(1+z)$ times the observed frequency. This makes the steep-spectrum jets relatively fainter and could cause the lack of radio quasars observed with prominent extended jets at high redshifts \citep[e.g.][]{2015IAUS..313..327G,2022ApJS..260...49K}.

\floattable
\begin{deluxetable}{ccccc}
\tablecaption{Information about the high-redshift ($z>5$) radio quasars with VLBI imaging observations from the literature}
\label{tab:highz-list}
\tablehead{
\colhead{Name} & \colhead{$z$}&\colhead{Ref. for $z$}&\colhead{Ref. for VLBI} & \colhead{Radio morphology} 
}
\colnumbers
\startdata
\multicolumn{5}{c}{$z>6$}\\
J1129$+$1846 & 6.82 & 1 & 16  & Compact quasar  \\
J2331$+$1129 & 6.57 &  2 & 17 & Compact quasar \\
J2318$-$3113 & 6.44 &  3 & 18 & Compact quasar \\
J1429$+$5447 & 6.18 & 4,5 & 19 & Compact quasar \\
J1427$+$3312 & 6.12 &  6 & 20,21 & CSO candidate \\
J0309$+$2717 & 6.10 &  7 & 22 & Core--jet blazar \\  \hline
\multicolumn{5}{c}{$z>5$}\\
J2228$+$0110 & 5.95 &  8 & 23 & Compact quasar \\
J2329$-$1520 & 5.84 &  9 & 24 & CSO candidate \\
J0836$+$0054 & 5.77 &  10 &25,26 & Compact quasar \\
J1530$+$1049 & 5.72 &  11 & 27 & MSO galaxy \\
J0906$+$6930 & 5.47 &  12 & 12,28,29 & Core--jet blazar \\
J1026$+$2542 & 5.25 &  13 & 30,31 & Core--jet blazar \\
J0131$-$0321 & 5.13 &  14 & 32 & Compact blazar \\
J0913$+$5919 & 5.18 &  15 & 33 & Compact quasar \\
\enddata
\tablecomments{Columns: (1) source name in J2000 format; (2) source redshift; (3) reference for the source redshift: 1--\citet{2021ApJ...909...80B}, 2--\citet{2022ApJ...929L...7K}, 3--\citet{2018ApJ...854...97D}, 4--\citet{2010AJ....139..906W}, 5--\citet{2011ApJ...739L..34W}, 6--\citet{2006ApJ...652..157M}, 7--\citet{2020AA...635L...7B}, 8--\citet{2011ApJ...736...57Z}, 9--\citet{2018ApJ...861L..14B}, 10--\citet{2001AJ....122.2833F}, 11--\citet{2018MNRAS.480.2733S}, 12--\citet{2004ApJ...610L...9R}, 13--\citet{2012ApJS..203...21A}, 14--\citet{2014ApJ...795L..29Y}, 15--\citet{2001AJ....122..503A}; (4) reference for VLBI image: 16--\citet{2021AJ....161..207M}, 17--Frey et al. 2022 (in prep.), 
18--\citet{2022AA...662L...2Z}, 
19--\citet{2011AA...531L...5F}, 20--\citet{2008AA...484L..39F}, 21--\citet{2008AJ....136..344M}, 22--\citet{2020AA...643L..12S}, 23--\citet{2014AA...563A.111C}, 24--\citet{2018ApJ...861...86M}, 25--\citet{2005AA...436L..13F}, 26--\citet{2003MNRAS.343L..20F}, 27--\citet{2018RNAAS...2..200G} , 28--\citet{2017MNRAS.468...69Z}, 29--\citet{2020NatCo..11..143A}, 30--\citet{2013MNRAS.431.1314F}, 31--\citet{2015MNRAS.446.2921F}, 32--\citet{2015MNRAS.450L..57G}, 33--\citet{2004AJ....127..587M}; (5) source morphology in VLBI images and classification.
}

\end{deluxetable}

The small sample significantly limits our knowledge of high-redshift jets. It was found that there are two distinct cosmological epochs of SMBH formation: the number density of non-jetted AGNs hosting SMBHs peaks at $z \sim 2-2.5$, while the formation of SMBHs in jetted AGNs took place earlier, at $z \sim 4$ \citep[][and references therein]{2015MNRAS.446.2483S,2021Galax...9...23S}. This seems to indicate that black holes with powerful jets grow faster than those of the same mass but without jets. Thus, observational studies of jetted AGNs at $z\sim4$ can bring us crucial information about the rapid growth of the first black holes. Moreover, the sample size of radio-loud AGNs in this cosmic period is much larger than during the cosmic reionization ($z\gtrsim 7$), facilitating statistical studies.

The kinematics of high-redshift jets provides key information for understanding the jet nature. In particular, VLBI can measure the jet proper motion and independently calculate the Doppler-boosting factor, which can be used to estimate the bulk Lorentz factor and viewing angle of the jet flow. These place strong observational constraints on the modeling of the spectral energy distribution (SED) of the jet \citep{2022arXiv220309527S}. Current detections of jet proper motions of high-redshift ($z\gtrsim4$) AGNs are limited to a few sources \citep{2010AA...521A...6V,2015MNRAS.446.2921F,2018MNRAS.477.1065P,2020NatCo..11..143A,2020SciBu..65..525Z}. All these are blazars, with apparent superluminal speeds distributed in a wide range from $2\,c$ to about $20\,c$, where $c$ denotes the speed of light. In contrast, the jets of high-redshift galaxies move at mildly relativistic speeds \citep{2022MNRAS.511.4572A}, consistent with the traditional notion of their large viewing angle, without strong beaming effects. More proper motion measurements help extend our understanding of the nature and geometry of high-redshift jets.

In the present work, we provide new high-quality VLBI images of nine high-redshift radio-loud quasars. By combining archival multi-frequency, multi-epoch VLBI data as well as optical data, we classify their source nature and determine for the first time their jet proper motions and other radio properties on pc scales. This work increases the size of the proper motion sample of $z>3.5$ jets by a factor of two. The subsequent sections of the paper are organized as follows. In Section~\ref{sec:obs}, we describe the method of sample selection, the new Very Long Baseline Array (VLBA) observations, and the data processing. Section~\ref{sec:results} presents the new VLBI images which are used as the reference for source classification. Section~\ref{sec:disc} contains the analysis and discussion of the jet kinematics. Section~\ref{sec:sum} summarizes the main results and conclusions of this paper. Comments on individual target sources are given in Appendix~\ref{sec:appa}, extended tables and figures can be found in Appendix~\ref{sec:appb}. Throughout this paper, we use the cosmological parameters derived from a flat $\Lambda$ Cold Dark Matter ($\Lambda$CDM) model \citep{2011ApJS..192...18K} with $\Omega_\mathrm{m} = 0.27$, $\Omega_{\Lambda} = 0.73$, and $H_{0} = 70$~km s$^{-1}$Mpc$^{-1}$.

\section{Target Selection and VLBI Observations} \label{sec:obs}

\subsection{The Selected High-redshift Sample}
\label{sec:sample}

Our focus is on the jet kinematics of a sample of high-redshift blazars at $z \sim 4$. To measure the proper motion of the jet, at least two epochs of VLBI observations at the same observing frequency are required, preferably with similar $(u,v)$ coverage at each epoch. The time interval between epochs should also be taken into account when measuring jet proper motions at high redshifts. The minimum time gap between epochs can be estimated as $t_\mathrm{gap} = \frac{D_\mathrm{min}}{\mu_\mathrm{comp}}$, where $D_\mathrm{min}$ is the minimum distance that can be distinguished between available epochs, usually depending on the observing resolution and frequency. Here $\mu_\mathrm{comp}$ denotes the proper motion of a well-identified component. In the case of VLBI measurements, $D_\mathrm{min}$ can be measured in mas, $\mu_\mathrm{comp}$ in \masyr, and thus $t_\mathrm{gap}$ in yr.
For high-redshift radio jets, their $\mu_\mathrm{comp}$ could be much smaller than their low-redshift counterparts due to the cosmological time dilation, thus a large $t_\mathrm{gap}$ is usually needed to obtain reliable measurements.
This is one of the reasons why only a few $z>4$ jet proper motions have been measured so far.

To select suitable AGNs, we first checked the VLBI archive database for radio sources with $3.5\lesssim z \lesssim 4.5$. We used the Astrogeo database\footnote{\url{http://astrogeo.org/}} to build our sample, which is currently the largest collection of VLBI imaging data available. It accumulates data from a series of astrometric and geodetic VLBI surveys such as the VLBA Calibrator Surveys \citep[VCS,][]{2002ApJS..141...13B,2003AJ....126.2562F,2005AJ....129.1163P,2006AJ....131.1872P,2007AJ....133.1236K,2008AJ....136..580P,2016arXiv161004951P}, the United States Naval Observatory surveys \citep[USNO;][]{2021AJ....162..121H}, and various other projects. 
The Astrogeo database now provides more than 100,000 VLBI images of more than 17,000 AGNs, mostly observed in snapshot mode at 2.3 and 8.4~GHz. 
Although the quality of the snapshot images is not very high, we only focus on compact radio structures from bright radio sources with flux densities above $50$~mJy, so these images are sufficient for our research purposes.
By cross-matching the list with the NASA/IPAC Extragalactic Database\footnote{
\url{http://ned.ipac.caltech.edu}} \citep{https://doi.org/10.26132/ned1} and some large optical spectroscopic projects \citep[e.g., Sloan Digital Sky Survey Data Releases 9 and 12,][]{2012ApJS..203...21A,2017A&A...597A..79P}, over 60 $z>3.5$ quasars are found (the full sample will be given in our subsequent study). Since VLBI is sensitive to compact sources with high brightness temperatures, the high-redshift extragalactic sources detected in the flux density-limited VLBI surveys are all radio-loud jetted AGNs. From this parent sample, we then selected a sub-sample for the jet proper motion study, according to the following criteria:
\begin{itemize}
    \item The redshift is $3.5\lesssim z \lesssim 4.5$.
    \item The flux density at 8.4~GHz is $S_\mathrm{8.4}>50$~mJy, to enable high dynamic range imaging. 
    \item The source shows resolved jet structure in the existing 8.4-GHz archival images, a premise for being able to measure jet component motion.
    \item Multiple epochs of VLBI imaging are available at 8.4~GHz. A minimum time gap $t_\mathrm{gap} \approx 7$~yr can be estimated, by assuming a moderate jet component speed $10\,c$ and $D_\mathrm{min}$ from 3-$\sigma$ position error. A rough estimate of $\approx 0.2$ for the position error is based on the 10\% restoring beam size of typical 8.4-GHz VLBA observations. The corresponding $\mu_\mathrm{comp}$ is $0.08-0.09$~\masyr for the given redshift range.

\end{itemize}

Based on the above criteria, we selected nine targets whose detailed information can be found in Table~\ref{tab:basic}. There is no other bias in this sample, except for the limitation on the flux density and the declination ($\mathrm{Dec.} > -40\degr$). The latter is because the archival data were mostly obtained with the VLBA located on the northern hemisphere. In particular, our sample selection did not restrict the AGN optical properties. At the time of selecting the present sample, J1606+3124 met our criteria. However, a subsequent study \citep{2022MNRAS.511.4572A} found that its optical classification and redshift are debatable and need further confirmation. \citet{2022MNRAS.511.4572A} suggest that the source is more likely a high-redshift radio galaxy rather than a quasar. 

We have collected images from the Astrogeo archive for a total of 39 epochs for these nine sources, with 3--6 epochs per source and a maximum time span of $\sim 24$~yr (which equals approximately $4$~yr in the source's rest frame). In snapshot mode, a target is observed in a few scans (observation time segments), each lasting for a few minutes. Scans of multiple sources are interleaved to ensure that each source has good $(u,v)$ coverage. Our selected target sources are routinely monitored in the VLBA calibrator surveys, the $(u,v)$ coverage at 8.4~GHz are similar from epoch to epoch, and the prominent components can be clearly imaged with similar quality, even with limited sensitivity.

In order to obtain high-quality images of high-$z$ AGNs for better identification of the radio components and classification of their radio structures, we initiated a new epoch of VLBA observations of the selected 9 sources. The new data are used together with the archival data for jet kinematic studies. Details of the new VLBA observations are presented in the next subsection. 

\floattable
\begin{deluxetable}{ccccccc}
\tablecaption{Information about the high-$z$ quasars selected for this jet proper motion study from Astrogeo}
\tablehead{
\colhead{Name} & \colhead{RA (J2000)}&\colhead{Dec (J2000)}&\colhead{$z$} & \colhead{$N_\mathrm{epoch}$} & \colhead{Epoch range} & \colhead{Reference} }
\colnumbers
\startdata
J0048$+$0640 & 00h48m58.7231s & $+$06d40m06.475s & 3.580  & 5 & 2004$-$2018 & \citet{1999AA...342..378G}  \\
J0753$+$4231 & 07h53m03.3375s & $+$42d31m30.765s & 3.595  & 4 & 1995$-$2018 &\citet{2010MNRAS.405.2302H} \\
J1230$-$1139 & 12h30m55.5560s & $-$11d39m09.795s & 3.528  & 3 & 1996$-$2018 &\citet{1997MNRAS.284...85D}  \\
J1316$+$6726 & 13h16m27.2015s & $+$67d26m24.262s & 3.515  & 4 & 1995$-$2018 &\citet{2009ApJS..180...67R}  \\
J1421$-$0643 & 14h21m07.7556s & $-$06d43m56.356s & 3.689  & 5 & 1997$-$2018 &\citet{2001AA...379..393E}  \\
J1445$+$0958 & 14h45m16.4653s & $+$09d58m36.073s & 3.552  & 6 & 1997$-$2018 &\citet{2010MNRAS.405.2302H}  \\
J1606$+$3124 & 16h06m08.5184s & $+$31d24m46.458s & 4.560  & 4 & 1996$-$2018 &\citet{2008ApJS..175...97H}  \\
J1939$-$1002 & 19h39m57.2566s & $-$10d02m41.521s & 3.787  & 5 & 1996$-$2019 &\citet{1991ApJS...77....1L}  \\
J2102$+$6015 & 21h02m40.2191s & $+$60d15m09.837s & 4.575  & 4 & 1994$-$2018 &\citet{2004ApJ...609..564S}  \\
\enddata
\tablecomments{Columns: (1) source name, (2)--(3) source equatorial coordinates right ascension and declination, (4) redshift, (5) number of available VLBI epochs at 8.4~GHz for this study, (6) time range spanned by the VLBI observations, (7) literaure reference for the redshift.}
\label{tab:basic}
\end{deluxetable}

\subsection{VLBA Observations and Data Reduction}
\label{sec:obs-data}

The VLBA observations (project code: BZ064, PI: Y. Zhang) were conducted on 2017 February 5 and 2017 March 19 at 8.4 GHz. Nine out of the ten VLBA antennas participated in the observations (see Table \ref{tab:obs}. The observations were performed using the DDC (digital downconverter) system with a baseband bandwidth of 128~MHz, divided into 256 spectral channels of 500~kHz each, using four baseband channels (IFs) of left and right circular polarization and 2-bit sampling. This configuration results in a total data rate of 2048~Mbit\,s$^{-1}$. Since the target sources are all compact and bright ($S_\mathrm{8.4}>50$~mJy), fringe search can be done using themselves as calibrators, so a standard continuum observation mode was used. A total of 6~h of observation time was divided into two sessions: BZ064A (2~h) and BZ064B (4~h), according to the distribution of the right ascensions of the 9 target sources (see Table~\ref{tab:basic}). During the observations, pointing on sources in the same group were alternated, each obtaining about 0.5~h on-source time. This allowed us to reach a fairly good $(u,v)$ coverage and high image quality. We refer to Table~\ref{tab:obs} for the details of the observations.

After observations, the data were correlated with 2-s integration time using the DiFX correlator \citep{2011PASP..123..275D} in Socorro (New Mexico, USA). The correlated data were then downloaded to the China Square Kilometre Array (SKA) Regional Centre prototype \citep{2019NatAs...3.1030A} via the internet for further calibration and imaging. We calibrated the data using the VLBI data processing pipeline developed by our group \citep{2022arXiv220613022A}. The main steps of the procedure are detailed below. 
We used the US National Radio Astronomy Observatory (NRAO) Astronomical Image Processing System (\textsc{AIPS}) software package \citep{2003ASSL..285..109G} to calibrate the amplitude and phase of the visibility data. Our pipeline is written in \texttt{Python} language\footnote{http://www.python.org}, using \texttt{ParselTongue} as an interface to convert \textsc{AIPS} tasks into \texttt{Python} scripts \citep{2006ASPC..351..497K}, so that most of the operations can be executed automatically.

The pipeline first loaded the data into \textsc{AIPS} and performed a couple of tasks to assist the user in inspecting the data quality. Input parameters (names for fringe finders, bandpass calibrators and target sources, reference antenna, solution intervals, etc.) were determined manually. Then the pipeline automatically conducted the amplitude, phase, and bandpass calibrations, and splitted the visibility data into single-source files. The procedures and parameters used in our experiment are described as follows. The visibility amplitudes were calibrated using the antenna gain curves and system temperatures measured at each station during the observations. The \textsc{AIPS} procedure \texttt{VLBATECR} was then used to correct for atmospheric opacity as well. Next, the bright calibrator sources (J2148$+$0657 for BZ064A and 3C\,273 for BZ064B; Table \ref{tab:obs}) were used to calibrate the delays and global phase errors of the instruments using the \texttt{FRING} task. This operation calibrated and aligned the delays and phases between different sub-bands. Next, global fringe fitting was performed on all sources and the resulting gain solutions were interpolated and applied to calculate and remove the phase errors. We checked and found that over 98\% of the phase solutions were successful. Finally, the antenna-based bandpass functions were solved by using data from the calibrators and applied to correct the target sources' visibility data. At this point, the initial calibration was complete. The calibrated data were exported to an external single-source \texttt{FITS} file by averaging over each sub-band (128~MHz each) and a time interval of 2~s.

After calibration, together with the archival calibrated visibility data obtained from the Astrogeo archive, all single-source data files exported from \textsc{AIPS} were loaded into the Caltech \textsc{Difmap} software package \citep{1997ASPC..125...77S} for final imaging and model-fitting.
The hybrid mapping process consists of several iterations of \texttt{CLEAN} and self-calibration. Final \texttt{CLEAN} images were obtained after a few iterations of phase and amplitude self-calibration, repeated by gradually reducing the solution intervals from a few hours to 1~min.
The data from the Astrogeo archive have been calibrated, so we just conducted the imaging and model-fitting following similar steps in \textsc{Difmap} as above.

To parameterize the core and the jet components, we conducted model fitting on the self-calibrated visibility data in \textsc{Difmap}. This utilizes the Levenberg--Marquardt non-linear least-squares minimization technique in the visibility domain \citep{1997ASPC..125...77S}. During the processing, we followed the rules of continuity and simplicity, assuming that most components move downstream of the jet and the number of jet components does not change much over the successive epochs. Since we care more about the change in the position of the jet components, most components are fitted with circular Gaussian brightness distribution models. Very few components whose structure was too compact were fitted with a point source model. In a few cases, the jet components showed extended structure and could not be adequately fitted by a single circular Gaussian model. These features could result from diffuse emission or newly ejected components along the jet trajectory. Regarding this, several point-source models which were not treated as physical components were applied to account for the extra flux density. These components are also listed in Table~\ref{tab:mod}.

\citet{1999ASPC..180..301F} introduced a method to estimate the uncertainty of the fitted model component parameters. This method considers only the statistical error of the image and is therefore closely related to the image quality, while the actual error in the observed data is often higher than the statistical error. For the peak brightness and integrated flux density of each component, we considered an additional 5\% calibration uncertainty originating from the measured gain curves and system temperatures. For the position of each component, we found that the \citet{1999ASPC..180..301F} method underestimates the actual error of the snapshot observations. 
Due to the short observing time, the sparse $(u,v)$ coverage of the archival Astrogeo data could result in side lobes. In such conditions, we chose to identify components based on our own long-track VLBA observations, and used the synthesized beam size instead of the fitted component size to calculate the position error,
i.e., $\frac{\theta_{\rm beam}}{\rm SNR}$, where SNR denotes the signal-to-noise ratio. In such snapshot observations, our conservative estimation of the errors provided a more reasonable assessment of the observed properties of the jet, and similar applications are common in large VLBA surveys \citep[e.g. the MOJAVE survey;][]{2009AJ....138.1874L,2016AJ....152...12L,2019ApJ...874...43L}.

\section{Results} \label{sec:results}

\begin{figure*}
	\centering
	\begin{tabular}{ccc}
		\includegraphics[width=0.3\textwidth]{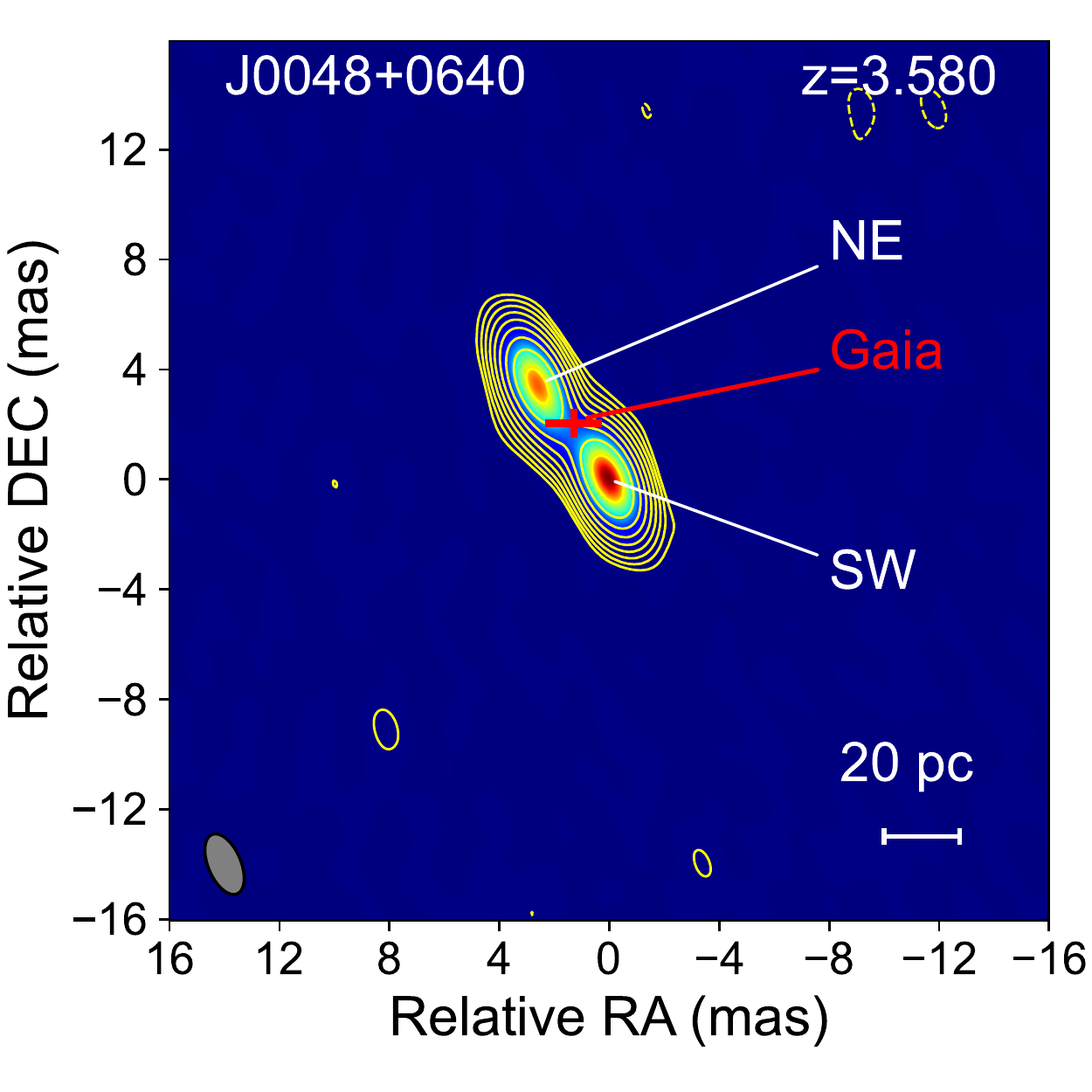}&
		\includegraphics[width=0.3\textwidth]{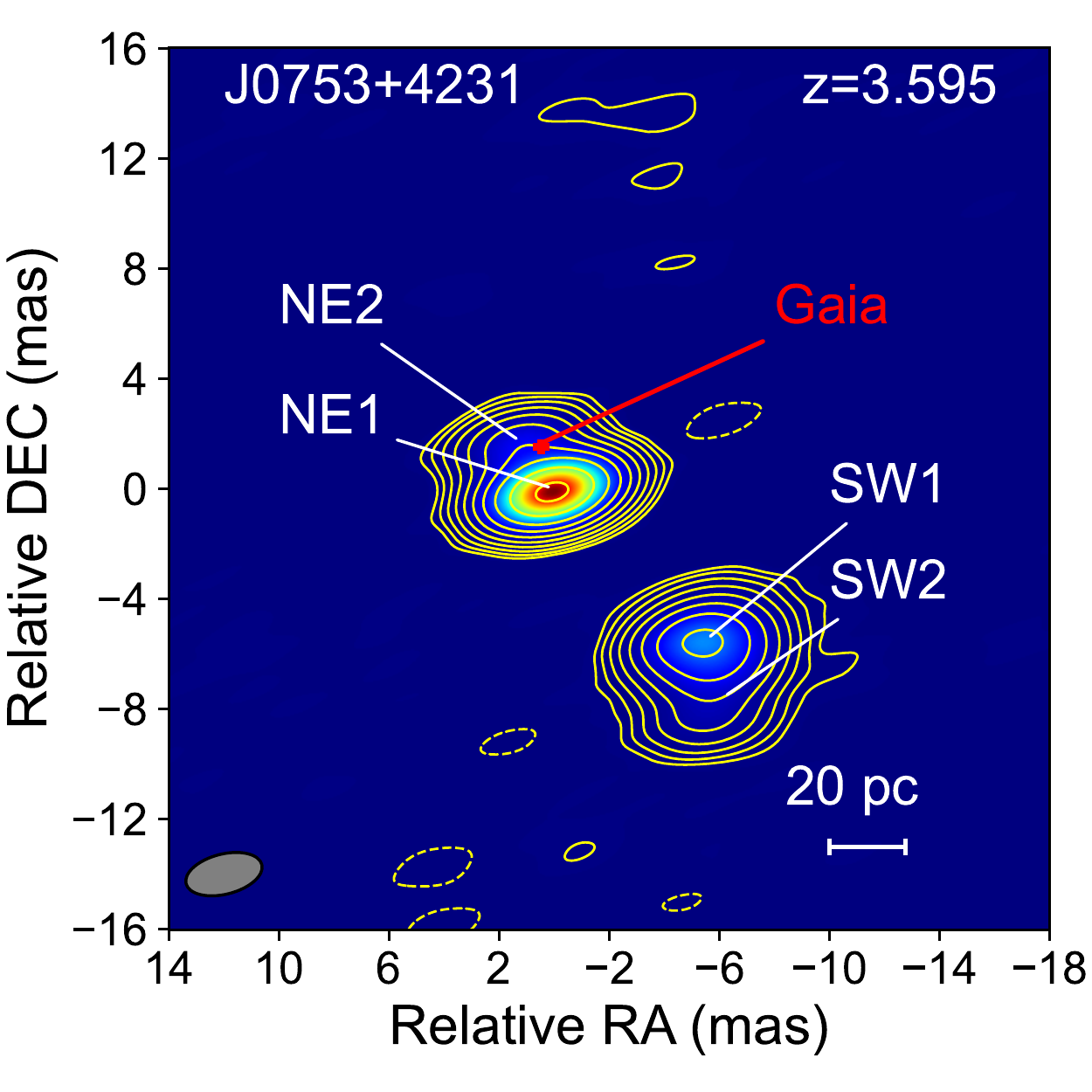}&
		\includegraphics[width=0.3\textwidth]{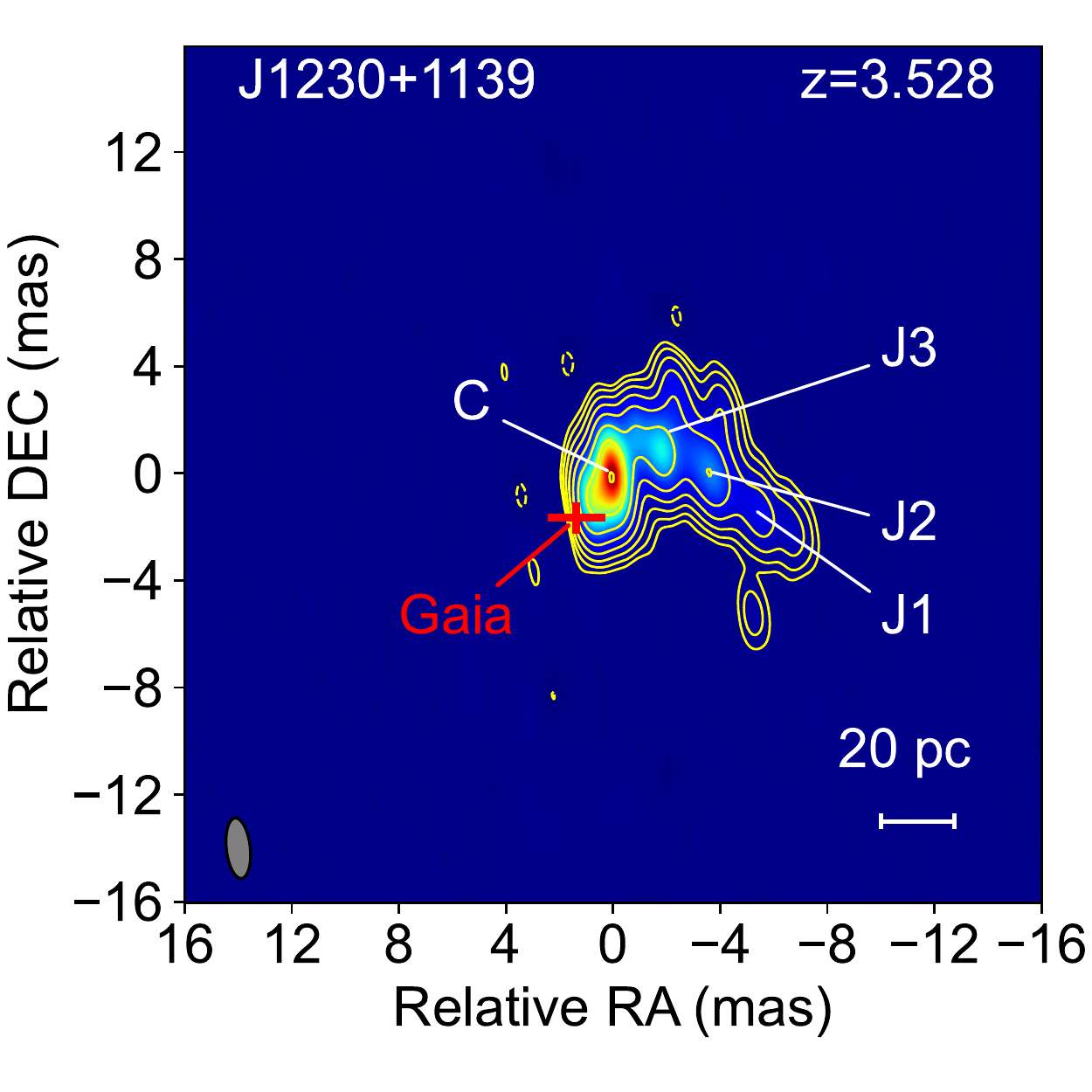}
	\end{tabular}
	\begin{tabular}{ccc}
		\includegraphics[width=0.3\textwidth]{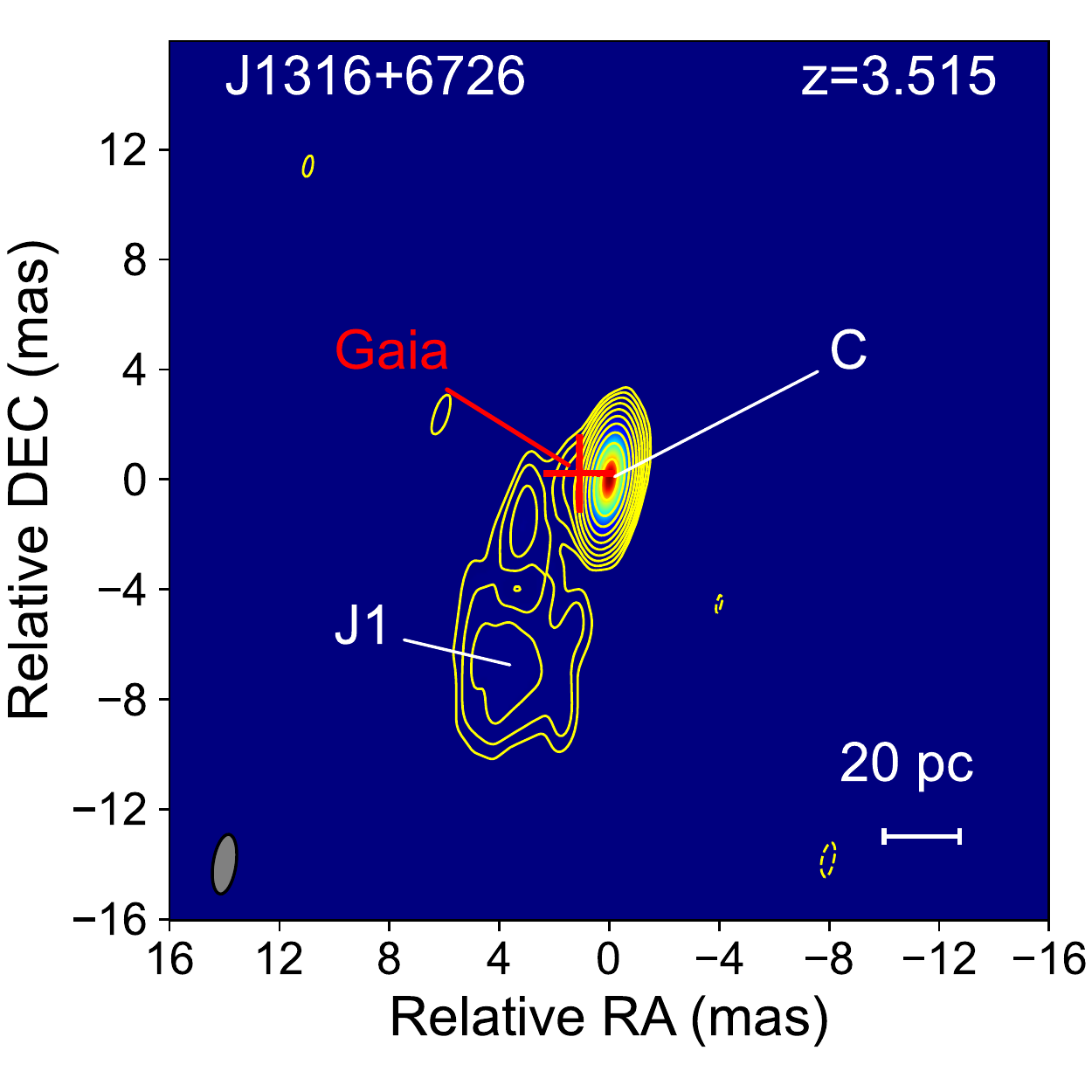}&
		\includegraphics[width=0.3\textwidth]{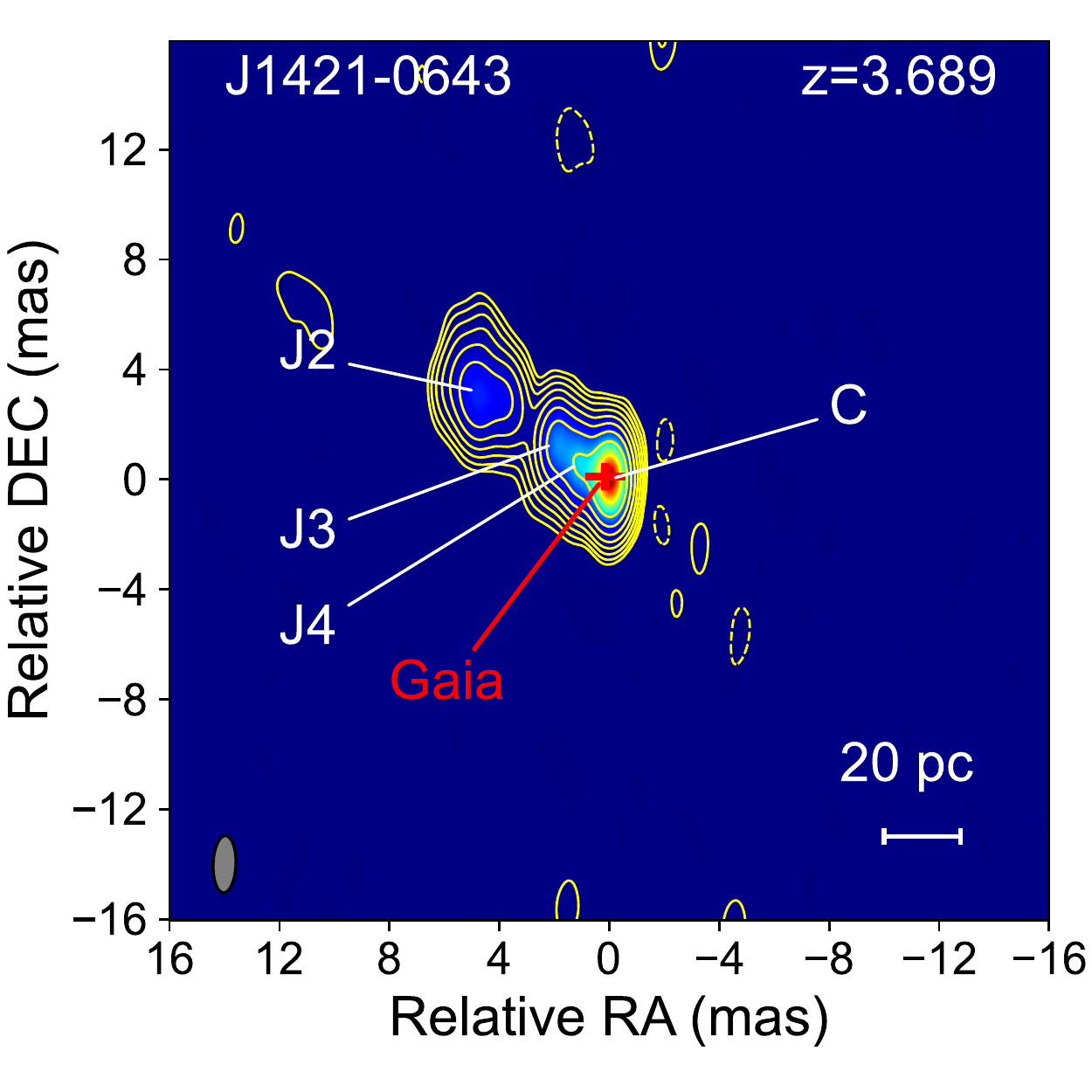}&
		\includegraphics[width=0.3\textwidth]{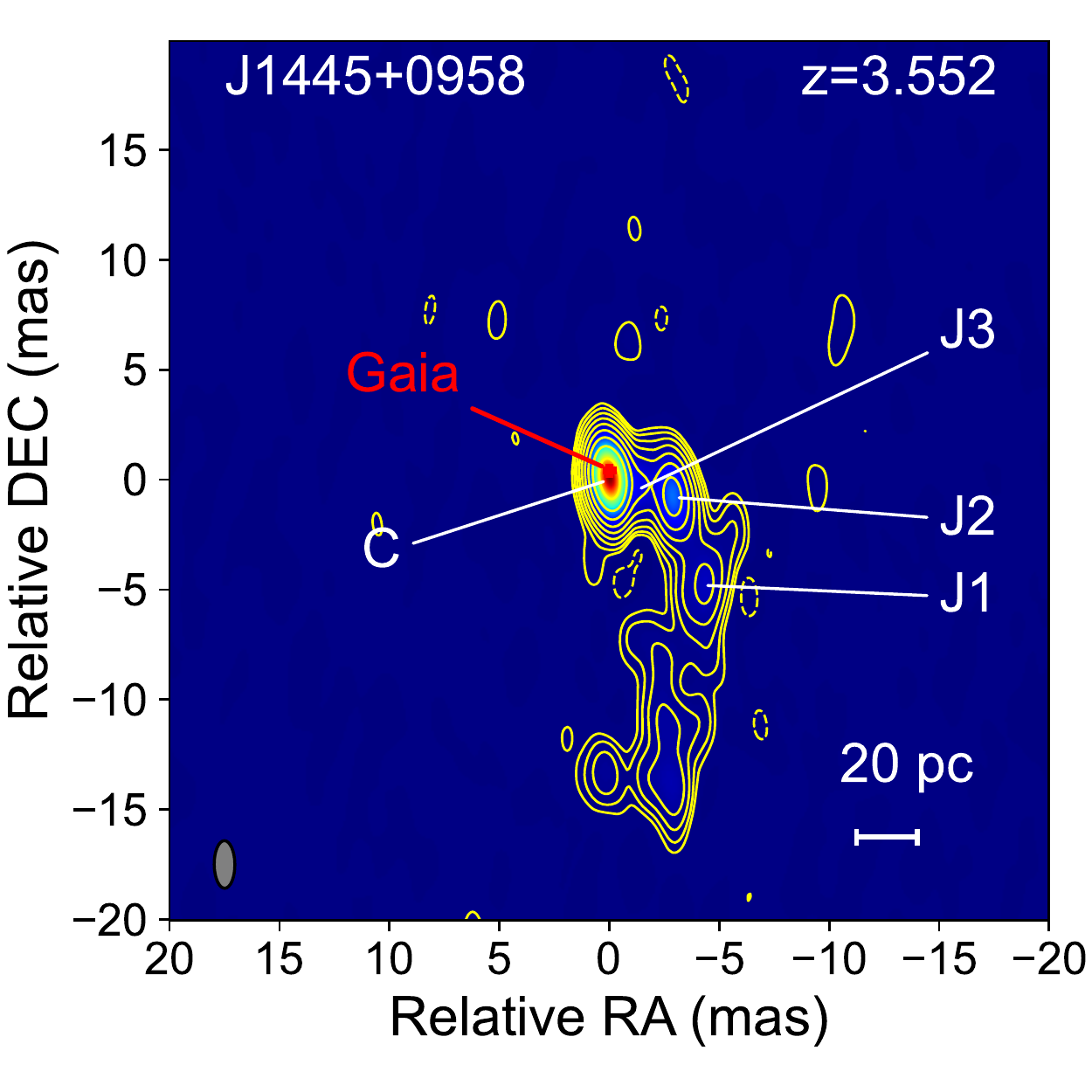}
	\end{tabular}
	\begin{tabular}{ccc}
		\includegraphics[width=0.3\textwidth]{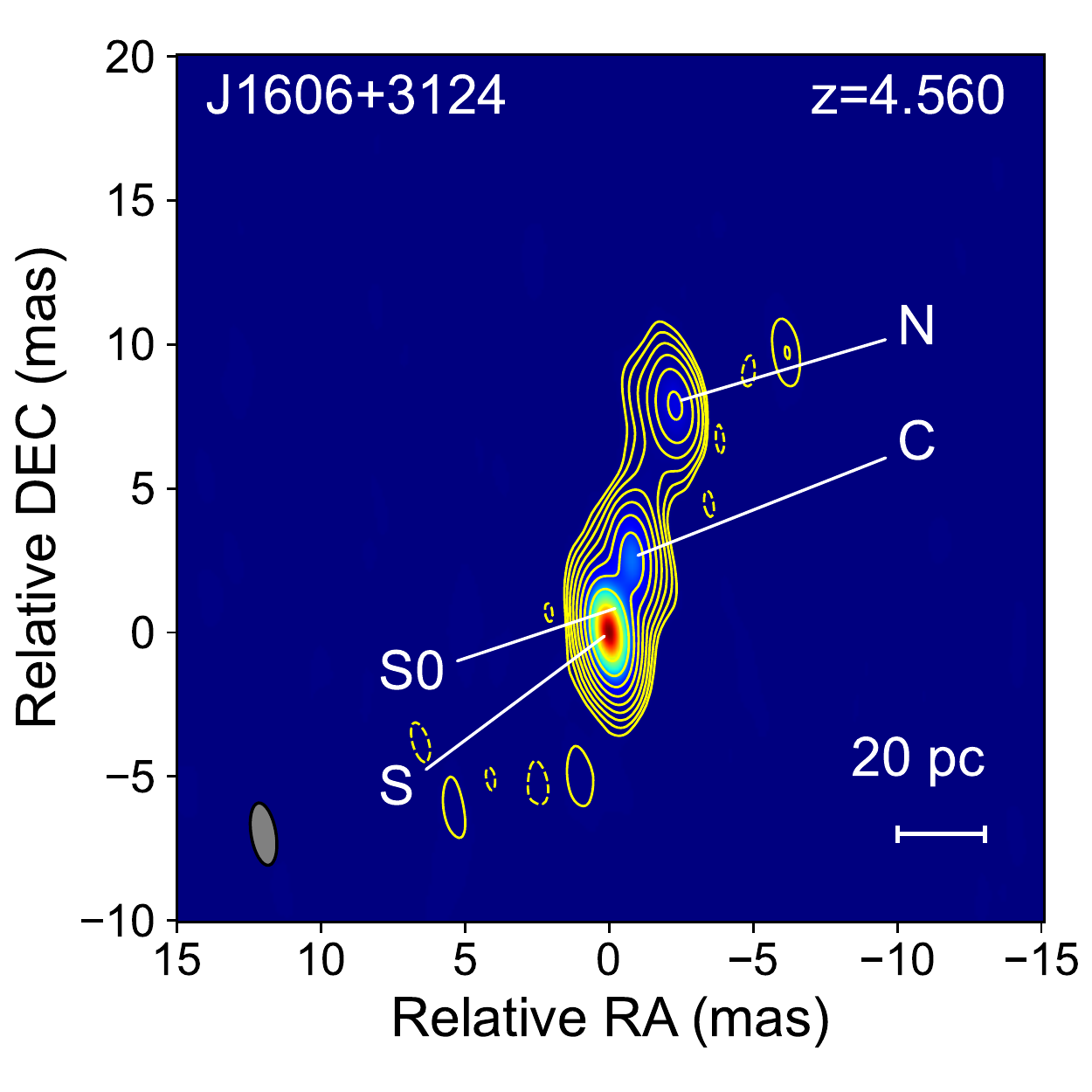}&
		\includegraphics[width=0.3\textwidth]{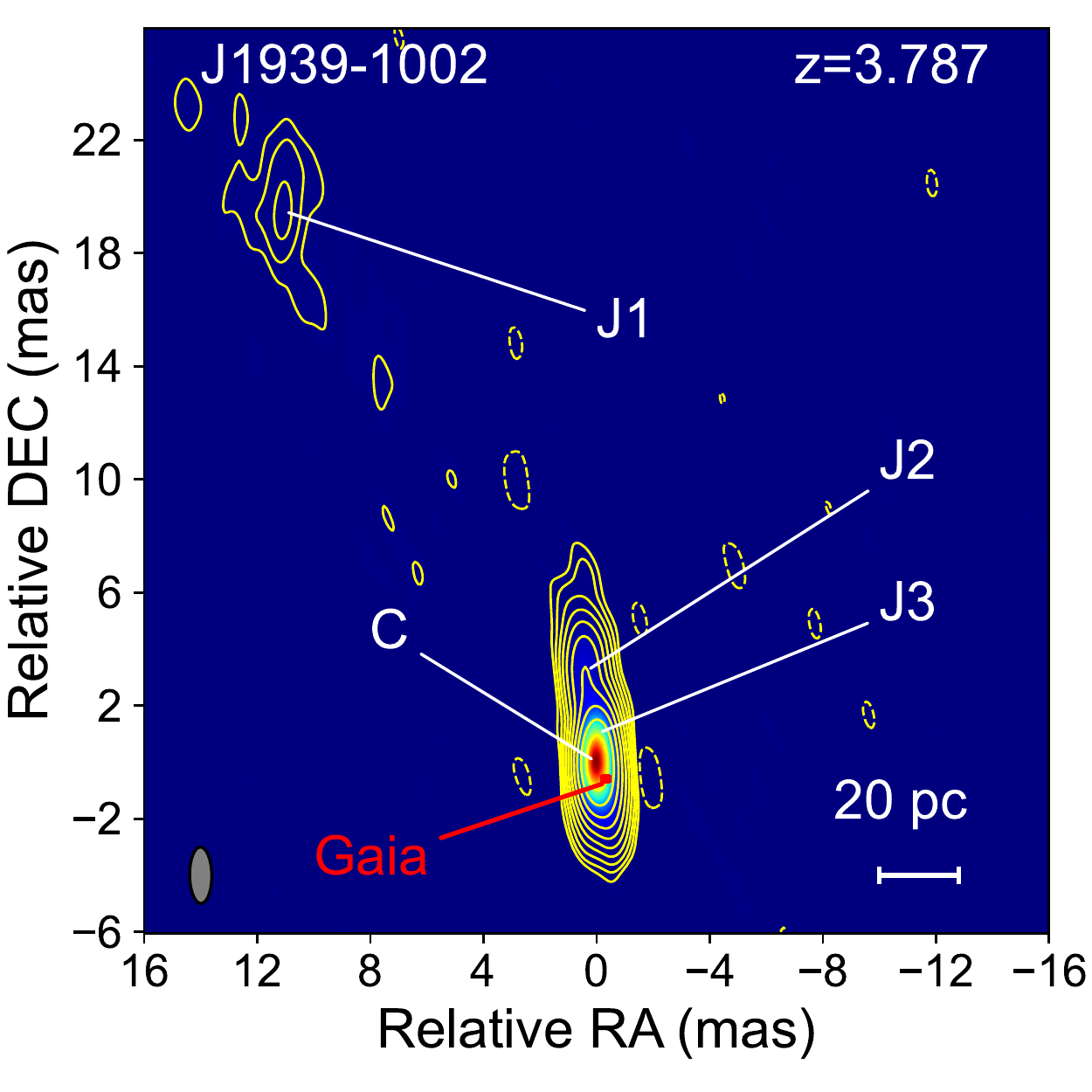}&
		\includegraphics[width=0.3\textwidth]{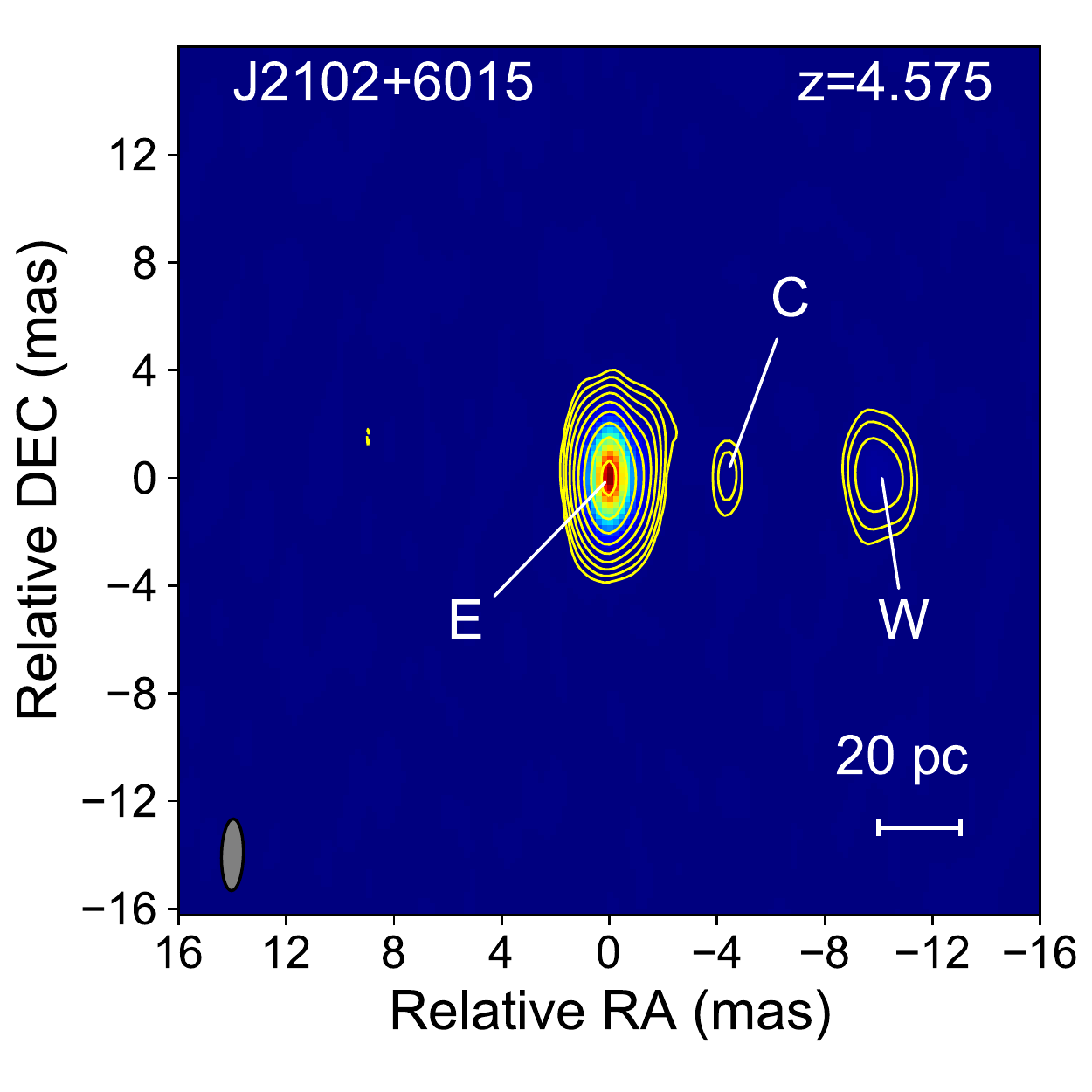}
	\end{tabular}
	\caption{Naturally weighted images of the 9 high-redshift sources after deep cleaning, derived from our VLBA observations at 8.4~GHz in 2017. The lowest contours represent $\pm 4$ times of the rms noise, and the positive contour levels increase by a factor of 2. The image parameters are listed in Table~\ref{tab:img}. The green crosses mark the optical positions detected by Gaia, the bar length are 3 times the Gaia positional errors reported in \texttt{\url{https://gea.esac.esa.int/archive}}.}
	\label{fig:col}
\end{figure*}

Here we show mas-resolution images of the nine high-$z$ radio-loud quasars and classify their radio structures by combining their radio morphology, radio spectrum, and Gaia optical astrometric data.
Figure~\ref{fig:col} displays the high-quality images obtained from our VLBA observations in 2017. The lowest noise level of the images is $\sim 0.1$~mJy\,beam$^{-1}$ (Table~\ref{tab:img}), close to the thermal noise level.
The components marked with labels are from the model fitting results from Section~\ref{sec:obs-data}; their parameters can be found in Table~\ref{tab:mod}. Our new VLBA observations are more sensitive than the archival astrometric snapshot data from Astrogeo, thus the models fitted to our own 2017 VLBA data are used as a reference. Certain model components are too weak or too diffuse to be detected in some lower-quality images. Those components are not used for proper motion determination.
Since these sources were intentionally chosen to have a resolved structure in VLBI images (see Section~\ref{sec:obs}), all images demonstrate a rich jet structure but present two different types of characteristics: 
\begin{itemize}
    \item In six sources (J1230+1139, J1316+6726, J1421$-$0643, J1445+0958, J1939$-$1002, and J2102+6015), the brightest component is located at one end of the radio structure. J1230+1139, J1316+6726, and J1445+0958 show a large jet bending ($\approx 90\degr$) within 2~mas. J1421$-$0643 and J1939$-$1002 display linear collimated structure. A faint diffuse feature can be seen at 20~mas from the brightest component in J1939$-$1002. In the image of J2102+6015, there are two faint discrete components to the west of the brightest component \citep[see also in][]{2021MNRAS.507.3736Z}. 
    \item The other three sources (J0048+0640, J0753+4231, and J1606+3124) show two compact components with comparable brightness at either end of the radio structure. 
\end{itemize}
The archival Astrogeo images show similar structures, with minor differences because faint components are not always detected in the less sensitive snapshot observations. An example is J1 component in J1939$-$1002. Basic information on the archival VLBI observations is given in Table~\ref{tab:obs-arc}.

A conclusive classification of radio structures from VLBI images alone is not always possible, and for this reason, we searched optical astrometric data of these sources from the Gaia \citep{2016A&A...595A...1G} Early Data Release 3 \citep{2021AA...649A...1G} via the Gaia archive \citep{gaia:edr3}. For quasars, the optical nucleus detected by Gaia corresponds to the accretion disk, with some contribution from the innermost part of the synchrotron-emitting jet \citep[e.g.][]{2019ApJ...871..143P}. It better represents the central black hole than the radio core, which is the synchrotron self-absorbed jet base at the given frequency. We found Gaia data for 7 of our 9 sources. The remaining two (J1606+3124 and J2102+6015) appear too weak in optical, which is consistent with their identification as a galaxy (see below for a detailed discussion). 

In Figure~\ref{fig:col}, we marked the Gaia positions with crosses whose size represents the 3-sigma positional errors. By comparing optical positions with the mas-scale radio structure of the 7 sources, we found the following:  
\begin{itemize}
    \item J0048+0640: the Gaia position lies between the two components, NE and SW. Considering the comparable brightness of NE and SW components and their almost equal distances from the optical nucleus, we can classify J0048+0640 as a compact symmetric object \citep[CSO;][]{1982AA...106...21P}.
    \item J0753+4231: the optical position is close to the northernmost component NE2, so this AGN may be a core--jet source. The brightest VLBI component is NE1, with a separation of about 2~mas downstream along the jet. NE1 may be a moving knot or a standing shock in the jet.
    \item J1230+1139: the optical position is close to the brightest component C, but with a significant deviation. The situation is similar to that of J0753+4231. Moreover, there is a clear extension from component C toward the optical position, indicating that the radio core is weaker than the bright jet knot.
    \item J1316+6726: the Gaia position corresponds to the brightest radio component C within the positional uncertainty. This AGN is classified as a core--jet source. 
    \item J1421$-$0643, J1445+0968, and J1939$-$1002: core--jet sources with their optical nuclei and radio cores in good positional agreement. 
\end{itemize} 

Since the astrometric/geodetic snapshot observations found in the Astrogeo database are carried out in dual frequency bands (either at around 2 and 8~GHz, or at 4 and 8~GHz), the simultaneous dual-frequency observations with the same pointing centers enable us to produce spectral index maps for each target source. This can assist in the classification of their radio structure.
In making the spectral index maps, we first selected a set of 2/8-GHz ($S/X$-band) data with relatively high-quality images for each source, and performed the same hybrid mapping procedure. We set the same image size, pixel size, and restoring beam size for both bands. Model fitting was also performed on the self-calibrated 2-GHz visibility data to identify the optically thin components that were used for calculating the offset between the $S/X$ image brightness peaks, in order to properly align them. The spectral index maps were finally made using our \texttt{Python} script which follows similar calculations as done in the \textsc{AIPS} task \texttt{COMB}. Features with brightness lower than 3 times the rms noise in the 2-GHz maps were omitted to make the spectral index images clearer and more reliable.

The resulting spectral index images are shown in Figure~\ref{fig:spx} (The spectral index $\alpha$ is defined as $S \propto \nu^\alpha$, where $S$ is the flux density and $\nu$ the frequency). J0048+0640 shows a steep spectrum throughout the whole emission structure (mainly NE and SW components), reinforcing its classification as a CSO. The southern component of J0753+4231 has a steep spectrum and the northern component has a flat spectrum, which is consistent with its core--jet classification. In J1230+1139, J1316+6726, J1421$-$0643, J1445+0958, and J1939$-$1002, the flattest spectrum (or rising spectrum with positive spectral index) appears at one end of the radio structure, corresponding to the position of the optical nucleus, and the rest of the structure shows a steep spectrum. This information supports that the flat-spectrum components are associated with the radio cores and these sources are core--jet quasars. J1606+3124 lacks a distinct flat-spectrum component, and it is most likely a CSO \citep[see the discussion in][]{2022MNRAS.511.4572A}. Based on its VLBI images, J2102+6015 is a rather peculiar object. Both its eastern and western parts have relatively flat spectra. In the higher-resolution images, the eastern component is resolved into three sub-components \citep{2018AA...618A..68F}. However, none of the VLBI components have detectable proper motion \citep{2021MNRAS.507.3736Z}. The spectral index map supports that J2102+6015 could be a CSO candidate \citep[see][]{2021MNRAS.507.3736Z}.

In summary, the mas-scale morphology, the radio spectral index map, and, where Gaia data are available, the cross-match between the optical nucleus and radio components all together suggest the CSO classification of the radio structure in J0048+0640, J1606+3124, and J2102+6015. The other 6 high-redshift quasars are core--jet sources.

\section{Discussion}\label{sec:disc}

In this section, we focus on the jet proper motion and discuss the properties of the relativistic jets in our sample of high-redshift jetted AGNs.

\subsection{Core Brightness Temperature and Doppler Boosting}

Extremely high brightness temperatures observed in AGN cores, close to or above the inverse-Compton limit \citep{1969ApJ...155L..71K}, are usually considered as due to the beaming effect of relativistic jets pointing close to the observer's line of sight. VLBI observations can be used directly to measure the brightness temperature: \begin{equation}
\label{equ:tb}
T_{\mathrm B}=1.22 \times 10^{12} (1+z) \frac{S_{\nu}}{\theta_\mathrm{comp}^{2}\nu^{2}} \, {\mathrm K},
\end{equation}
where S$_{\nu}$ is the flux density of the VLBI component (in Jy), $\nu$ the observing frequency (in GHz), and $\theta_\mathrm{comp}$ is the diameter (full width at half-maximum, FWHM) of the circular Gaussian component (in mas).
The estimated $T_\mathrm{B,obs}$ values for the VLBI components in each source are listed in Table~\ref{tab:mod}.

The radio core is conventionally defined as the optically-thick section of the jet base in the vicinity of the central SMBH. For radio-loud quasars with a core-jet structure, the core is usually the brightest and most compact component in the VLBI image. Assuming that the magnetic field energy density and the particle energy density are in an equipartition state, the brightness temperature will have a maximum value called equipartition brightness temperature, $T_\mathrm{B,eq}$ \citep{1994ApJ...426...51R}. This can be considered as the intrinsic brightness temperature of the relativistic jet. Observed brightness temperatures above the limit ($T_\mathrm{B,eq} \approx 5\times 10^{10}$~K) are generally considered to be caused by the Doppler boosting effect of beamed jets. The Doppler factor can be estimated from the observed $T_\mathrm{B,obs}$ values as $\delta = T_\mathrm{B,obs}/T_\mathrm{B,eq}$. 

We must be aware that the equipartition estimate of the Doppler factor is valid for the optically thick parts of the jet, where the frequency of the $T_\mathrm{B,obs}$ measurement should be close to the spectral peak of the source. For the high-redshift quasars in our sample, the observed 8.4~GHz corresponds to a rest-frame frequency of $\sim 40$~GHz, which most likely exceeds the spectral peak frequencies. To estimate the Doppler factors of our sample, we adopted the following approach to obtain appropriate values. Firstly, we used the $T_\mathrm{B,obs}$ values measured from all the 8-GHz ($X$-band) epochs to calculate an average brightness temperature for each source. During the averaging calculations, the extreme $T_\mathrm{B,obs}$ values were omitted, reducing the impact of possible model-fitting biases due to some potentially poor-quality data. The average $T_\mathrm{B,obs}$ is likely more characteristic to the quiescent states of the AGN core. The resulting average $T_\mathrm{B,obs}$ for each source can be found among the individual comments on the sources (Appendix~\ref{sec:appa}).

Recently, \citet{2020ApJS..247...57C} estimated the correlation between the observed $T_\mathrm{B,obs}$ and the frequencies, based on large VLBA surveys of compact radio AGNs at multiple frequencies. They found that the spectral peaks are at around 7~GHz in the rest frame of the sources. Based on the empirical correlation from \citet{2020ApJS..247...57C}, and using the equipartition brightness temperature as the intrinsic brightness temperature that is valid at the spectral peak frequency, we can then estimate the intrinsic brightness temperature $T_\mathrm{B,int}$ for each of our observed sources, taking the actual rest-frame frequencies into account. The extrapolated $T_\mathrm{B,int}$ values are in the range of $(1.1-1.4) \times 10^{10}$~K for the sources in our sample.

From the estimates above, we find that all the six core--jet sources in our sample have high core brightness temperatures exceeding the extrapolated $T_\mathrm{B,int}$, confirming that they contain highly relativistic jets with Doppler-boosted radio emission. We calculated their Doppler factors that can be found in Table~\ref{tab:kparm}.
The brightness temperatures of J1316+6726, J1445+0958, and J1939$-$1002 are high and these values have prominent temporal variations. For J0753+4231, the brightest jet component NE1 has the highest brightness temperature. The three CSOs, J0048+0640, J1606+3124, and J2102+6015, do not have identifiable cores, but the brightness temperatures of their hot spots exceed $T_\mathrm{B,eq}$, suggesting that the equipartition assumption probably does not apply in these regions. The brightness temperatures of J0048+0640 hot spots show a significant decrease at epochs 2017 September 18 and 2018 January 18, while a significant increase in the component size occurs. This can be explained by the adiabatic expansion of the hot spots.

\subsection{Jet Proper Motion}

Since the cosmological time dilation is proportional to $(1+z)$, the jets in the high-redshift sample appear changing slower. Thus reliable component proper motion measurements require VLBI observations spanning a longer time interval in the observer's frame \citep[e.g.][]{2015MNRAS.446.2921F, 2018MNRAS.477.1065P,2020NatCo..11..143A,2020SciBu..65..525Z}. 

Using our fitted Gaussian model components, we measured apparent proper motions based on the time evolution of the separation of core and jet components. For components that are too close to the core, their fitted positions may be affected by the finite restoring beam size and perhaps newly ejected features. On the other hand, components too far away from the core ($\gtrsim 10$~mas) are usually weak and diffuse, their positional uncertainties are too large. Therefore, we excluded these components from our proper motion determination.

Finally, we derived proper motions for 18 jet features from 9 target sources. The results are shown in Table~\ref{tab:pm}.
We fitted the linear proper motion along the RA ($\mu_x$) and Dec ($\mu_y$) directions separately, using a least-squares method, and then calculated the total proper motion as $\sqrt{\mu_x^2+\mu_y^2}$. For the core--jet sources, we calculated the rate of change of the jet component position relative to the core with time. For CSO sources, we calculated the separation speed of the two opposite hot spots using a particular terminal hot spot as a reference. Assuming that both hot spots advance with the same speed, the advance speed of a particular hot spot is half of the calculated separation speed.

Figure~\ref{fig:pm} demonstrates the jet component trajectories and proper motion fitting results of the VLBI components in the sample. For each source, we selected the most appropriate jet components for proper motion measurements: they have data from the most epochs, are clearly distinguishable in VLBI images, and have the smallest positional errors. 

The detailed radio properties of two CSOs, J1606+3124 and J2102+6015, have been presented before \citep{2022MNRAS.511.4572A,2021MNRAS.507.3736Z}, so we only briefly summarize the results here. The separation speed between terminal hot spots S and N in J1606+3124 is $0.013 \pm 0.002$~mas\,yr$^{-1}$ ($1.60\pm0.25\,c$), and between the hot spot S and the inner jet knot C is $0.006 \pm 0.002$~mas\,yr$^{-1}$ ($0.74 \pm 0.25\,c$). This leads to a hot spot advance speed of $0.8\,c$ \citep{2022MNRAS.511.4572A}. The separation speed between the eastern and western components in J2102+6015 is $0.023 \pm 0.011$~mas\,yr$^{-1}$ ($2.8 \pm 1.4\,c$) \citep{2021MNRAS.507.3736Z}. In the third CSO in our sample, J0048+0640, we obtained a separation speed of $0.005\pm0.002$~mas\,yr$^{-1}$ ($0.6\pm0.2\,c$), giving a hot spot advance speed of $\sim0.3\,c$.

The apparent jet component speeds in the core--jet sources are in the range of $1.4-17.5\,c$, consistent with the known proper motions in low-redshift radio-loud quasars \citep[e.g.][]{2012ApJ...758...84P,2019ApJ...874...43L}. The maximum speed is close to that of the fastest high-$z$ jet observed before \citep{2020SciBu..65..525Z}, but lower than the maximum value found in low-redshift AGNs. The quasars J1230$-$1139, J1316+6726, and J1445+0958 contain fast-moving components in the outer part of the jets. These large proper motions could be caused by the projection effect of a large  jet bending.

\subsection{Lorentz Factors and Viewing Angles}

High-redshift blazar sources are potentially valuable for studying the cosmological evolution of radio source number density and SMBH accretion. Understanding the distribution of the jet Lorentz factors is fundamental for assessing the number density of AGNs with jets misaligned with respect to the line of sight. These constitute the mostly hidden parent population of highly-beamed jetted sources (blazars). The latter objects are more easily detectable in flux density-limited observations, because of their Doppler-boosted emission. Kinematic measurements of high-redshift AGN jets and estimates of jet Lorentz factors such as presented here are still very rare. For sources at lower redshifts, \citet{2019ApJ...874...43L} found that the distribution of Lorentz factors peaks between $\Gamma=5-15$, with a shallow tail reaching $\Gamma \approx 50$. The misaligned parent population of low-$\Gamma$ jetted sources is larger because those jets require very small viewing angles to be detected as blazars.

The Lorentz factor can be obtained by fitting the broad-band spectral energy distribution (SED) of a blazar \citep[e.g.][]{1997A&A...324..395B}, or directly from VLBI observations. Based on the Doppler factor and the apparent superluminal speed of the jet derived from the VLBI observations, we can estimate the bulk Lorentz factor and viewing angle \citep[see Eqs. B5 and B7 in][]{1993ApJ...407...65G}. The bulk Lorentz factors and jet viewing angles are calculated for five core--jet sources (with the exception of J1316$+$6726, due to the lack of valid proper motion measurements). The results are presented in Table~\ref{tab:kparm}, along with values for other high-$z$ radio quasars taken from the literature. 
Two sources, J1230$-$1139 and J1421$-$0643, show relatively large Lorentz factors, which exceed the typical maximum values \citep[$\Gamma>20$, e.g.][]{2004ApJ...609..539K} based on the $\beta_\mathrm{app,max}$ for the studied high-redshift sources. In the relativistic beaming model \citep[see Appendix A in][]{1995PASP..107..803U}, this can happen when the viewing angle is very small (i.e. $\theta < \theta_\mathrm{crit}$, where the critical angle $\theta_\mathrm{crit} = \arcsin (\Gamma^{-1})$). 
In such cases, a minor change in the viewing angle would greatly increase the jet apparent speed. An alternative possibility could be that the sizes of the VLBI cores are overestimated, which could be caused by the overlap of the emitting components or variability. This will lead to  a decrease in the estimated Doppler factor. Both cases above are common in blazars with core--jet structures.
These observational evidences, including higher Lorentz factors and smaller viewing angles, support that these sources belong to the blazar class.

\floattable
\begin{deluxetable}{ccccccc}
\tablecaption{VLBI kinematic parameters measured for high-$z$ radio quasars\label{tab:kparm}}
\tablehead{
\colhead{Name} & \colhead{$z$} & \colhead{$\beta$} & \colhead{$\delta$} & \colhead{$\Gamma$}& \colhead{$\theta$} & \colhead{Ref.}  }

\colnumbers
\startdata
J0753$+$4231&3.59&1.8$\pm$0.1&6.0$\pm$1.2 &3.4$\pm$0.5 &5.3$\pm$2 & 1 \\
J1230$-$1139&3.53&10.3$\pm$0.8&2.4$\pm$1.9 &23.3$\pm$16.9 &10.5$\pm$1.1 & 1 \\
J1421$-$0643&3.69&15.9$\pm$0.3&3.0$\pm$1.5 &44.2$\pm$20.9 &6.9$\pm$0.3 & 1 \\
J1445$+$0958&3.55&12.9$\pm$0.6&11.1$\pm$4.6 &13.1$\pm$1.1 &5.1$\pm$1.8 & 1 \\
J1939$-$1002&3.79&0.5$\pm$0.1&16.5$\pm$5.2 &8.3$\pm$2.6 &0.2$\pm$0.2 & 1 \\
\hline 
J0906$+$6930& 5.47& 2.5$\pm$0.8  & 6.0$\pm$0.8 & 3.6$\pm$0.5 &6.8$\pm$2.2  &2\\
J1026$+$2542& 5.27& 12.5$\pm$ 3.9  & $\gtrsim$ 5 &$\gtrsim$ 12.5  &4.6& 3\\
J1430$+$4204& 4.72& 19.5$\pm$1.9  & 22.2$\pm$10.6 & 14.6$\pm$3.8 & 2.2$\pm$1.6 & 4\\ 
J2134$-$0419& 4.33& 4.1$\pm$ 2.7  & 3--8 &2--7  & 3--20 & 5 \\
\enddata
\tablecomments{Columns: (1) J2000 source name; (2) redshift; (3) apparent transverse speed of the fastest jet component, in the unit of $c$; (4) Doppler factor estimated from the VLBI core brightness temperature; (5) bulk Lorentz factor; (6) jet viewing angle in $\degr$; (7) reference for the estimated parameters: 1--this paper, 2--\citet{2020NatCo..11..143A}, 3--\citet{2015MNRAS.446.2921F},  4--\citet{2020SciBu..65..525Z}, 5--\citet{2018MNRAS.477.1065P}}
\end{deluxetable}

\subsection{High-redshift Quasars in the Apparent Proper Motion--Redshift Diagram}

Except for J0906+6930 \citep{2017MNRAS.468...69Z,2020NatCo..11..143A}, J1026+2542 \citep{2015MNRAS.446.2921F}, J1430+4204 \citep{2020SciBu..65..525Z}, and J2134$-$0419 \citep{2018MNRAS.477.1065P}, there are no robust proper motion measurements in other quasar jets at $z>3.5$. The main reason is the lack of known sources that are sufficiently bright with prominent well-resolved mas-scale radio jet structure. Also, because of the cosmological time dilation, reliably detecting positional changes in jet components requires decades-long history of VLBI monitoring observations. Thanks to the accumulated astrometric snapshot VLBI data, here we could derive jet proper motions for another 6 high-redshift blazars. This increases the available proper motion sample in the early Universe substantially (Table~\ref{tab:kparm}).

\begin{figure*}
	\centering
	\includegraphics[width=0.8\textwidth]{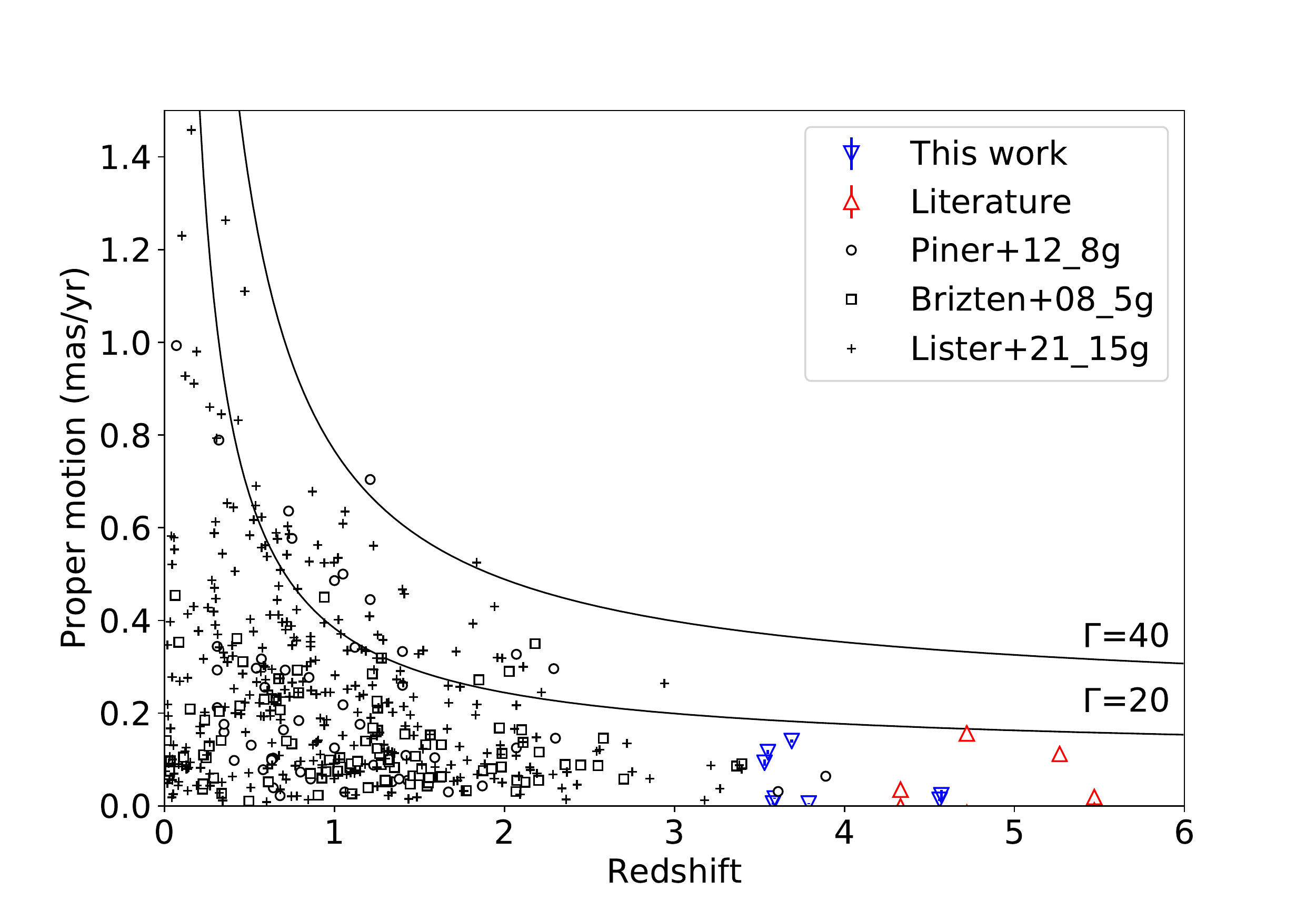}
	\caption{The apparent proper motion--redshift ($\mu-z$) diagram based on the latest VLBI observations of radio jets. Down triangles: this work; up triangles: from the literature, see Table~\ref{tab:kparm};  circles:   \citet{2012ApJ...758...84P} measured at 8.4~GHz; squares: \citet{2008AA...484..119B} measured at 5~GHz; plus signs: \citet{2021ApJ...923...30L} measured at 15~GHz.
		The solid lines represent the maximum apparent speed $\beta_\mathrm{app,max}$ with the given bulk Lorentz factors $\Gamma = 20$ and 40.}
	\label{fig:pm-z}
\end{figure*}

An even larger sample of high-redshift jet proper motions could eventually become useful to constrain cosmological model parameters through the apparent proper motion--redshift ($\mu-z$) relation \citep[e.g.][]{1988ApJ...329....1C,1994ApJ...430..467V,1999NewAR..43..757K}. Earlier studies based on large but lower-redshift source samples (Figure~\ref{fig:pm-z}) found that the upper bound on the $\mu-z$ relation is consistent with the predictions of the $\Lambda$CDM cosmology and a distribution of jet Lorentz factors where the vast majority of the jets have $\Gamma \lesssim 25$. For a given Lorentz factor, the apparent jet component speed cannot exceed $\sqrt{\Gamma^2-1} \approx \Gamma$ \citep[e.g.][]{1995PASP..107..803U}. To better populate the high-redshift region, we added our own proper motion measurements to the $\mu-z$ diagram constructed from literature data (Figure~\ref{fig:pm-z}). Large VLBI surveys \citep{2008AA...484..119B,2012ApJ...758...84P,2021ApJ...923...30L} at lower redshifts, as well as measurements for individual high-redshift quasars are considered. From our sample studied here, only the highest jet component speeds of each source are plotted. Note that VLBI observations made at different frequencies (ranging from 5 to 15~GHz) are collected in Figure~\ref{fig:pm-z}. This may result in systematically  different apparent proper motion values \citep[e.g.][]{2004ApJ...609..539K}. Nevertheless, the general trend in which our new measurements also fit is clearly seen: jets with the fastest apparent proper motion, $\beta_\mathrm{app} \gtrsim 40$, are only found at $z \lesssim 2$. Given that our estimated Lorentz factors reach about 40 (Table~\ref{tab:kparm}), it is expected that with larger samples of high-redshift quasars, their apparent proper motions could be as high as $\mu \approx 0.4$~mas\,yr$^{-1}$ (Figure~\ref{fig:pm-z}), without violating the current cosmological paradigm and without requiring extremely high bulk Lorentz factors in the jet.
Our study helps populating the high-redshift end of the apparent proper motion--redshift diagram with reliable jet proper motions measured with VLBI.

\section{Summary}
\label{sec:sum}

In this paper, we reported new 8.4-GHz VLBA observations of 9 high-redshift ($z>3.5$) jetted quasars. Based on archival dual-band (usually 2- and 8-GHz) astrometric VLBI data and our high-quality VLBA data taken in 2017, we presented high-resolution radio images and spectral index maps of each source in the sample. Accurate optical positions from Gaia were also considered for 7 out of the 9 sources, in the context of their mas-scale radio structure.
By fitting the source visibility data with circular Gaussian brightness distribution model components, and using appropriate component identifications across multiple observing epochs, kinematic properties of the jets were derived.

In our sample, six sources (J0753+4231, J1230$-$1139, J1316+6726, J1421$-$0643, J1445+0958, and J1939$-$1002) were classified as core--jet blazars. These sources have flat-spectrum (or inverted-spectrum) features at the brighter end of the radio structure (the so-called cores), which also closely correspond to the  position of the optical nucleus where Gaia measurements are available. The brightness temperatures of their VLBI cores all exceed the estimated equipartition threshold. In the framework of the relativistic beaming model, we also estimated the jet kinematic and geometric parameters (Doppler factor, Lorentz factor, and jet viewing angle) of these core--jet blazars (except for J1316+6726 due to the lack of valid proper motion measurements). The results are presented in Table~\ref{tab:kparm}.

Three sources in our sample (J0048+0640, J1606+3124, and J2102+6015) were classified as CSOs or CSO candidates.
The spectral index maps of J0048+0640 and J1606+3124 show steep-spectrum emission at both sides of their jet ends. For J2102+6015, its eastern and western features have relatively flat (or inverted) spectra, and none of the VLBI components have detectable proper motion \citep[see also ][]{2021MNRAS.507.3736Z}. 
The maximum jet proper motion detected in our sample is $\sim 0.15$~mas\,yr$^{-1}$, corresponding to an apparent jet speed of $\sim18\,c$. The range of our jet proper motions shows good consistency with low-redshift quasars, where large values of $\beta_\mathrm{app}$ only appear at low redshifts \citep[e.g.][]{2012ApJ...758...84P,2019ApJ...874...43L}.
Our study substantially increases the sample of high-redshift radio quasars with reliable jet proper motions measured with VLBI. It may serve as an important starting point for accumulating data for future studies of high-redshift AGN jets.

\begin{acknowledgments}
This research was funded by National Key R\&D Programme of China under grant number 2018YFA0404603 and the China--Hungary project funded by the the Ministry of Science and Technology of China and Chinese Academy of Sciences. YKZ was sponsored by Shanghai Sailing Program under grant number 22YF1456100. This research was supported by the Hungarian National Research, Development and Innovation Office (NKFIH), grant number OTKA K134213. This work is supported by the Ministry of science and higher education of Russia under the contract 075-15-2022-262.
The National Radio Astronomy Observatory is a facility of the National Science Foundation operated under cooperative agreement by Associated Universities, Inc. We acknowledge the use of data from the Astrogeo Center database maintained by Leonid Petrov.
\end{acknowledgments}

\facility{VLBA}

\software{\textsc{AIPS} \citep{2003ASSL..285..109G}, \textsc{Difmap} \citep{1997ASPC..125...77S}}

\appendix

\section{Comments on Individual Sources}
\label{sec:appa}

\subsection{J0048$+$0640}

The first VLBI image of this source was made at 5~GHz from a European VLBI Network (EVN) observation in 1996. It showed a double-component source extended along the northeast--southwest direction \citep{1999AA...344...51P}. The two components were separated by $\sim 3.8$~mas,  consistently with our 8.4-GHz imaging results.
The source is marginally resolved at 2.3~GHz and clearly resolved into two components at 8.4~GHz (Figure \ref{fig:col}).
In producing the spectral index image, we used the midpoints of the 2.3- and 8.4-GHz images for the alignment (the two components can be extracted from the best-quality 2.3-GHz data with model fitting).
Both the NE and SW components exhibit flat spectra and high brightness temperatures. The average $T_\mathrm{B}$ is $7.0 \pm 1.3 \times 10^{10}$~K for NE and $4.6 \pm 3.9 \times 10^{10}$~K for SW, with the Gaia nucleus positioned in between. The possibility of gravitational lensing can be ruled out because the two components would violate the preservation of surface brightness \citep[see e.g.][]{2019A&A...630A.108S,2021MNRAS.507L...6C}.
Considering the source compactness, it is very unlikely to be a dual AGN, but this scenario is difficult to rule out with observations at other wavebands because of the insufficient resolution.
The most likely explanation is that J0048$+$0640 is a CSO.
No significant proper motion was detected between the two components over a 14-yr time baseline.
Recent multi-frequency total flux density measurements of $z>3$ AGNs with the RATAN-600 radio telescope show a peaked spectrum for J0048$+$0640 with $\nu_\mathrm{peak} \approx 4$~GHz in the observer's frame \citep{2021MNRAS.508.2798S}, reinforcing that it is a young CSO source.

\subsection{J0753$+$4231}

The source was included in the CSO candidate sample of CSOs observed in the northern sky (COINS) and VLBA Imaging and Polarimetry Survey (VIPS) \citep{2000ApJ...534...90P,2007ApJ...658..203H,2016MNRAS.459..820T}. In recent studies of the radio spectra of high-$z$ quasars, this source was also identified as a candidate MHz-peaked spectrum source with $\nu_\mathrm{peak} < 1$~GHz in the observer's frame \citep[e.g.][]{2013AstBu..68..262M,2021MNRAS.508.2798S}.
In our study, J0753$+$4231 shows a double structure in both the 2.3- and 8.4-GHz images. The Gaia position and the spectral index map support that the source is a one-sided core--jet quasar rather than a CSO. The core should be located in the outermost NE2 component. 
The average brightness temperatures $T_\mathrm{B}$ of NE1 and NE2 are $8.4 \pm 1.7 \times 10^{10}$~K and $1.4 \pm 0.4 \times 10^{10}$~K, respectively.
We estimate the maximum proper motion of the jet to be $0.018 \pm 0.001$~mas\,yr$^{-1}$ ($1.8 \pm 0.1\,c$) using NE2 (core) as the reference point.
In a previous study, \citet{2008AA...484..119B} also detected four components at 5~GHz, and conducted proper motion measurements based on three VLBI epochs, leading to a maximum speed $\sim 0.1$~mas\,yr$^{-1}$. Compared to their proper motions, our measured value is much smaller, but it is based on a longer time period (22~yr) and has higher accuracy.

\subsection{J1230$-$1139}

The 8.4-GHz image of this source shows a prominent curved jet that bends from the west to the southwest at a projected distance of about 45~pc. Two components, J1 and J2, are detected in the curved southwest jet section, and J3 is located where the jet bends. The brightest component is C, but it is not a radio core. The extension east of C shows a flat spectrum and should host the core. The maximum proper motion is $\sim 0.10$~mas\,yr$^{-1}$, corresponding to an apparent transverse speed of $\sim 12\,c$. The mean $T_\mathrm{B}$ of this source is $3.4 \pm 2.7 \times 10^{10}$~K. The relatively lower brightness temperature may be due to the presence of self-absorption in the core, resulting in an underestimate of the flux density and an overestimate of its size. The estimated viewing angle of the innermost jet is around $10\degr$, derived from the Doppler factor and maximum apparent jet speed, classifying it as a blazar. 
The source also shows significant variability, consistent with its blazar identification. It shares similar high-resolution jet properties with other high-$z$ blazars \citep[e.g.][]{2020SciBu..65..525Z,2020NatCo..11..143A}. 

\subsection{J1316$+$6726}

The redshift of this source is from photometric measurements, but with a very high probability \citep[$\sim 95\%$,][]{2009ApJS..180...67R}. Since the jet is well-separated from the core, we included it in our sample and tried to check its high-$z$ nature by determining its jet proper motions following the method introduced by \citet{2020MNRAS.497.2260A}.
In the VLBI images of this source, a weak and extended jet can be found at around 8~mas southeast of the bright core component, with a sharp bending at a projected distance of 60~pc from the core. The jet is only marginally detected at 3 available epochs, and we did not detect a significant (i.e. $\ge 3\sigma$) outward motion due to the significant positional errors and the short time period.
The mean $T_\mathrm{B}$ of the core component is $13.3 \pm 5.5 \times 10^{10}$~K, indicating strong non-thermal emission.
The apparent inward motion needs to be verified by future observations. If this proper motion is caused by the motion of a newly-formed jet component in the core, then the new jet speed along the jet direction is $14.2\,c$, which does not exceed the expected proper motion upper limit for a high-$z$ jet \citep[see an example of the opposite situation in][]{2020MNRAS.497.2260A}, and thus its photometric redshift can be considered plausible.

\subsection{J1421$-$0643}

This source shows a typical core--jet radio structure. The mean $T_\mathrm{B}$ is $4.2 \pm  2.1 \times 10^{10}$~K.
This source is a prominent high-$z$ blazar. Previous studies have discovered large kpc-scale radio and X-ray jets extending towards the northeast of this nucleus \citep{2020MNRAS.497..988W}.
Our VLBI images reveal the pc-scale jet structure, showing three jet components moving away from the core radially toward the northeast direction. The position angle of the pc-scale jet is consistent with that of the kpc-scale jet (about $30\degr$). The estimated maximum jet proper motion is $\sim 0.15$~mas\,yr$^{-1}$ ($\sim 16\,c$).

\subsection{J1445$+$0958}

This object, also known as OQ\,172, has a rich jet structure starting toward the west and turning the south. It was once identified as a GHz-peaked spectrum (GPS) galaxy due to the less variable radio and optical emission, the non-flat radio spectrum and the lack of pc-scale structural evolution \citep[see][and the references therein]{2015ApJ...812...79P}.
Previous multi-band VLBI observations revealed a clockwise jet trajectory from high-frequency to low-frequency images, which was attributed to the interaction between the relativistic jet and the narrow-line region medium \citep{2017MNRAS.468.2699L}.
In our image at 8.4~GHz, the emission of the jet that bends from west to south was modeled and analyzed. Since the jet is becoming weak and diffuse in the southern tail, barely detected in the less sensitive archival snapshot observations, we only fit the brighter jet section in the northern part (Figure~\ref{fig:col}).
From seven epochs spanning 21~yr, we were able to obtain jet proper motion in J1445$+$0958 for the first time. Among the three jet components, J1 stands out as the fastest one, with a proper motion of 0.13 $\pm$ 0.01~mas\,yr$^{-1}$.
The mean $T_\mathrm{B}$ of this source is $15.5 \pm 6.4 \times 10^{10}$~K, suggesting a highly beamed jet.
From the spectral index map, the high brightness temperature, and the Gaia optical position, the source could be a blazar. Although its GPS-type radio spectrum is inconsistent with the typically flat spectrum of low-redshift blazars, it is not unprecedented in high-redshift quasars \citep[e.g.][]{2021MNRAS.508.2798S}.

\subsection{J1606$+$3124}

This source was previously identified as a flat-spectrum radio quasar
\citep[][]{2007ApJS..171...61H,2007AA...469..451T,2016MNRAS.463.3260C}. However, further radio spectral studies based on simultaneous multi-frequency observations suggested that the source is a GPS radio galaxy in the early Universe \citep[RATAN-600,][]{2012AA...544A..25M,2019AstBu..74..348S}.
In our study, by analyzing the spectral index map and the archival radio spectra, we conclude that the source could be the a high-redshift CSO source.
Additional results and a detailed discussion of the CSO identification have been presented elsewhere \citep[][]{2022MNRAS.511.4572A}.

\subsection{J1939$-$1002}
The VLBI morphology resembles a core--jet radio source. The two compact jet components lie close to the bright core, and a distant jet component (J1) is in the northeastern direction at a distance of about 22~mas. This distant feature shows a slight change in the position angle. It could possibly be explained by the rise of the Doppler-boosted region caused by the helical jet path, which is usually seen in flat-spectrum radio quasars (FSRQs) \citep[e.g.][]{2000A&A...361..529A,2004A&A...417..887H}. However, the possibility of jet interaction with the surrounding interstellar medium (ISM) cannot be ruled out either \cite[e.g.][]{2000Sci...289.2317G,2020NatCo..11..143A}. The mean $T_\mathrm{B}$ of this source is $21.4 \pm 6.7 \times 10^{10}$~K, which well exceeds the equipartition value and indicates strong Doppler boosting with $\delta \ge 4$. 

Since the jet component J1 is weak and diffuse at most available epochs, we just measured proper motions of J2 and J3.
The resulting jet proper motion is $-0.014 \pm 0.001$~mas\,yr$^{-1}$ for J2 and $0.005\pm 0.001$~mas\,yr$^{-1}$ for J3.
The apparent inward jet motion is physically meaningless. Despite position errors, the curved jet motion across the line of sight and the newly-emerging features that move beyond the time separations among the available epochs
could be the possible reasons for the negative proper motion of J2 \citep[see e.g.][]{2013AJ....146..120L,2016AJ....152...12L}.

\subsection{J2102$+$6015}

Previous studies of this source claim it as an FSRQ with moderate Doppler boosting effects \citep{2016MNRAS.463.3260C}, but recent works prefer J2102$+$6015 to be a GPS radio source \citep[e.g.][]{2017MNRAS.467.2039C,2018AA...618A..68F}. 
We find further interesting characteristics of this source. 
Considering the high-resolution images, the spectral index behavior, and the component separation speeds, we believe this high-$z$ AGN is most likely a CSO.
From the deep imaging and model-fitting results, we found that neither the E nor the W features are unresolved. To further constrain the source nature, we collected more high-resolution VLBI data from further epochs, and used somewhat more conservative positional error estimates (i.e. one-tenth of the restoring beam size) to conduct component identification and proper motion estimates.
A more robust proper motion estimate of 0.023~$\pm$~0.011~mas\,yr$^{-1}$ was obtained. The details are presented in a separate paper \citep{2021MNRAS.507.3736Z}.

\listofchanges

\section{Tables and Figures}
\label{sec:appb}
\begin{figure*}
  \centering
  \begin{tabular}{ccc}
	\includegraphics[width=0.3\textwidth]{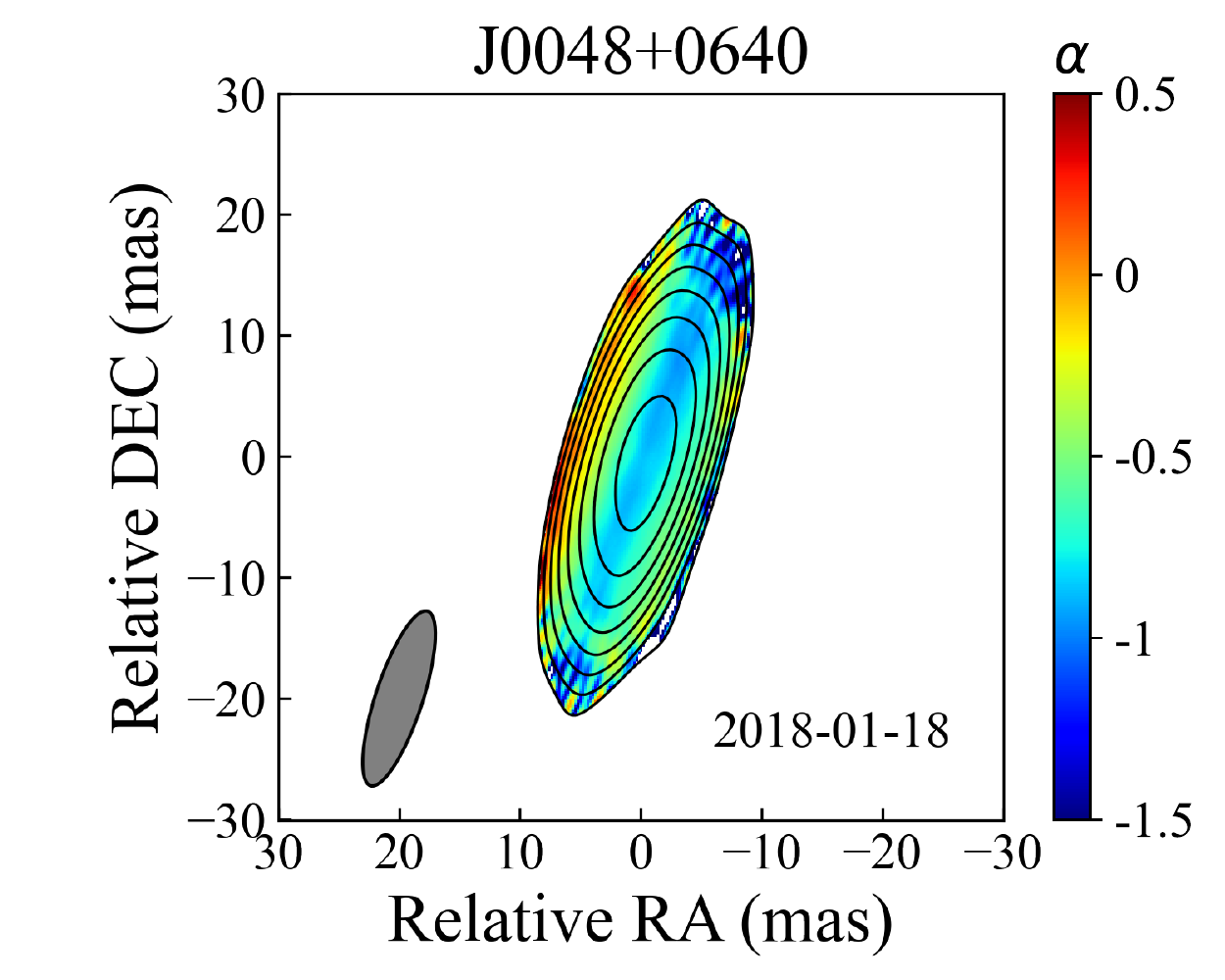}&
	\includegraphics[width=0.3\textwidth]{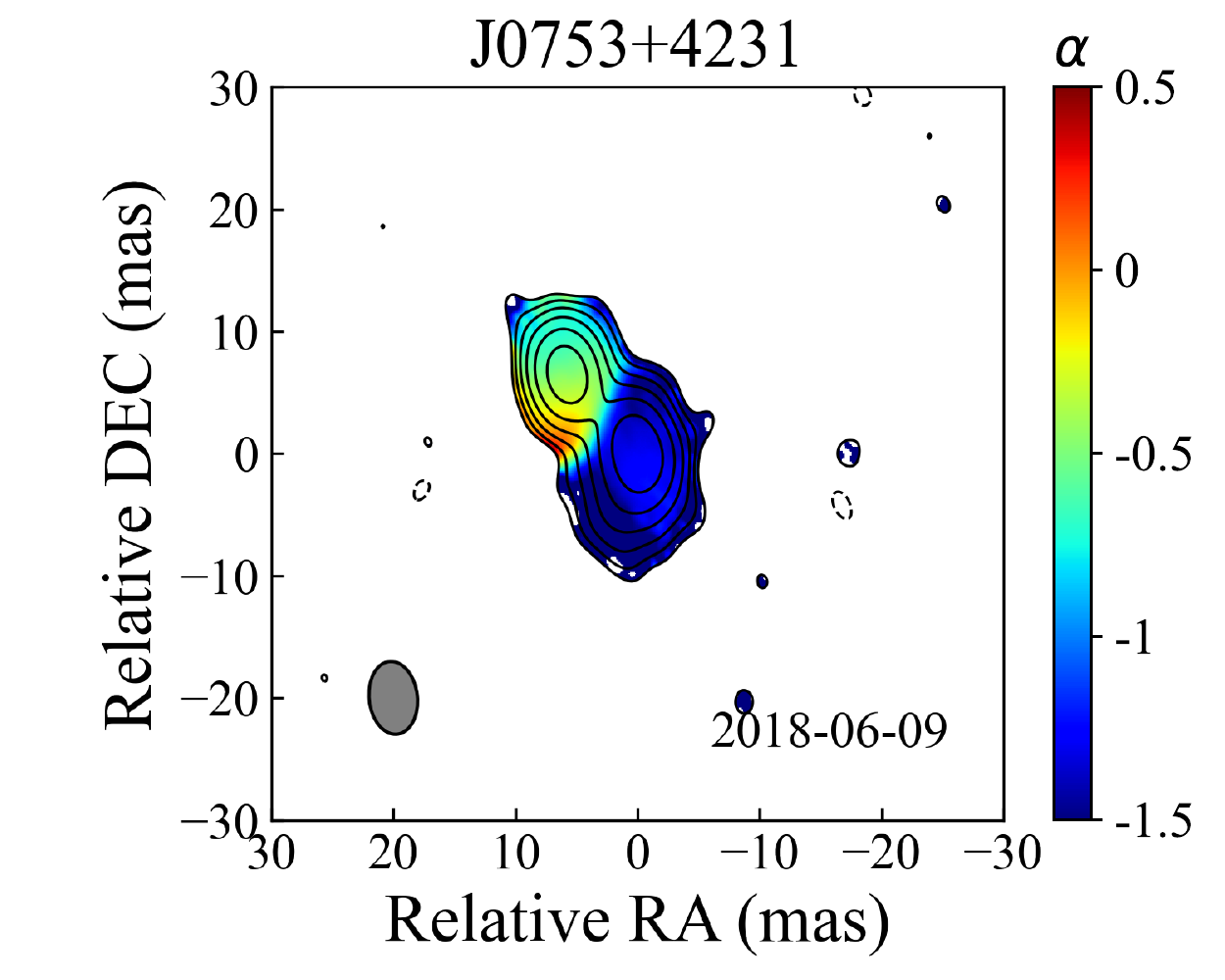}&
	\includegraphics[width=0.3\textwidth]{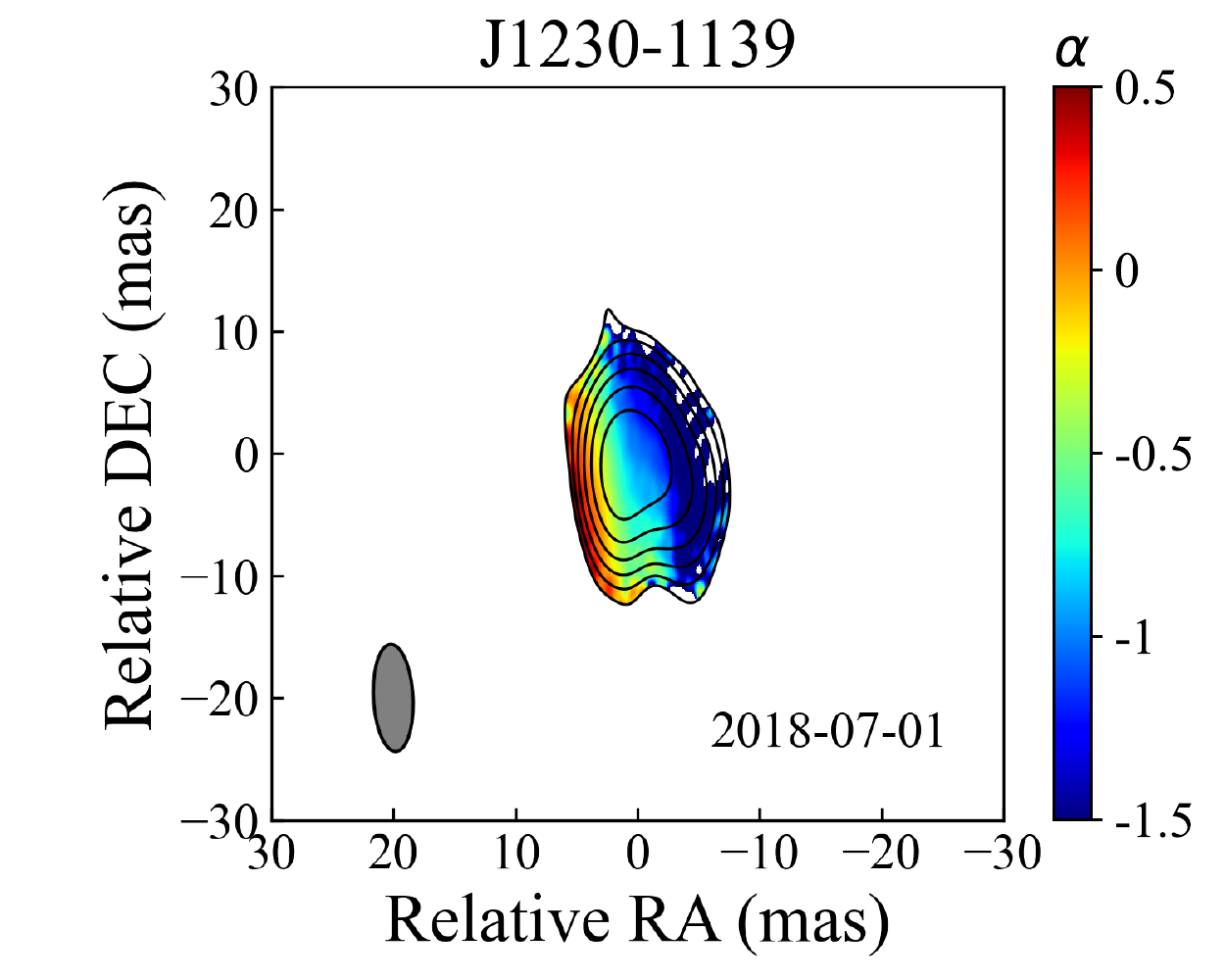}
  \end{tabular}
  \begin{tabular}{ccc}
	\includegraphics[width=0.3\textwidth]{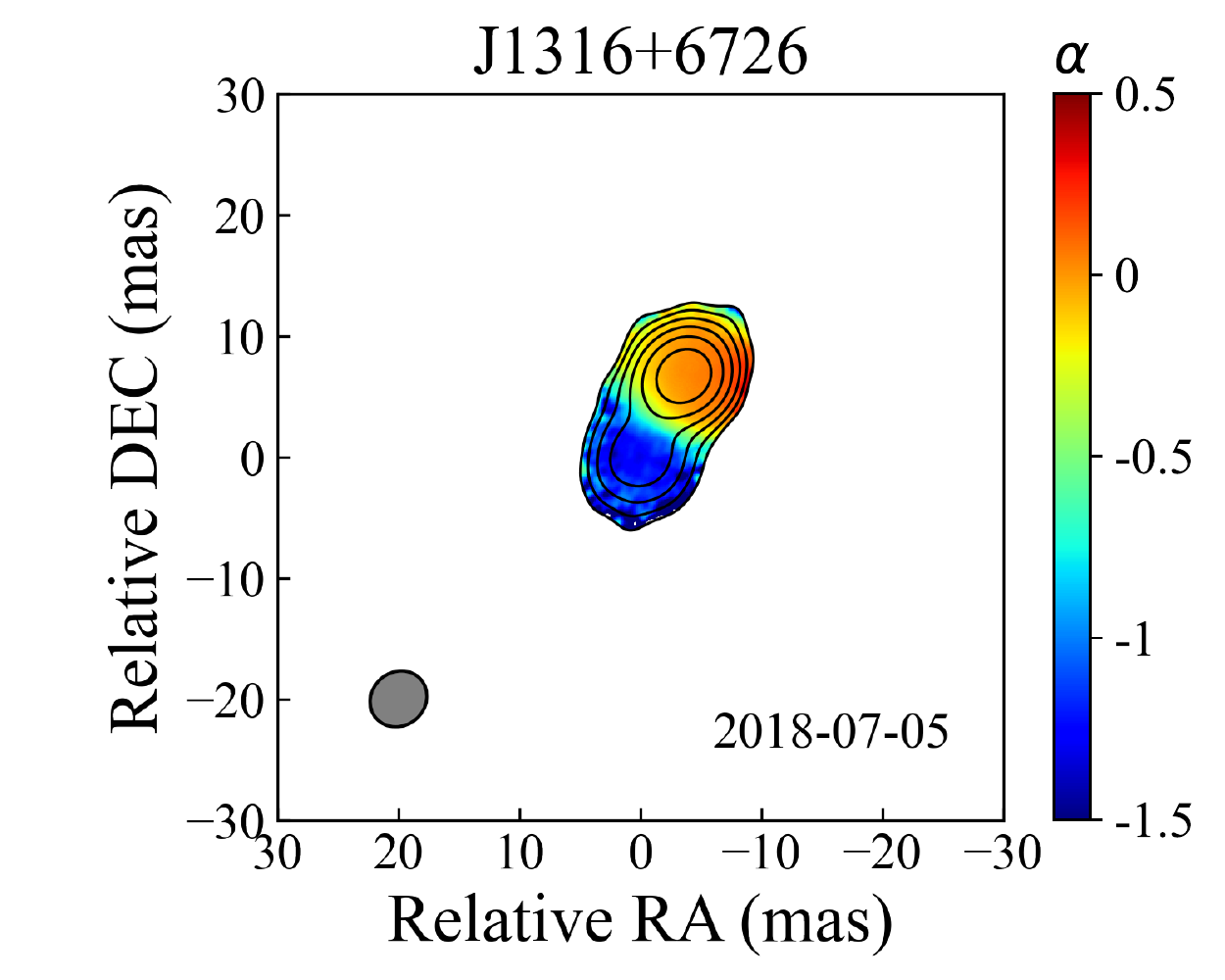}&
	\includegraphics[width=0.3\textwidth]{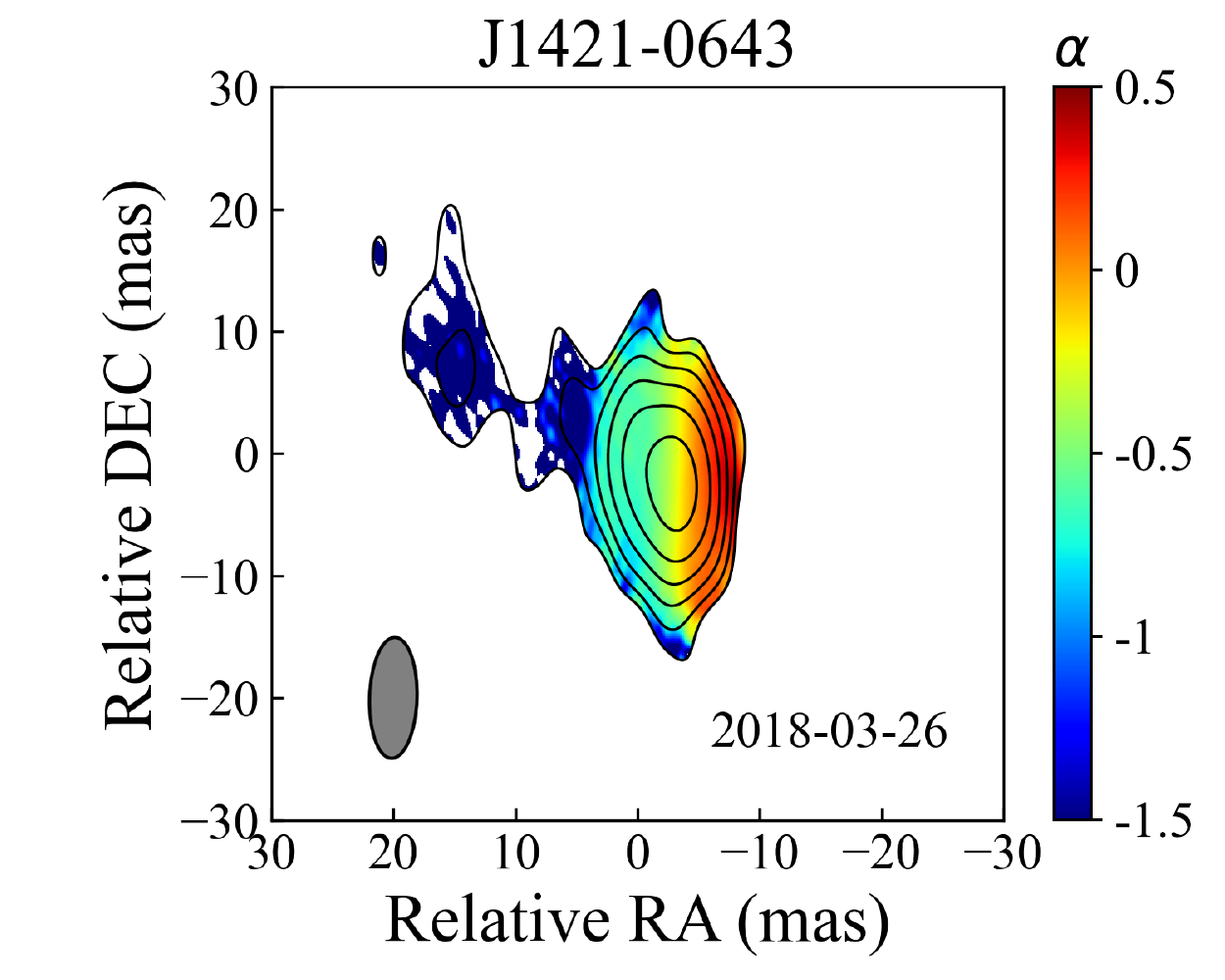}&
	\includegraphics[width=0.3\textwidth]{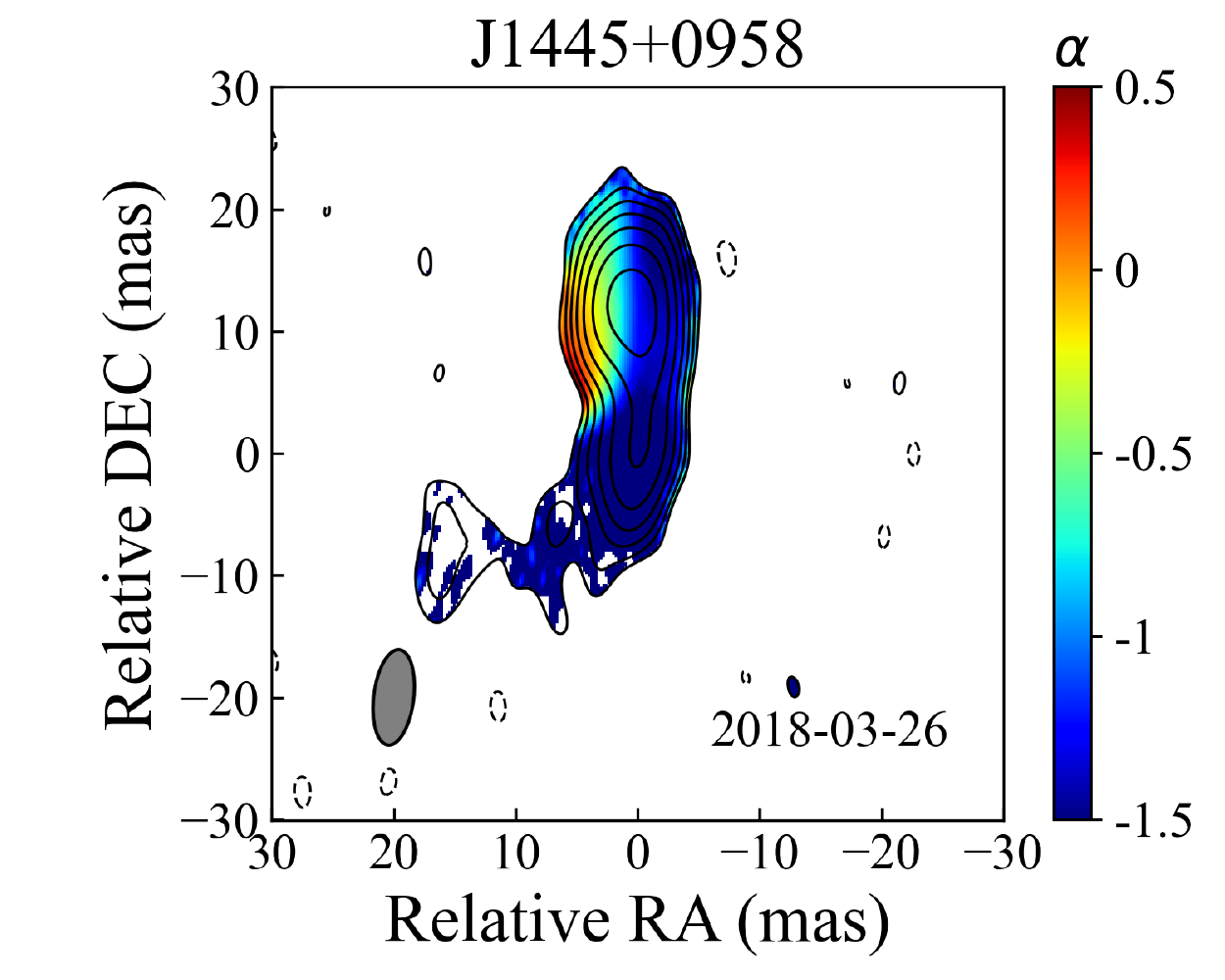}
  \end{tabular}
  \begin{tabular}{ccc}
	\includegraphics[width=0.3\textwidth]{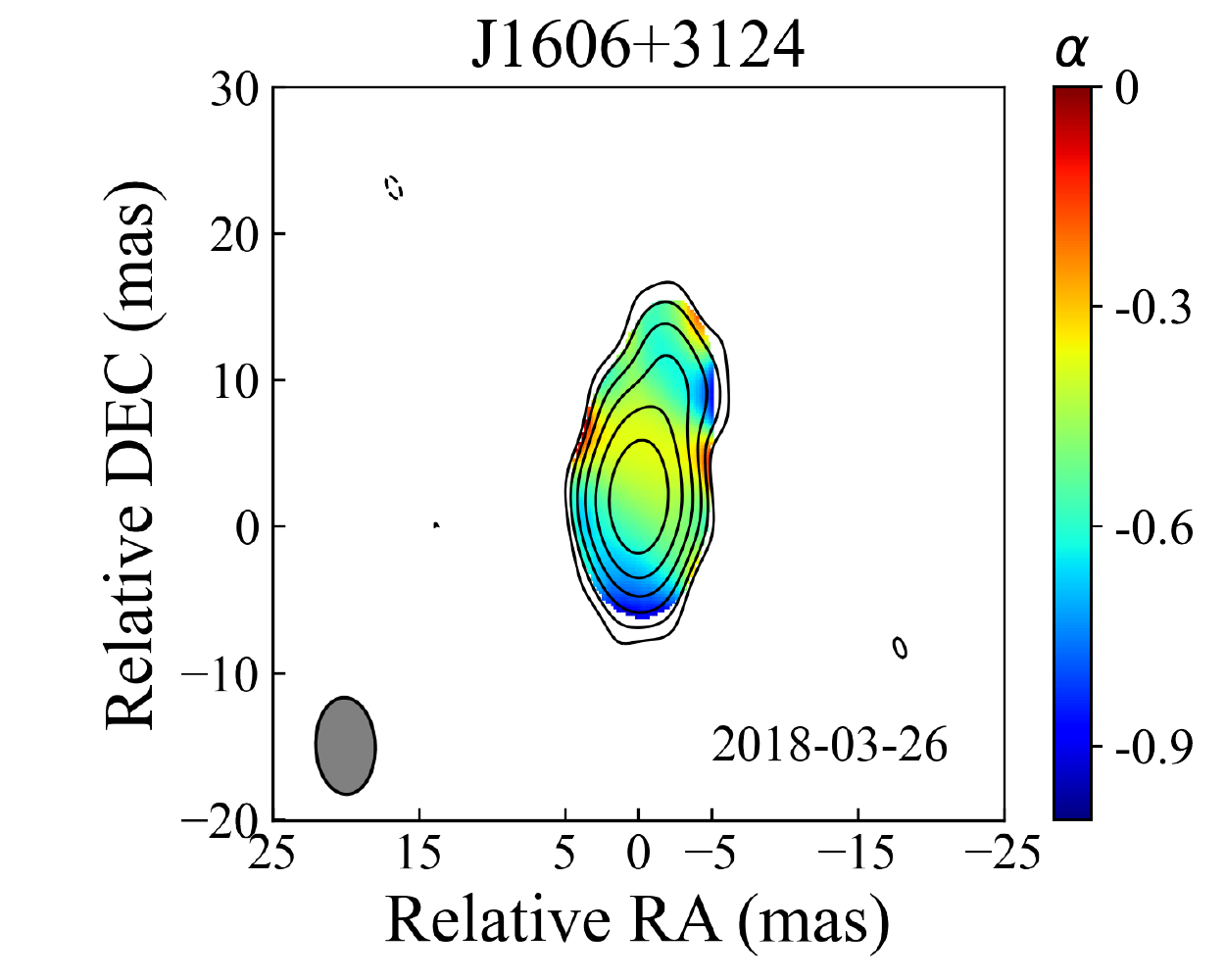}&
	\includegraphics[width=0.3\textwidth]{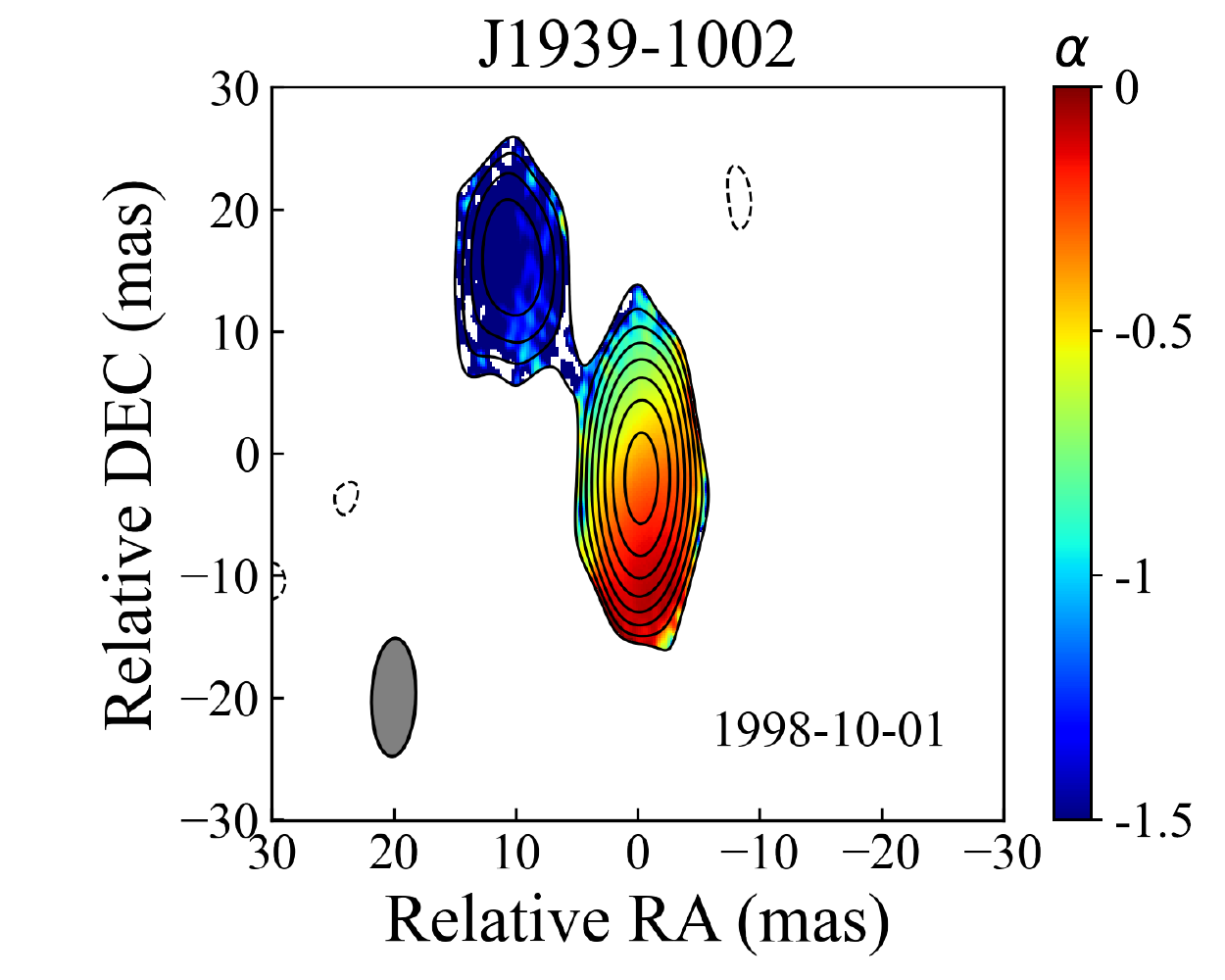}&
	\includegraphics[width=0.3\textwidth]{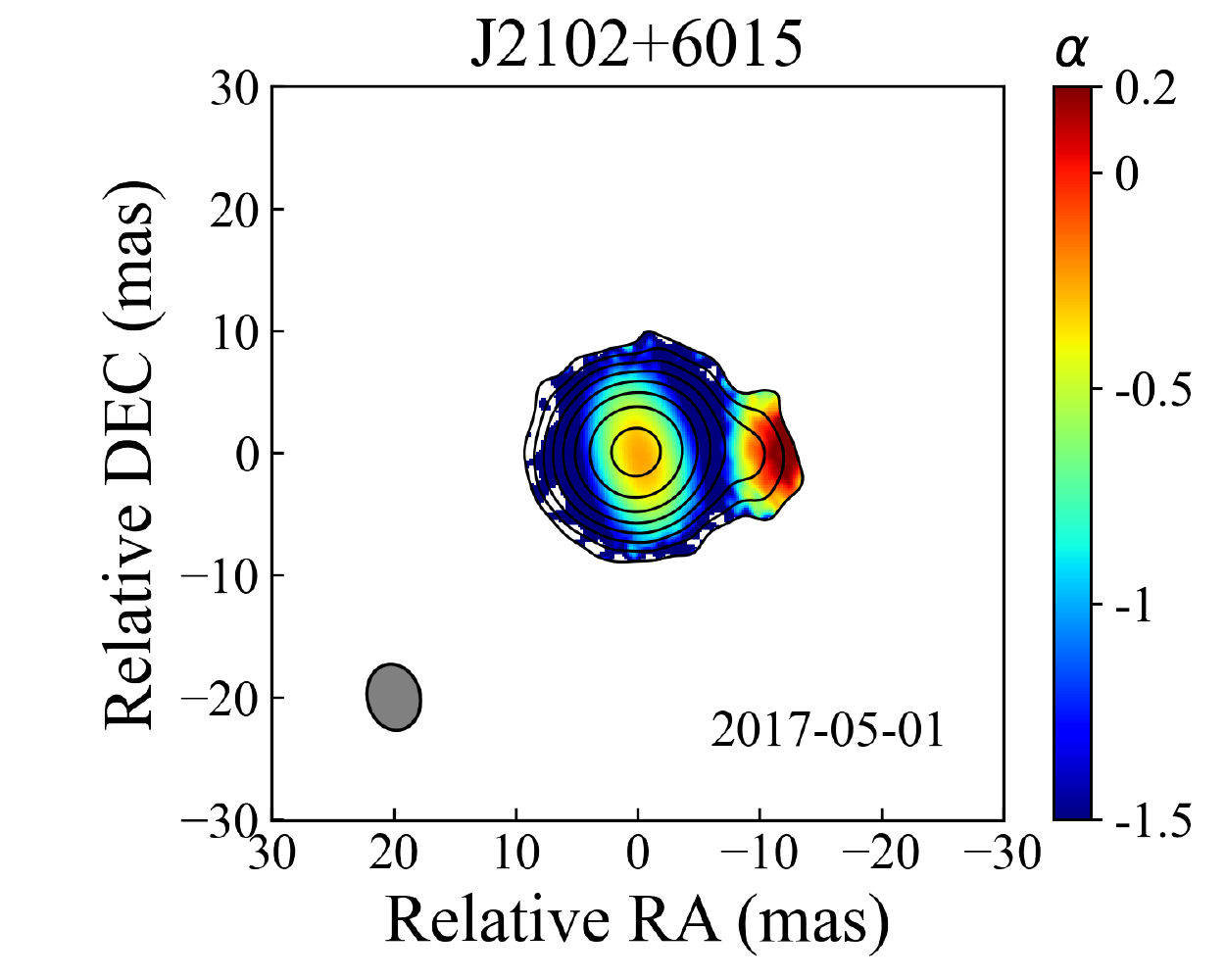}
  \end{tabular}

	\caption{Spectral index maps of our sample, based on simultaneous 2.3- and 8.4-GHz observations. In each map, the background contours represent the 2.3-GHz image. They start from 3 times of its rms noise and increase by a factor of 2. The bottom left ellipse are the restored beams of the 2.3-GHz map and the epochs used for the imaging are shown at bottom right.}
	\label{fig:spx}
\end{figure*}

\begin{figure*}
\gridline{
    \includegraphics[width=0.24\textwidth]{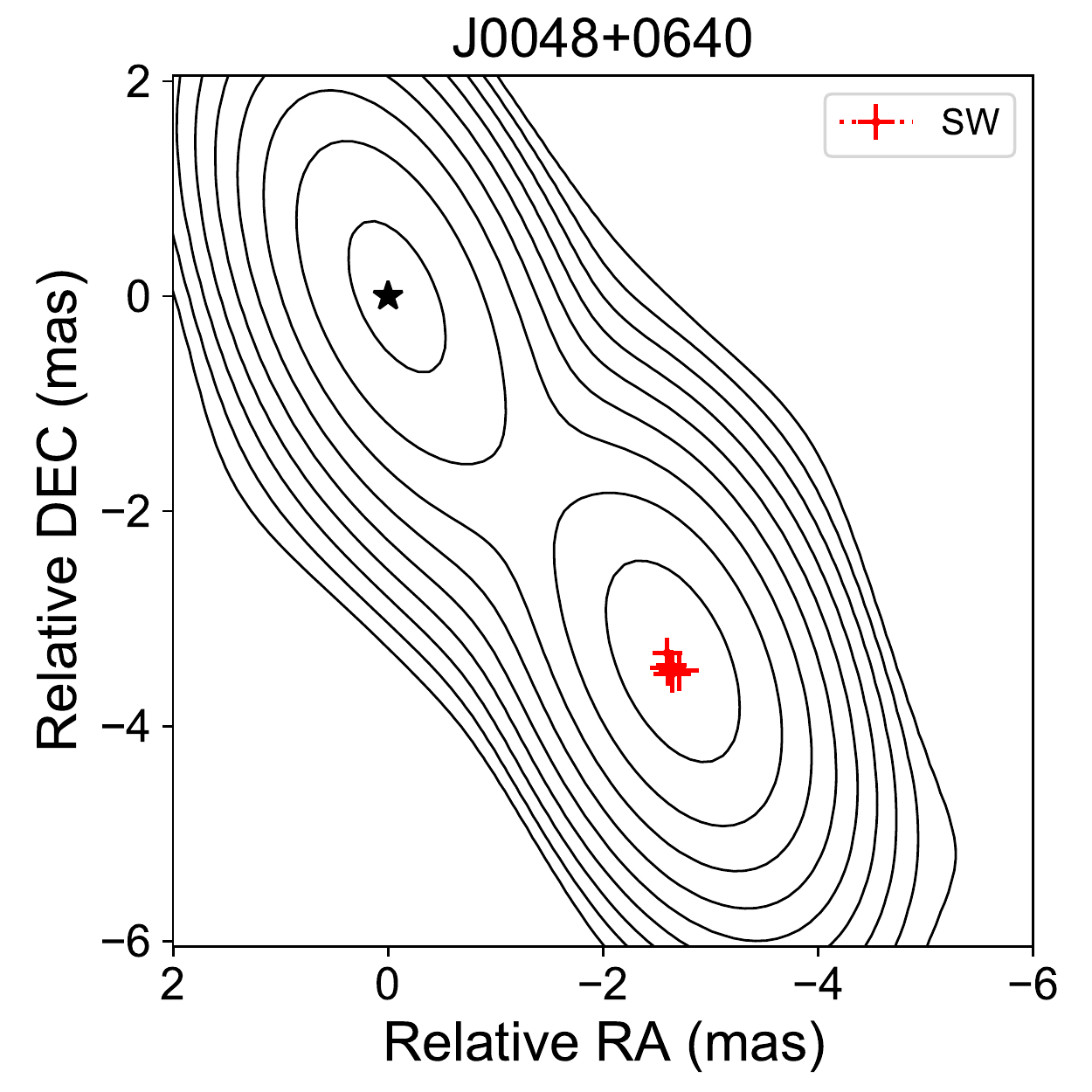}
	\includegraphics[width=0.3\textwidth]{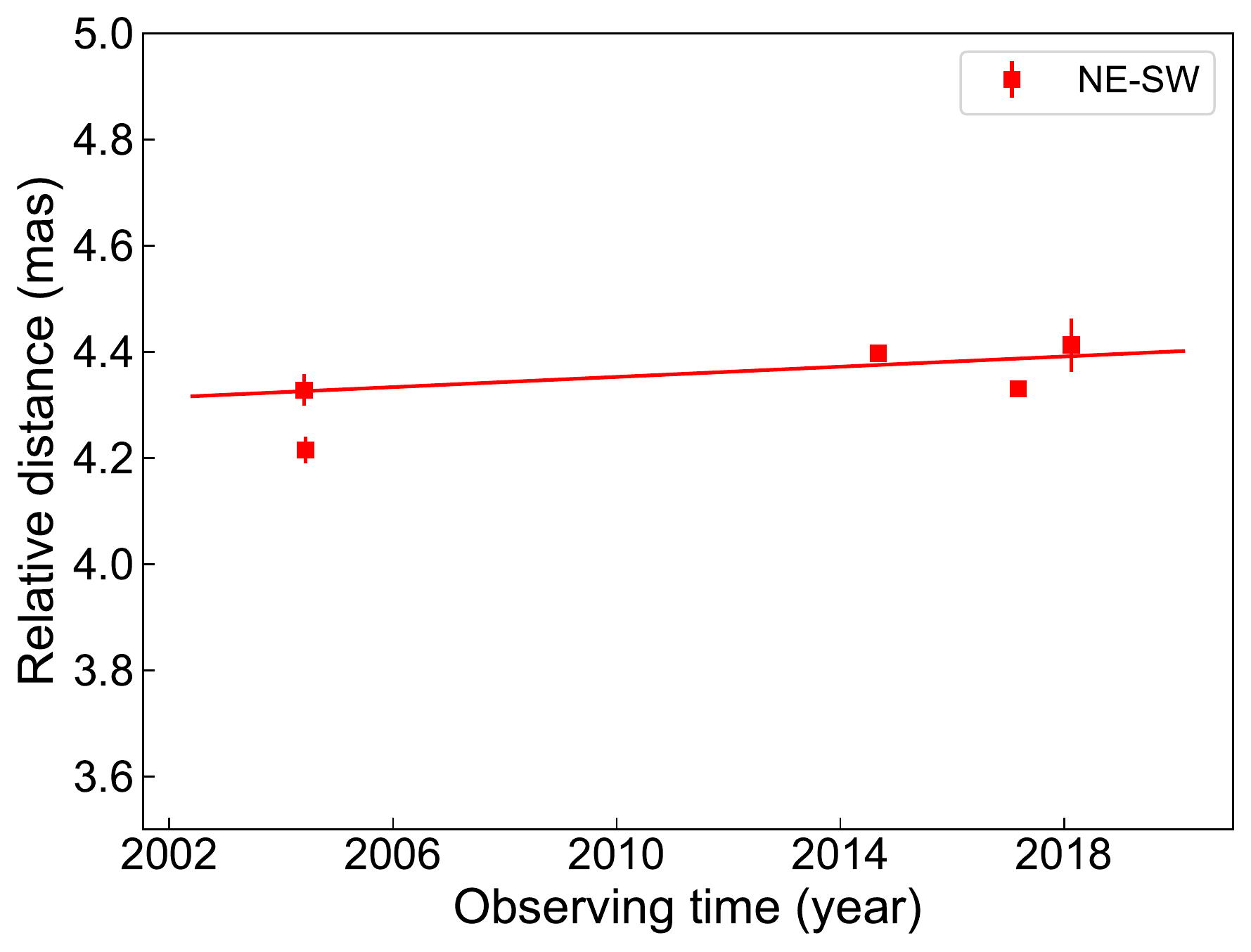}
	\includegraphics[width=0.3\textwidth]{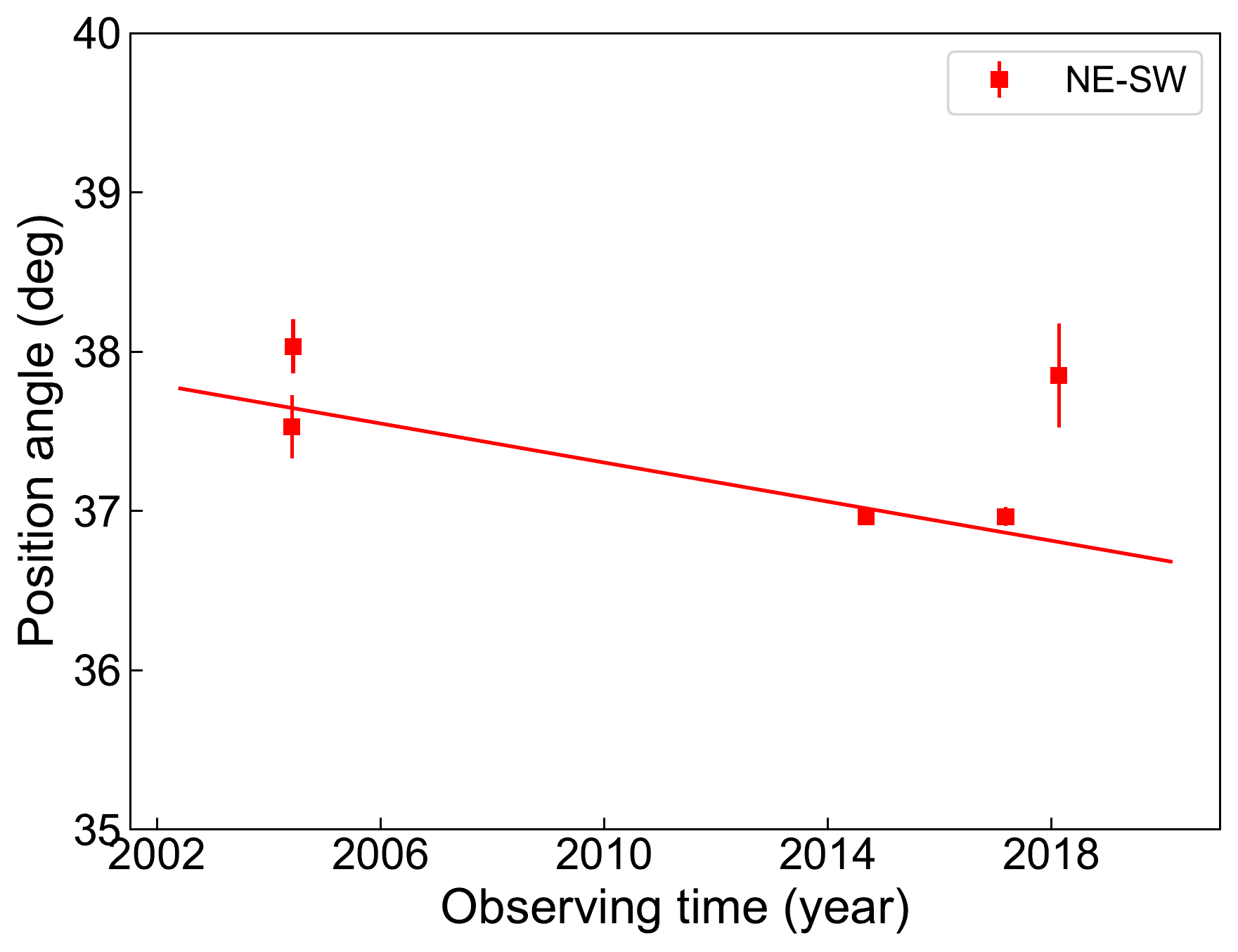}
	}
\gridline{
	\includegraphics[width=0.24\textwidth]{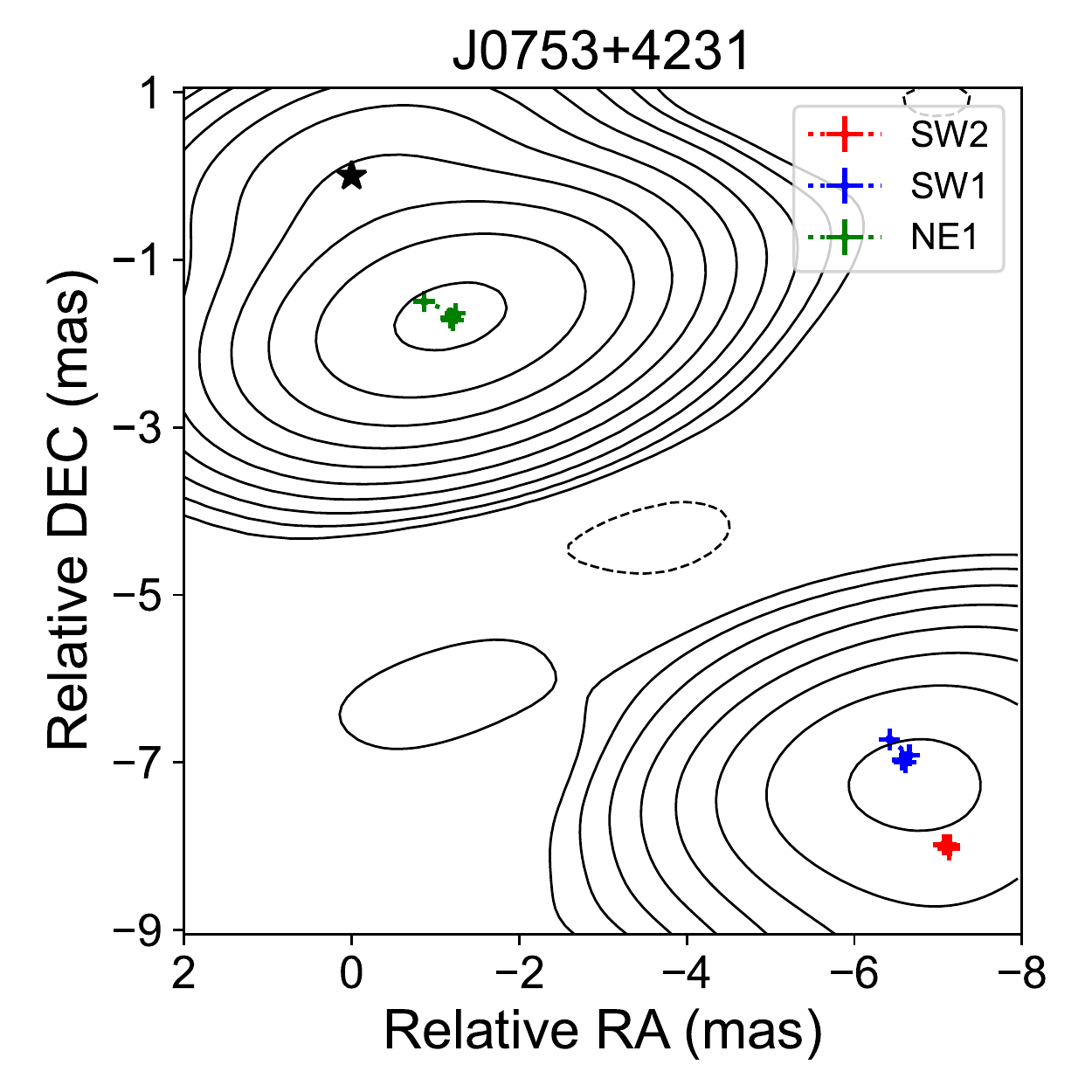}
	\includegraphics[width=0.3\textwidth]{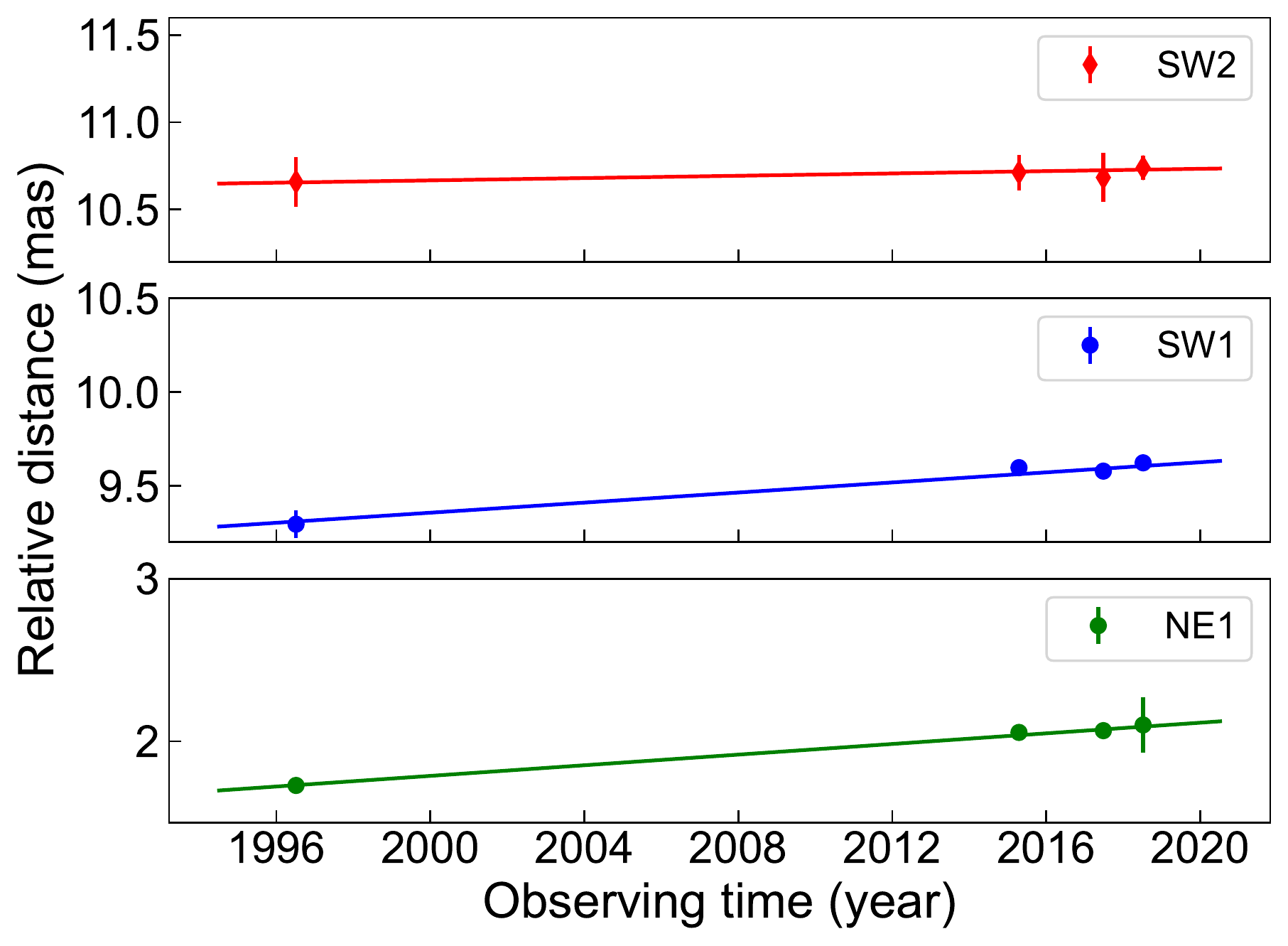}
	\includegraphics[width=0.3\textwidth]{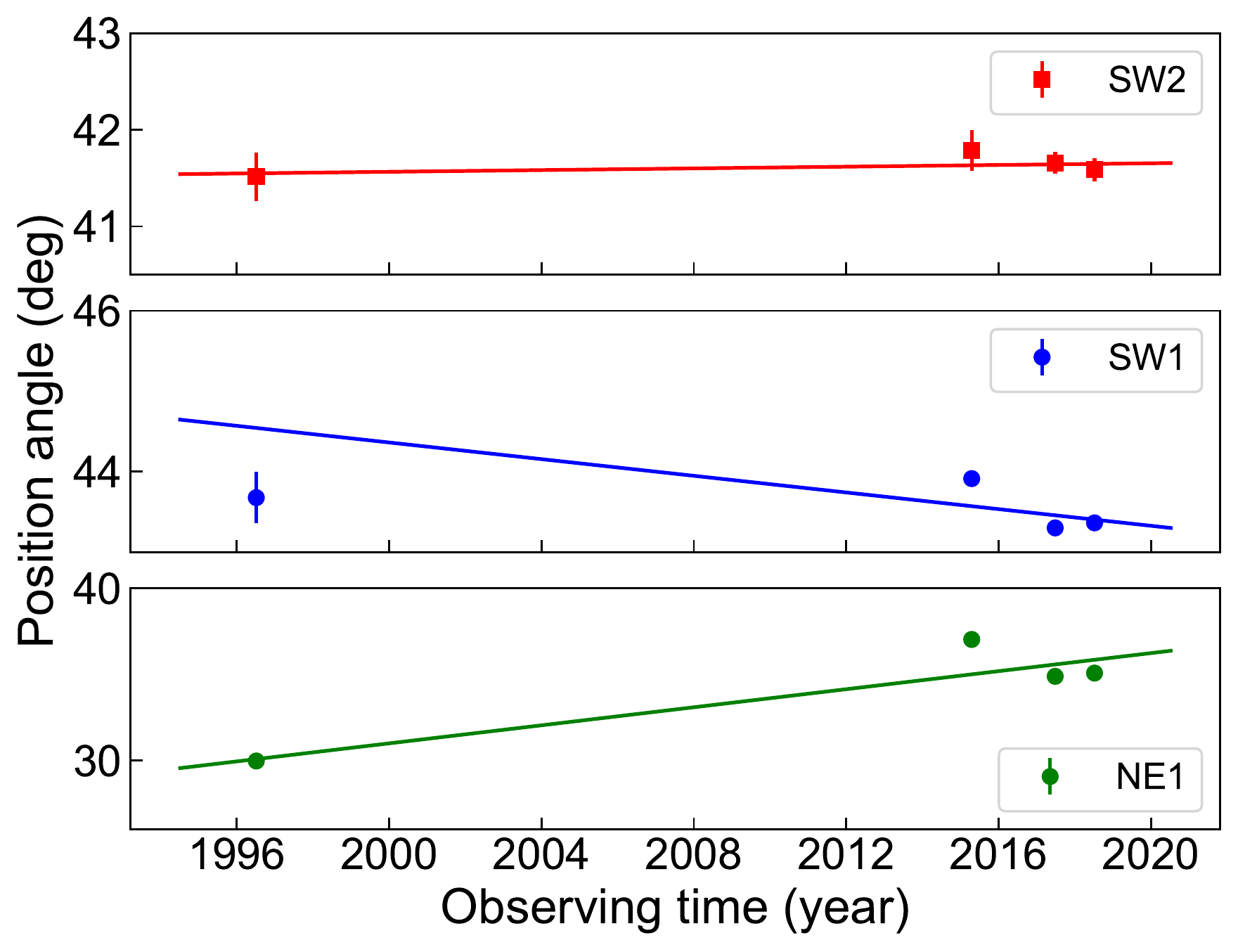}
}
\gridline{
	\includegraphics[width=0.24\textwidth]{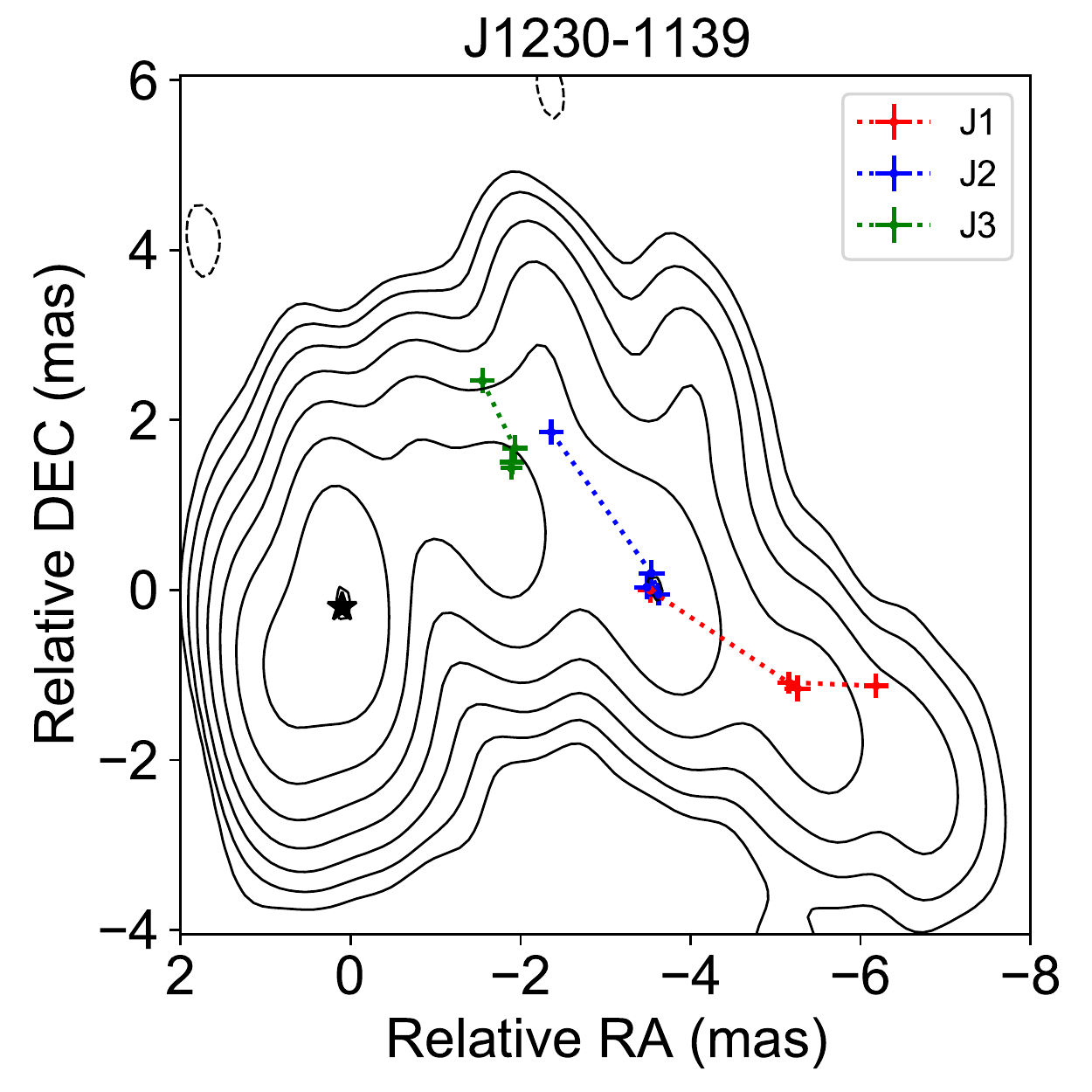}
	\includegraphics[width=0.3\textwidth]{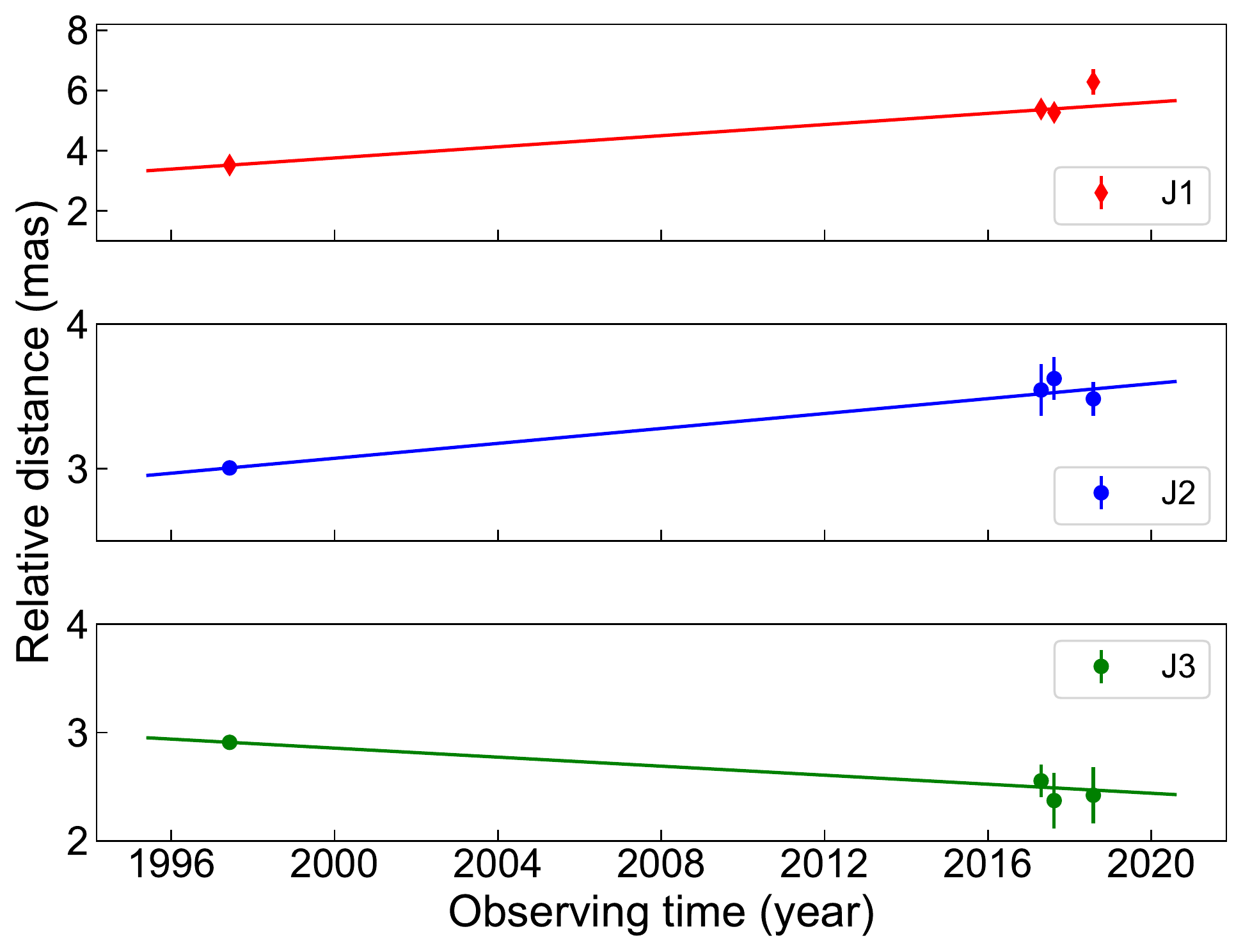}
	\includegraphics[width=0.3\textwidth]{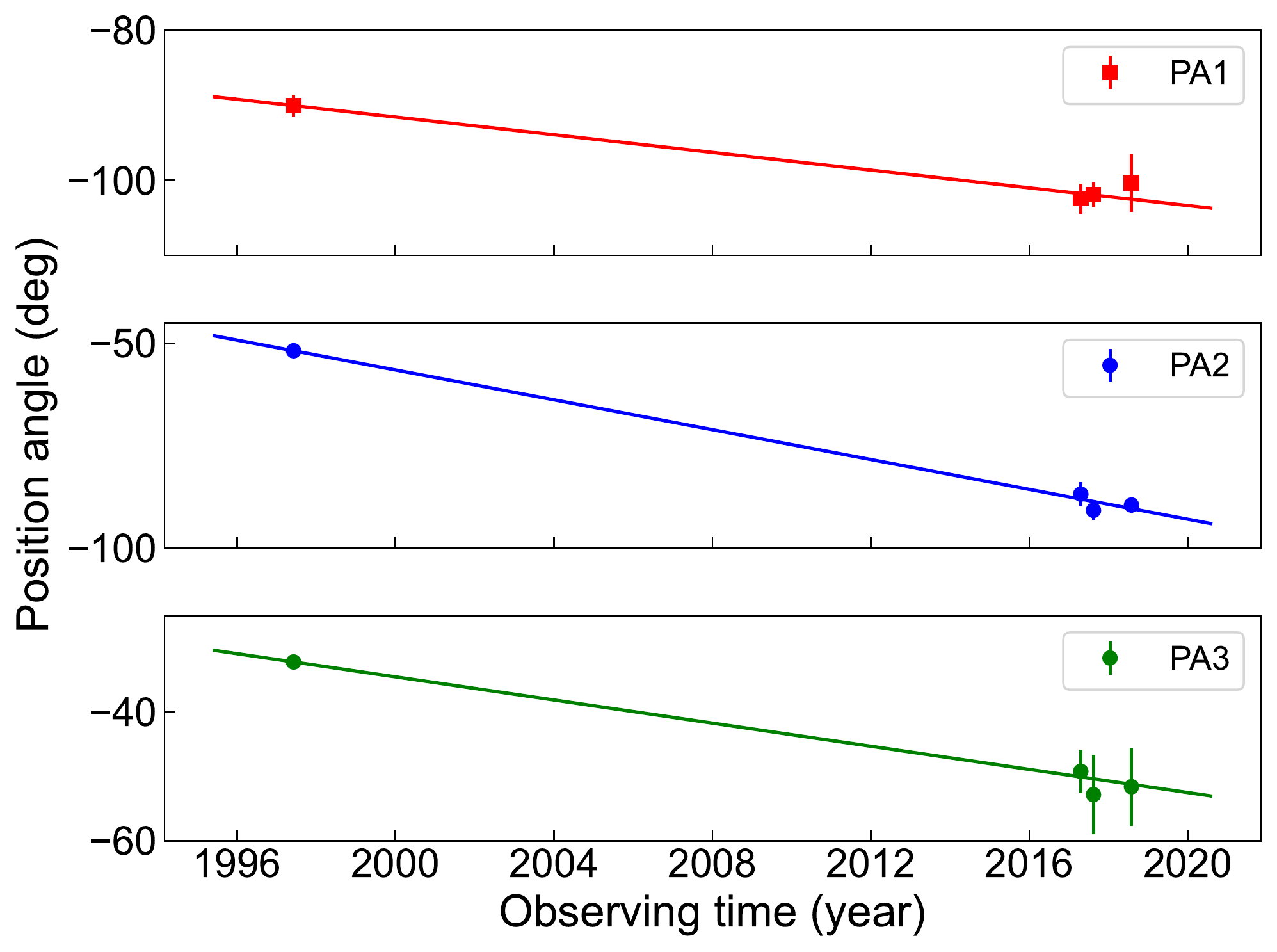} 
}
\caption{Proper motions of the fitted jet components. For each line, the left panel is a sketch map of the component motions with respect to the core component (marked with a star). The middle panel represents the change of relative radial distances of the components. The right panel is the corresponding position angle change. \label{fig:pm}}
\figurenum{3}
\end{figure*}

Table \ref{tab:obs} lists the basic observing information for the proposed VLBA session BZ064. Table \ref{tab:img} presents the parameters of the VLBI CLEAN images for each target source in our VLBA observation BZ064. Table \ref{tab:obs-arc} lists the 8-GHz VLBI imaging parameters of the target sources based on their archival VLBA experiments. Table \ref{tab:pm} is showing the jet proper motion parameters of the major jet components within each source in our sample, which were estimated based on our multi-epoch 8.4 GHz VLBA observations. Table \ref{tab:mod} catalogued the parameters of the brightness distribution models (Gaussian or point-source models) that are used to fit the target sources in the visibility domain. 
Figure \ref{fig:spx} shows the spectral index maps of the sources in our sample, made from their simultaneous 2.3 and 8,4 GHz VLBA images. Figure \ref{fig:pm} demonstrates how the jet components move with respect to their "cores" (optically thick radio cores or referencing components) in our sample and exhibits the fitted results of their radial motions and position angle changes.
\floattable
\begin{deluxetable}{cccccc}
\tablecaption{Information on our VLBA observations \label{tab:obs}}
\tablehead{
\colhead{Code} & \colhead{Freq.} & \colhead{Date} & \colhead{Target} & \colhead{Calibrators}& \colhead{Participating telescopes$^a$}  \\
 & \colhead{(GHz)} & \colhead{(yyyy-mm-dd)} & & &  }
\colnumbers
\startdata
BZ064A & 8.4 & 2017-02-05 & J0048$+$0640 & 2145$+$0657 &  BR, FD, HN, LA, NL, OV, PT, SC \\
... & ... & ...& J1939$-$1002 & ... & BR, FD, HN, LA, MK, NL, OV, PT, SC \\
... & ... & ...& J2102$+$6015 & ... &  BR, FD, HN, LA, MK, NL, OV, PT, SC \\
BZ064B & 8.4 & 2017-03-19 & J0753$+$4231 & 3C273, 3C345 &  BR, FD, HN, LA, MK, NL, OV, PT, SC  \\
... & ... & ... & J1230$-$1139 & ... & BR, FD, HN, LA, MK, NL, OV, PT, SC \\
... & ... & ... & J1316$+$6726 & ... &  BR, FD, HN, LA, MK, NL, OV, PT, SC \\
... & ... & ... & J1421$-$0643 & ... & BR, FD, HN, LA, MK, NL, OV, PT, SC \\
... & ... & ... & J1445$+$0958 & ... & BR, FD, HN, LA, MK, NL, OV, PT, SC \\
... & ... & ... & J1606$+$3124 & ... &  BR, FD, HN, LA, MK, NL, OV, PT, SC  \\
\enddata
\tablenotetext{a}{VLBA telescopes participating in the observations: BR (Brewster), FD (Fort Davis), HN (Hancock), KP (Kitt Peak), LA (Los Alamos), MK (Mauna Kea), NL (North Liberty), OV (Owens Valley), PT (Pie Town), SC (Saint Croix)}
\end{deluxetable}

\begin{deluxetable}{cccccrc}
\tablecaption{Parameters of the \texttt{CLEAN} images from the present VLBA observations at 8.4 GHz \label{tab:img}}
\tablehead{
		\colhead{Name} & 
		\colhead{Date} & \colhead{$S_\mathrm{peak}$}& \colhead{$B_\mathrm{maj}$} & \colhead{$B_\mathrm{min}$} & \colhead{PA}            & 
		\colhead{$\sigma$} \\
		 & 
		\colhead{yyyy-mm-dd}   & 
		\colhead{(mJy\,beam$^{-1}$)} & \colhead{(mas)} & 
		\colhead{(mas)} & 
		\colhead{($^{\circ}$)} & 
		\colhead{(mJy beam$^{-1}$)}
		}
	\colnumbers
\startdata
J0048$+$0640 &   2017-02-05   &       53.9        &   2.6    &   1.2    &   24.8   &    0.1  \\
J0753$+$4231 &   2017-03-19   &       164.6       &   2.9    &   1.6    &$-$74.2   &    0.1  \\	
J1230$-$1139 &   2017-03-19   &       55.0        &   2.4    &   1.0    &   6.0    &    0.1  \\	
J1316$+$6726 &   2017-03-19   &       92.6        &   2.3    &   0.9    &$-$8.7    &    0.1  \\
J1421$-$0643 &   2017-03-19   &       85.6        &   2.2    &   0.9    &$-$2.1    &    0.1  \\
J1445$+$0958 &   2017-03-19   &       244.0       &   1.9    &   0.7    &$-$0.5    &    0.2  \\    
J1606$+$3124 &   2017-03-19   &       326.4       &   2.4    &   0.9    &   9.4    &    0.2  \\
J1939$-$1002 &   2017-02-05   &       355.5       &   2.1    &   0.8    &   0.7    &    0.1  \\    
J2102$+$6015 &   2017-02-05   &       99.9        &   2.7    &   0.8    &$-$1.6    &    0.1  \\  
\enddata
\tablecomments{Columns: (1) J2000 source name; (2) date of the observation; (3) peak intensity of the image; (4)--(5) major and minor axes from the FWHM size of the elliptical Gaussian restoring beam; (6) position angle of the beam major axis, measured from north to east; (7) the rms noise of the post-fit image.}
\end{deluxetable}

\begin{deluxetable*}{cccccccccc}
\tablecaption{Basic information on the archival VLBA experiments at 7.6--8.7~GHz ($X$ band)\label{tab:obs-arc}}
\tablehead{
\colhead{Name} & \colhead{Date} & \colhead{Code} & \colhead{Freq.} & \colhead{$S_\mathrm{peak}$}& \colhead{$\sigma$}& \colhead{$B_\mathrm{maj}$}& \colhead{$B_\mathrm{min}$}& \colhead{PA}& \colhead{BW} \\ 
\colhead{} & \colhead{yyyy-mm-dd} &  & \colhead{(GHz)} & \colhead{(mJy beam$^{-1}$)} & \colhead{(mJy\,beam$^{-1}$)}  & \colhead{(mas)}& \colhead{(mas)}&  \colhead{($^\circ$)}& \colhead{(MHz)}  }
\colnumbers
\startdata
J0048$+$0640 & 2004-04-30 & bp110a & 8.6  & 67.4 & 1.15 & 2.2  & 1.0  & 4.2  & 8 $\times$ 4  \\
.. & 2004-05-08 & bp110 & 8.6  & 72.3 & 1.19 & 2.1  & 0.9  & 7.6  & 8 $\times$ 4  \\
.. & 2014-08-09 & bg219e & 8.7  & 53.4 & 0.25 & 2.6  & 1.2  & 18.9  & 32 $\times$ 12  \\
.. & 2017-09-18 & uf001q & 8.7  & 49.0 & 0.47 & 4.9  & 1.7  & $-$13.9  & 32 $\times$ 12  \\
.. & 2018-01-18 & ug002a & 8.7  & 54.5 & 0.22 & 3.4  & 1.0  & $-$18.1  & 32 $\times$ 12  \\
J0753$+$4231 & 1996-06-07 & bb023 & 8.3  & 163.5 & 1.03 & 1.9  & 0.9  & -6.5  & 8 $\times$ 4  \\
.. & 2015-03-17 & bg219i & 8.7  & 166.1 & 0.46 & 1.4  & 1.1  & $-$18.9  & 32 $\times$ 12  \\
.. & 2017-05-27 & uf001i & 8.7  & 135.0 & 0.35 & 1.5  & 1.0  & 12.1  & 32 $\times$ 12  \\
.. & 2018-06-09 & ug002j & 8.7  & 118.5 & 0.31 & 1.6  & 1.1  & 9.9  & 32 $\times$ 12  \\
J1230$-$1139 & 1997-05-07 & bb023 & 8.3  & 114.6 & 1.37 & 2.4  & 0.9  & 1.4  & 8 $\times$ 4  \\
.. & 2017-07-16 & uf001n & 8.7  & 55.6 & 0.90 & 2.0  & 0.9  & $-$6.4  & 32 $\times$ 12  \\
.. & 2018-07-01 & ug002k & 8.7  & 56.9 & 0.78 & 2.2  & 1.0  & $-$2.0  & 32 $\times$ 12  \\
J1316$+$6726 & 1995-04-19 & bb023 & 8.3  & 131.9 & 6.10 & 2.5  & 0.9  & $-$9.3  & 4 $\times$ 4  \\
.. & 2005-07-20 & bk124 & 8.6  & 81.2 & 0.45 & 2.2  & 1.3  & 40.8  & 8 $\times$ 4  \\
.. & 2014-12-20 & bg219f & 8.7  & 63.4 & 0.19 & 1.5  & 1.0  & $-$86.7  & 32 $\times$ 12  \\
J1421$-$0643 & 1997-05-07 & bb023 & 8.3  & 137.1 & 0.86 & 2.3  & 0.9  & $-$3.2  & 8 $\times$ 4  \\
.. & 2015-01-22 & bg219g & 8.7  & 61.2 & 0.28 & 2.1  & 0.9  & $-$1.7  & 32 $\times$ 12  \\
.. & 2015-03-17 & bg219i & 8.7  & 87.6 & 0.41 & 2.1  & 0.9  & 0.6  & 32 $\times$ 12  \\
.. & 2017-02-19 & uf001c & 8.7  & 70.8 & 0.28 & 2.0  & 0.8  & 4.0  & 32 $\times$ 12  \\
.. & 2018-03-26 & ug002d & 8.7  & 86.3 & 0.22 & 2.3  & 1.0  & 1.6  & 32 $\times$ 12  \\
J1445$+$0958 & 1997-01-10 & bf025 & 8.3  & 161.3 & 1.42 & 1.9  & 1.0  & 1.0  & 8 $\times$ 4  \\
.. & 1999-05-10 & rdv15 & 8.6  & 182.3 & 0.72 & 1.2  & 0.9  & $-$8.6  & 8 $\times$ 4  \\
.. & 2001-07-05 & rdv29 & 8.6  & 207.4 & 1.47 & 1.6  & 0.9  & 0.7  & 8 $\times$ 4  \\
.. & 2005-06-22 & bl129 & 8.3  & 304.3 & 0.81 & 2.2  & 1.0  & $-$4.6  & 8 $\times$ 4  \\
.. & 2008-04-02 & rdv68 & 8.6  & 224.6 & 2.03 & 0.7  & 0.5  & 9.0  & 8 $\times$ 4  \\
.. & 2018-03-26 & ug002d & 8.7  & 216.5 & 0.92 & 2.0  & 0.9  & $-$2.9  & 32 $\times$ 12  \\
J1606$+$3124 & 1996-05-15 & bb023 & 8.3  & 304.4 & 0.90 & 2.1  & 0.9  & $-$18.2  & 8 $\times$ 4  \\
.. & 2014-08-09 & bg219e & 8.7  & 337.0 & 0.33 & 2.1  & 1.1  & 20.9  & 32 $\times$ 12  \\
.. & 2017-09-18 & uf001q & 8.7  & 323.6 & 0.88 & 3.2  & 1.6  & $-$25.5  & 32 $\times$ 12  \\
.. & 2018-03-26 & ug002d & 8.7  & 270.1 & 0.36 & 1.8  & 1.1  & 4.2  & 32 $\times$ 12  \\
J1939$-$1002 & 1997-01-10 & bf025 & 8.3  & 246.7 & 0.80 & 2.1  & 0.9  & $-$1.2  & 8 $\times$ 4  \\
.. & 1997-05-07 & bb023 & 8.3  & 318.7 & 0.85 & 2.3  & 0.9  & $-$1.0  & 8 $\times$ 4  \\
.. & 1998-10-01 & rdv11 & 8.6  & 318.6 & 0.55 & 2.5  & 0.9  & $-$1.6  & 4 $\times$ 4  \\
.. & 2006-07-11 & rdv57 & 8.6  & 285.6 & 2.16 & 2.3  & 0.7  & $-$5.6  & 8 $\times$ 4  \\
.. & 2019-02-04 & sb072d2 & 7.6  & 350.2 & 0.39 & 2.4  & 1.1  & 5.4  & 32 $\times$ 8  \\
J2102$+$6015 & 1994-08-12 & bb023 & 8.3  & 136.9 & 3.06 & 2.2  & 0.9  & $-$65.4  & 4 $\times$ 4  \\
.. & 2006-12-18 & bp133 & 8.6  & 92.4 & 0.71 & 1.4  & 1.2  & $-$56.7  & 8 $\times$ 4  \\
.. & 2017-05-01 & uf001h & 8.7  & 99.0 & 0.17 & 1.3  & 1.1  & 13.2  & 32 $\times$ 12  \\
.. & 2018-06-09 & ug002j & 8.7  & 94.4 & 0.17 & 1.5  & 1.1  & $-$14.0  & 32 $\times$ 12  \\
\enddata
\tablecomments{Columns: (1) J2000 source name; (2) observing date; (3) project code; (4) observing frequency; (5) peak intensity of the image; (6) rms noise, (7)-(8) major and minor FWHM axes of the elliptical Gaussian restoring beam; (9) position angle of the beam major axis, measured from north to east; (10) observing bandwidth of each IF $\times$ number of IFs.}
\end{deluxetable*}

\floattable
\begin{deluxetable*}{ccccccccc}
\tablecaption{Jet component proper motions \label{tab:pm}}
\tablehead{
\colhead{Name} & \colhead{Comp.} & \colhead{$\mu_\mathrm{rad}$} & \colhead{$\mu_ x$} & \colhead{$\mu_y$}& \colhead{$|\mu_\mathrm{tot}|$}& \colhead{$|\beta_\mathrm{rad}|$}& \colhead{$|\beta_\mathrm{tot}|$} & \colhead{$\mathrm{PA}_\mathrm{var}$}
}  
\colnumbers
\startdata
J0048+0640&SW&0.005$\pm$0.002&0.001$\pm$0.001&$-$0.005$\pm$0.002&0.005$\pm$0.002&0.6$\pm$0.2&0.6$\pm$0.2&$-$0.1$\pm$0.1 \\
J0753+4231&SW2&0.003$\pm$0.005&$-$0.003$\pm$0.004&$-$0.002$\pm$0.009&0.004$\pm$0.006&0.3$\pm$0.6&0.4$\pm$0.7&0.0$\pm$0.0 \\
…&SW1&0.013$\pm$0.003&$-$0.006$\pm$0.002&$-$0.013$\pm$0.004&0.014$\pm$0.004&1.5$\pm$0.3&1.6$\pm$0.4&$-$0.1$\pm$0.0 \\
…&NE1&0.016$\pm$0.001&$-$0.016$\pm$0.001&$-$0.009$\pm$0.002&0.018$\pm$0.001&1.8$\pm$0.1&2.1$\pm$0.1&0.3$\pm$0.1 \\
J1230$-$1139&J1&0.093$\pm$0.007&$-$0.087$\pm$0.005&$-$0.055$\pm$0.011&0.103$\pm$0.007&10.3$\pm$0.8&11.4$\pm$0.8&$-$0.6$\pm$0.1 \\
…&J2&0.026$\pm$0.004&$-$0.057$\pm$0.003&$-$0.088$\pm$0.006&0.105$\pm$0.005&2.9$\pm$0.4&11.7$\pm$0.6&$-$1.8$\pm$0.1 \\
…&J3&$-$0.021$\pm$0.006&$-$0.018$\pm$0.004&$-$0.044$\pm$0.010&0.048$\pm$0.009&$-$2.3$\pm$0.7&5.3$\pm$1.0&$-$0.9$\pm$0.1 \\
J1316+6726&J1&$-$0.128$\pm$0.115&0.192$\pm$0.142&0.231$\pm$0.101&0.300$\pm$0.119&$-$14.2$\pm$12.8&33.4$\pm$13.3&$-$1.9$\pm$0.8 \\
J1421$-$0643&J2&0.140$\pm$0.003&0.126$\pm$0.002&0.065$\pm$0.005&0.142$\pm$0.003&15.9$\pm$0.3&16.1$\pm$0.3&0.3$\pm$0.1 \\
…&J3&0.081$\pm$0.001&0.064$\pm$0.001&0.051$\pm$0.001&0.082$\pm$0.001&9.2$\pm$0.1&9.3$\pm$0.1&$-$0.6$\pm$0.1 \\
…&J4&0.140$\pm$0.009&0.102$\pm$0.006&0.116$\pm$0.015&0.154$\pm$0.012&15.9$\pm$1.0&17.5$\pm$1.4&$-$3.5$\pm$0.5 \\
J1445+0958&J1&0.116$\pm$0.005&$-$0.049$\pm$0.004&$-$0.122$\pm$0.008&0.131$\pm$0.008&12.9$\pm$0.6&14.7$\pm$0.8&$-$1.5$\pm$0.1 \\
…&J2&0.085$\pm$0.001&$-$0.089$\pm$0.001&$-$0.005$\pm$0.002&0.089$\pm$0.001&9.5$\pm$0.1&10.0$\pm$0.1&0.7$\pm$0.1 \\
J1606+3124&S-N&0.013$\pm$0.002&0.002$\pm$0.002&0.014$\pm$0.004&0.014$\pm$0.004&1.6$\pm$0.2&1.7$\pm$0.5&0.1$\pm$0.0 \\
…&S-C&0.006$\pm$0.002&0.004$\pm$0.002&0.008$\pm$0.003&0.009$\pm$0.003&0.7$\pm$0.2&1.1$\pm$0.3&0.1$\pm$0.0 \\
J1939$-$1002&J2&$-$0.014$\pm$0.001&$-$0.007$\pm$0.001&$-$0.013$\pm$0.002&0.015$\pm$0.002&$-$1.6$\pm$0.1&1.7$\pm$0.2&$-$0.1$\pm$0.0 \\
…&J3&0.005$\pm$0.001&$-$0.013$\pm$0.001&0.007$\pm$0.001&0.015$\pm$0.001&0.5$\pm$0.1&1.7$\pm$0.1&$-$0.7$\pm$0.1 \\
J2102+6015&W&0.020$\pm$0.010&$-$0.021$\pm$0.010&0.000$\pm$0.010&0.021$\pm$0.010&2.5$\pm$1.2&2.6$\pm$1.2&0.0$\pm$0.0 \\
\enddata
\tablecomments{Columns: (1) J2000 source name; (2) component identifier; (3)--(6) fitted proper motion in mas\,yr$^{-1}$: $\mu_\mathrm{rad}$ -- the jet radial motion; $\mu_x$, $\mu_y$ -- proper motion components along the right ascension and declination axes, respectively; $\mu_\mathrm{tot}$: total proper motion calculated from $\mu_x$ and $\mu_y$; (7)--(8) apparent transverse speed of jet radial ($\beta_\mathrm{rad}$) and total ($\beta_\mathrm{tot}$) motions, respectively, in units of $c$; (9) change of the jet position angle, in $\degr$\,yr$^{-1}$.}
\end{deluxetable*}

\begin{deluxetable}{ccccccccc}
\tablecaption{Model fitting parameters \label{tab:mod}}
\tablehead{
		\colhead{Name}   &  \colhead{Epoch} &  \colhead{Comp.}  &  \colhead{$S_\mathrm{peak}$}  &  \colhead{$S_\mathrm{int}$}  &   \colhead{$R$}  &  \colhead{PA}  &  \colhead{$\theta_\mathrm{comp}$} &  \colhead{$T_\mathrm{B}$} \\
		\colhead{}  & \colhead{yyyymmdd}   &    &  \colhead{(mJy\,beam$^{-1}$)}  &  \colhead{(mJy)}  &  \colhead{(mas)}  &  \colhead{($\degr$)} & \colhead{(mas)}  & \colhead{($\times 10^{10}$\,K)}  
	}
\colnumbers
\startdata
J0048$+$0640 & 20040430 & NE & 66.8$\pm$3.5  & 71.5$\pm$3.9  & ... & ... & 0.31$\pm$0.05 & 5.5$\pm$1.2 \\
... & ... & SW & 39.1$\pm$2.1  & 40.7$\pm$2.3  & 4.33$\pm$0.03 & $-$142.5$\pm$0.4  & 0.27$\pm$0.06 & 4.2$\pm$1.3\\
... & 20040508 & NE & 71.7$\pm$3.7  & 75.8$\pm$4.1  & ... & ... & 0.27$\pm$0.04 & 7.8$\pm$1.7 \\
... & ... & SW & 42.1$\pm$2.2  & 42.7$\pm$2.4  & 4.22$\pm$0.03 & $-$142.0$\pm$0.3  & 0.18$^{\star}$ & 9.9$\pm$3.9\\
... & 20140809 & NE & 50.3$\pm$2.5  & 50.4$\pm$2.5  & ... & ... & 0.22$\pm$0.01 & 7.8$\pm$0.7 \\
... & ... & P1$^a$ & 15.7$\pm$0.8  & 5.7$\pm$0.3  & 1.93$\pm$0.02 & $-$148.2$\pm$0.5  & 0.17$^{\star}$ & ...\\
... & ... & SW & 53.5$\pm$2.7  & 58.2$\pm$2.9  & 4.40$\pm$0.00 & $-$143.0$\pm$0.1  & 0.45$\pm$0.01 & 2.1$\pm$0.1\\
... & 20170205 & NE & 44.2$\pm$2.2  & 42.5$\pm$2.2  & ... & ... & 0.22$\pm$0.03 & 6.9$\pm$1.2 \\
... & ... & P2$^a$ & 22.7$\pm$1.2  & 6.5$\pm$0.4  & 1.42$\pm$0.02 & $-$154.7$\pm$0.6  & 0.15$^{\star}$ & ...\\
... & ... & SW & 58.5$\pm$2.9  & 64.6$\pm$3.3  & 4.33$\pm$0.01 & $-$143.0$\pm$0.1  & 0.49$\pm$0.02 & 2.2$\pm$0.2\\
... & 20170918 & NE & 20.2$\pm$1.5  & 18.7$\pm$1.8  & ... & ... & 0.64$^{\star}$ & 0.3$\pm$0.2 \\
... & ... & SW & 48.6$\pm$2.7  & 51.6$\pm$3.2  & 4.08$\pm$0.08 & $-$133.6$\pm$1.1  & 0.66$\pm$0.15 & 0.9$\pm$0.3\\
... & 20180118 & NE & 44.6$\pm$3.2  & 56.3$\pm$4.6  & ... & ... & 0.75$\pm$0.19 & 0.8$\pm$0.3 \\
... & ... & SW & 54.5$\pm$3.1  & 65.0$\pm$4.0  & 4.41$\pm$0.05 & $-$142.1$\pm$0.7  & 0.62$\pm$0.10 & 1.2$\pm$0.3\\
J0753$+$4231 & 19960607 &NE2(core)& 13.1$\pm$0.9  & 12.1$\pm$1.1  & ... & ... & 0.27$^{\star}$ & 1.4$\pm$0.9 \\
... & ... & NE1 & 162.7$\pm$8.2  & 182.0$\pm$9.2  & 1.73$\pm$0.06 & $-$150.0$\pm$0.2  & 0.39$\pm$0.12 & 9.7$\pm$0.6\\
... & ... & SW1 & 19.4$\pm$1.2  & 14.0$\pm$1.2  & 9.30$\pm$0.08 & $-$136.3$\pm$0.3  & 0.24$^{\star}$ & ...\\
... & ... & SW2 & 16.2$\pm$1.0  & 43.0$\pm$2.7  & 10.66$\pm$0.08 & $-$138.5$\pm$0.3  & 1.56$\pm$0.09 & ...\\
... & 20150317 & NE2(core) & 13.6$\pm$1.5  & 15.5$\pm$2.1  & ... & ... & 0.51$\pm$0.02 & 0.4$\pm$0.3 \\
... & ... & NE1 & 165.6$\pm$8.4  & 180.2$\pm$9.2  & 2.06$\pm$0.12 & $-$143.0$\pm$0.3  & 0.37$\pm$0.24 & 9.6$\pm$0.9\\
... & ... & SW1 & 32.9$\pm$1.7  & 31.7$\pm$1.6  & 9.60$\pm$0.12 & $-$136.1$\pm$0.1  & 0.51$\pm$0.02 & ...\\
... & ... & SW2 & 20.1$\pm$1.2  & 36.4$\pm$2.2  & 10.71$\pm$0.13 & $-$138.2$\pm$0.2  & 1.45$\pm$0.08 & ...\\
... & 20170319 & NE2(core) & 19.2$\pm$1.1  & 17.3$\pm$1.2  & ... & ... & 0.34$^{\star}$ & 1.2$\pm$0.6 \\
... & ... & NE1 & 165.6$\pm$8.3  & 172.6$\pm$8.7  & 2.09$\pm$0.06 & $-$144.7$\pm$0.2  & 0.38$\pm$0.13 & 9.5$\pm$0.7\\
... & ... & SW1 & 41.1$\pm$2.3  & 59.0$\pm$3.4  & 9.82$\pm$0.08 & $-$137.3$\pm$0.3  & 1.25$\pm$0.10 & ...\\
... & ... & SW2 & 8.6$\pm$0.7  & 11.6$\pm$1.1  & 11.84$\pm$0.15 & $-$141.6$\pm$0.6  & 1.38$\pm$0.27 & ...\\
... & 20170527 & NE2(core) & 15.6$\pm$1.0  & 16.0$\pm$1.1  & ... & ... & 0.27$\pm$0.02 & 1.6$\pm$0.8 \\
... & ... & NE1 & 134.7$\pm$6.8  & 148.9$\pm$7.6  & 2.07$\pm$0.04 & $-$145.1$\pm$0.2  & 0.39$\pm$0.09 & 7.4$\pm$0.6\\
... & ... & SW1 & 31.3$\pm$1.6  & 29.5$\pm$1.5  & 9.58$\pm$0.04 & $-$136.7$\pm$0.1  & 0.48$\pm$0.02 & ...\\
... & ... & SW2 & 20.2$\pm$1.1  & 31.9$\pm$1.7  & 10.68$\pm$0.05 & $-$138.3$\pm$0.1  & 1.35$\pm$0.04 & ...\\
... & 20180609 & NE2(core) & 14.4$\pm$1.0  & 14.1$\pm$1.1  & ... & ... & 0.27$^{\star}$ & 1.4$\pm$0.9 \\
... & ... & NE1 & 118.3$\pm$6.0  & 129.2$\pm$6.6  & 2.10$\pm$0.06 & $-$144.9$\pm$0.3  & 0.40$\pm$0.12 & 6.0$\pm$0.6\\
... & ... & SW1 & 30.1$\pm$1.5  & 27.6$\pm$1.4  & 9.62$\pm$0.06 & $-$136.6$\pm$0.1  & 0.49$\pm$0.02 & ...\\
... & ... & SW2 & 20.8$\pm$1.1  & 27.8$\pm$1.5  & 10.74$\pm$0.06 & $-$138.4$\pm$0.1  & 1.31$\pm$0.05 & ...\\
J1230$-$1139 & 19970507 & C(core) & 117.8$\pm$6.1  & 128.8$\pm$6.8  & ... & ... & 0.37$\pm$0.04 & 7.4$\pm$1.2 \\
... & ... & J3 & 33.7$\pm$2.1  & 21.8$\pm$1.8  & 2.91$\pm$0.05 & $-$32.2$\pm$1.0  & 0.26$^{\star}$ & ...\\
... & ... & J2 & 66.2$\pm$3.4  & 76.2$\pm$3.9  & 3.01$\pm$0.01 & $-$51.8$\pm$0.3  & 0.55$\pm$0.03 & ...\\
... & ... & J1 & 14.3$\pm$1.1  & 20.1$\pm$1.8  & 3.52$\pm$0.09 & $-$90.0$\pm$1.4  & 0.80$\pm$0.18 & ...\\
... & 20170319 & C(core) & 65.6$\pm$3.8  & 76.2$\pm$4.9  & ... & ... & 0.55$\pm$0.09 & 2.0$\pm$0.5 \\
... & ... & J3 & 14.9$\pm$1.6  & 41.1$\pm$4.8  & 2.56$\pm$0.15 & $-$49.2$\pm$3.4  & 1.80$\pm$0.30 & ...\\
... & ... & J2 & 14.3$\pm$1.8  & 12.7$\pm$2.3  & 3.55$\pm$0.18 & $-$86.8$\pm$2.9  & 0.51$^{\star}$ & ...\\
... & ... & J1 & 10.6$\pm$1.4  & 10.5$\pm$1.9  & 5.39$\pm$0.19 & $-$102.4$\pm$2.0  & 0.52$^{\star}$ & ...\\
... & 20170716 & C(core) & 57.1$\pm$3.1  & 65.2$\pm$3.8  & ... & ... & 0.43$\pm$0.06 & 2.6$\pm$0.5 \\
... & ... & J3 & 8.5$\pm$1.7  & 30.5$\pm$6.4  & 2.37$\pm$0.26 & $-$52.8$\pm$6.2  & 1.82$\pm$0.51 & ...\\
... & ... & J2 & 7.8$\pm$1.0  & 8.1$\pm$1.3  & 3.62$\pm$0.15 & $-$90.8$\pm$2.3  & 0.47$\pm$0.30 & ...\\
... & ... & J1 & 7.7$\pm$1.0  & 7.5$\pm$1.3  & 5.27$\pm$0.15 & $-$101.9$\pm$1.6  & 0.43$^{\star}$ & ...\\
... & 20180701 & C(core) & 58.1$\pm$3.4  & 69.1$\pm$4.5  & ... & ... & 0.56$\pm$0.09 & 1.6$\pm$0.4 \\
... & ... & J3 & 7.9$\pm$1.4  & 24.6$\pm$4.8  & 2.42$\pm$0.26 & $-$51.6$\pm$6.1  & 1.75$\pm$0.51 & ...\\
... & ... & J2 & 17.0$\pm$1.6  & 19.8$\pm$2.3  & 3.48$\pm$0.12 & $-$89.5$\pm$1.9  & 0.62$\pm$0.23 & ...\\
... & ... & J1 & 2.2$\pm$0.7  & 4.6$\pm$1.5  & 6.28$\pm$0.43 & $-$100.3$\pm$3.9  & 1.12$\pm$0.86 & ...\\
J1316$+$6726 & 19950419 & C(core) & 125.3$\pm$6.8  & 128.6$\pm$7.5  & ... & ... & 0.20$^{\star}$ & 25.5$\pm$11.6 \\
... & 20050720 & C(core) & 84.9$\pm$4.5  & 88.2$\pm$4.8  & ... & ... & 0.29$\pm$0.06 & 7.5$\pm$2.0 \\
... & 20141220 & C(core) & 63.6$\pm$3.2  & 64.0$\pm$3.2  & ... & ... & 0.10$\pm$0.01 & 48.8$\pm$5.4 \\
... & ... & J1 & 0.9$\pm$0.2  & 4.2$\pm$0.9  & 8.06$\pm$0.26 & 157.3$\pm$1.8  & 3.24$\pm$0.51 & ...\\
... & 20170319 & C(core) & 92.6$\pm$4.6  & 93.6$\pm$4.7  & ... & ... & 0.20$\pm$0.00 & 18.0$\pm$1.0 \\
... & ... & J3 & 37.3$\pm$1.9  & 1.3$\pm$0.1  & 0.78$\pm$0.01 & 110.3$\pm$0.3  & 0.79$\pm$0.01 & ...\\
... & ... & J2 & 1.1$\pm$0.1  & 1.3$\pm$0.1  & 3.33$\pm$0.08 & 112.3$\pm$1.3  & 0.68$\pm$0.15 & ...\\
... & ... & J1 & 1.4$\pm$0.1  & 5.3$\pm$0.4  & 7.76$\pm$0.07 & 152.8$\pm$0.5  & 2.89$\pm$0.15 & ...\\
... & 20180705 & C(core) & 86.9$\pm$4.4  & 89.7$\pm$4.5  & ... & ... & 0.21$\pm$0.01 & 14.3$\pm$1.0 \\
... & ... & J1 & 0.3$\pm$0.2  & 5.9$\pm$3.7  & 7.77$\pm$0.75 & 151.0$\pm$5.5  & 3.60$\pm$1.50 & ...\\
J1421$-$0643 & 19970507 & C(core) & 136.6$\pm$6.8  & 146.6$\pm$7.4  & ... & ... & 0.38$\pm$0.01 & 8.2$\pm$0.5 \\
... & ... & J3 & 43.4$\pm$2.2  & 14.0$\pm$0.8  & 0.79$\pm$0.01 & 68.7$\pm$1.0  & 0.13$^{\star}$ & ...\\
... & ... & J2 & 20.8$\pm$1.2  & 19.7$\pm$1.4  & 2.68$\pm$0.05 & 50.2$\pm$1.0  & 0.25$^{\star}$ & ...\\
... & ... & J1 & 27.2$\pm$1.7  & 47.2$\pm$3.1  & 4.08$\pm$0.05 & 57.5$\pm$0.8  & 1.12$\pm$0.11 & ...\\
... & 20150122 & C(core) & 60.2$\pm$3.0  & 66.5$\pm$3.4  & ... & ... & 0.44$\pm$0.01 & 2.6$\pm$0.2 \\
... & ... & J4 & 22.0$\pm$1.1  & 14.9$\pm$0.8  & 0.96$\pm$0.02 & 68.2$\pm$1.2  & 0.48$\pm$0.04 & ...\\
... & ... & J3 & 18.1$\pm$1.0  & 21.8$\pm$1.3  & 2.20$\pm$0.04 & 57.8$\pm$0.9  & 0.63$\pm$0.07 & ...\\
... & ... & J2 & 10.9$\pm$0.8  & 27.8$\pm$2.1  & 5.15$\pm$0.07 & 56.4$\pm$0.8  & 1.56$\pm$0.14 & ...\\
... & 20150317 & C(core) & 85.0$\pm$4.3  & 92.3$\pm$4.6  & ... & ... & 0.43$\pm$0.01 & 3.9$\pm$0.2 \\
... & ... & J4 & 32.1$\pm$1.6  & 24.7$\pm$1.3  & 0.89$\pm$0.01 & 72.2$\pm$0.8  & 0.61$\pm$0.02 & ...\\
... & ... & J3 & 24.7$\pm$1.3  & 31.7$\pm$1.6  & 2.26$\pm$0.01 & 57.3$\pm$0.3  & 0.73$\pm$0.02 & ...\\
... & ... & J2 & 14.0$\pm$0.9  & 37.4$\pm$2.4  & 5.15$\pm$0.05 & 56.3$\pm$0.6  & 1.58$\pm$0.10 & ...\\
... & 20170219 & C(core) & 70.0$\pm$3.6  & 81.4$\pm$4.3  & ... & ... & 0.46$\pm$0.04 & 3.0$\pm$0.4 \\
... & ... & J4 & 23.4$\pm$1.3  & 25.0$\pm$1.5  & 1.22$\pm$0.03 & 64.0$\pm$1.4  & 0.54$\pm$0.06 & ...\\
... & ... & J3 & 22.2$\pm$1.2  & 23.3$\pm$1.4  & 2.30$\pm$0.03 & 59.0$\pm$0.8  & 0.43$\pm$0.07 & ...\\
... & ... & J2 & 9.7$\pm$1.0  & 30.7$\pm$3.3  & 5.36$\pm$0.12 & 56.4$\pm$1.2  & 1.66$\pm$0.23 & ...\\
... & 20170319 & C(core) & 82.8$\pm$4.2  & 92.8$\pm$4.7  & ... & ... & 0.42$\pm$0.02 & 4.3$\pm$0.4 \\
... & ... & J4 & 29.1$\pm$1.5  & 36.7$\pm$2.0  & 1.19$\pm$0.02 & 61.7$\pm$1.1  & 0.68$\pm$0.05 & ...\\
... & ... & J3 & 20.0$\pm$1.1  & 17.7$\pm$1.1  & 2.39$\pm$0.03 & 56.0$\pm$0.8  & 0.20$^{\star}$ & ...\\
... & ... & J2 & 12.5$\pm$1.2  & 29.8$\pm$3.1  & 5.79$\pm$0.11 & 56.6$\pm$1.1  & 1.47$\pm$0.22 & ...\\
... & 20180326 & C(core) & 79.0$\pm$4.4  & 97.3$\pm$5.7  & ... & ... & 0.50$\pm$0.07 & 3.0$\pm$0.6 \\
... & ... & J4 & 23.0$\pm$1.4  & 28.6$\pm$2.0  & 1.28$\pm$0.06 & 65.9$\pm$2.6  & 0.61$\pm$0.11 & ...\\
... & ... & J3 & 20.7$\pm$1.3  & 23.3$\pm$1.7  & 2.35$\pm$0.06 & 57.6$\pm$1.5  & 0.47$\pm$0.12 & ...\\
... & ... & J2 & 8.3$\pm$1.3  & 30.3$\pm$4.8  & 5.40$\pm$0.22 & 56.4$\pm$2.3  & 1.66$\pm$0.44 & ...\\
J1445$+$0958 & 19970110 & C(core) & 155.0$\pm$7.9  & 161.1$\pm$8.3  & ... & ... & 0.38$\pm$0.02 & 9.0$\pm$0.9 \\
... & ... & J3 & 22.3$\pm$2.5  & 12.5$\pm$2.7  & 1.29$\pm$0.14 & 46.6$\pm$6.0  & 0.41$^{\star}$ & ...\\
... & ... & J2 & 84.4$\pm$4.5  & 100.3$\pm$5.6  & 1.20$\pm$0.03 & $-$118.1$\pm$1.2  & 0.69$\pm$0.05 & ...\\
... & ... & J1 & 38.0$\pm$2.9  & 54.6$\pm$4.8  & 3.32$\pm$0.08 & $-$110.8$\pm$1.4  & 0.83$\pm$0.16 & ...\\
... & 19990510 & C(core) & 178.3$\pm$9.4  & 193.3$\pm$10.5  & ... & ... & 0.34$\pm$0.03 & 12.1$\pm$1.8 \\
... & ... & J2 & 93.1$\pm$5.1  & 142.8$\pm$8.0  & 1.17$\pm$0.02 & $-$119.3$\pm$1.1  & 0.77$\pm$0.04 & ...\\
... & ... & J1 & 24.1$\pm$3.2  & 73.8$\pm$10.3  & 3.38$\pm$0.13 & $-$111.0$\pm$2.2  & 1.46$\pm$0.26 & ...\\
... & 20010705 & C(core) & 206.6$\pm$10.7  & 215.0$\pm$11.4  & ... & ... & 0.29$\pm$0.03 & 18.4$\pm$2.9 \\
... & ... & J2 & 83.8$\pm$4.5  & 112.6$\pm$6.3  & 1.18$\pm$0.02 & $-$116.7$\pm$1.1  & 0.70$\pm$0.05 & ...\\
... & ... & J1 & 32.7$\pm$2.0  & 51.1$\pm$3.3  & 3.27$\pm$0.04 & $-$109.1$\pm$0.7  & 0.82$\pm$0.08 & ...\\
... & 20050622 & C(core) & 300.4$\pm$15.0  & 278.1$\pm$14.0  & ... & ... & 0.15$\pm$0.01 & 99.7$\pm$10.1 \\
... & ... & J3 & 192.8$\pm$9.7  & 32.4$\pm$1.8  & 0.61$\pm$0.01 & 32.9$\pm$0.6  & 0.09$^{\star}$ & ...\\
... & ... & J2 & 76.1$\pm$4.0  & 99.1$\pm$5.4  & 1.31$\pm$0.03 & $-$109.1$\pm$1.2  & 0.87$\pm$0.05 & ...\\
... & ... & J1 & 50.5$\pm$3.0  & 69.9$\pm$4.5  & 3.21$\pm$0.05 & $-$109.1$\pm$0.9  & 0.88$\pm$0.10 & ...\\
... & ... & JF2 & 10.2$\pm$1.7  & 27.2$\pm$4.7  & 6.09$\pm$0.24 & $-$137.7$\pm$2.2  & 1.99$\pm$0.48 & ...\\
... & ... & JF1$^a$ & 5.0$\pm$1.5  & 46.5$\pm$13.8  & 12.12$\pm$0.45 & $-$166.7$\pm$2.1  & 4.59$\pm$0.89 & ...\\
... & 20080402 & C(core) & 222.5$\pm$11.8  & 224.7$\pm$12.5  & ... & ... & 0.17$\pm$0.02 & 59.1$\pm$11.0 \\
... & ... & J3 & 110.0$\pm$6.2  & 67.8$\pm$4.8  & 0.51$\pm$0.02 & 40.9$\pm$1.8  & 0.09$^{\star}$ & ...\\
... & ... & J2 & 21.0$\pm$2.6  & 91.7$\pm$11.7  & 2.05$\pm$0.07 & $-$105.0$\pm$2.0  & 1.03$\pm$0.14 & ...\\
... & ... & J1 & 11.6$\pm$1.2  & 18.7$\pm$2.2  & 4.17$\pm$0.05 & $-$116.4$\pm$0.8  & 0.37$\pm$0.11 & ...\\
... & 20170319 & C(core) & 247.9$\pm$12.5  & 213.9$\pm$10.8  & ... & ... & 0.26$\pm$0.01 & 24.9$\pm$2.3 \\
... & ... & J3 & 204.1$\pm$10.3  & 122.9$\pm$6.3  & 0.72$\pm$0.01 & 40.7$\pm$0.6  & 0.10$^{\star}$ & ...\\
... & ... & J3 & 24.2$\pm$2.0  & 27.8$\pm$2.8  & 1.12$\pm$0.09 & $-$108.1$\pm$4.5  & 0.78$\pm$0.18 & ...\\
... & ... & J2 & 57.7$\pm$3.5  & 74.6$\pm$4.9  & 2.82$\pm$0.05 & $-$103.4$\pm$0.9  & 0.62$\pm$0.09 & ...\\
... & ... & J1 & 15.8$\pm$1.7  & 35.7$\pm$4.2  & 6.16$\pm$0.13 & $-$138.7$\pm$1.2  & 1.39$\pm$0.26 & ...\\
... & ... & JF1$^a$ & 4.7$\pm$1.6  & 42.9$\pm$14.5  & 12.26$\pm$0.45 & $-$169.9$\pm$2.1  & 3.44$\pm$0.89 & ...\\
... & 20180326 & C(core) & 199.2$\pm$10.0  & 170.2$\pm$8.6  & ... & ... & 0.31$\pm$0.01 & 13.0$\pm$1.0 \\
... & ... & J3 & 174.6$\pm$8.8  & 110.8$\pm$5.7  & 0.75$\pm$0.01 & 39.7$\pm$0.6  & 0.09$^{\star}$ & ...\\
... & ... & J3 & 25.6$\pm$1.7  & 47.6$\pm$3.4  & 1.32$\pm$0.06 & $-$99.4$\pm$2.7  & 1.28$\pm$0.13 & ...\\
... & ... & J2 & 55.7$\pm$3.0  & 74.4$\pm$4.2  & 2.85$\pm$0.03 & $-$104.3$\pm$0.6  & 0.77$\pm$0.06 & ...\\
... & ... & J1 & 8.3$\pm$1.4  & 32.6$\pm$5.7  & 5.40$\pm$0.22 & $-$137.3$\pm$2.4  & 2.26$\pm$0.45 & ...\\
... & ... & JF1$^a$ & 3.2$\pm$0.8  & 39.0$\pm$10.3  & 11.48$\pm$0.36 & $-$168.6$\pm$1.8  & 4.34$\pm$0.71 & ...\\
J1606$+$3124 & 19960515 & S & 372.8$\pm$19.1  & 314.5$\pm$16.0  & ... & ... & 0.58$\pm$0.01 & 10.8$\pm$0.6 \\
... & ... & C & 70.0$\pm$4.0  & 79.4$\pm$4.9  & 2.49$\pm$0.01 & $-$19.9$\pm$0.3  & 0.39$\pm$0.02 & 5.2$\pm$0.5\\
... & ... & N & 34.8$\pm$2.2  & 48.5$\pm$3.3  & 7.97$\pm$0.02 & $-$17.3$\pm$0.2  & 0.75$\pm$0.03 & ...\\
... & 20140809 & S & 383.3$\pm$19.4  & 444.3$\pm$22.7  & ... & ... & 0.58$\pm$0.01 & 11.7$\pm$0.8 \\
... & ... & C & 77.6$\pm$4.3  & 99.4$\pm$5.8  & 2.60$\pm$0.02 & $-$17.7$\pm$0.6  & 0.64$\pm$0.03 & 2.2$\pm$0.2\\
... & ... & N & 26.9$\pm$1.9  & 39.9$\pm$3.0  & 8.29$\pm$0.03 & $-$17.0$\pm$0.4  & 0.87$\pm$0.07 & ...\\
... & 20170319 & S & 334.2$\pm$16.8  & 347.4$\pm$17.6  & ... & ... & 0.44$\pm$0.02 & 17.3$\pm$1.4 \\
... & ... & S0 & 9.0$\pm$1.7  & 21.2$\pm$4.3  & 1.12$\pm$0.05 & $-$24.8$\pm$1.4  & 0.13$^{\star}$ & ...\\
... & ... & C & 67.0$\pm$3.7  & 80.9$\pm$4.7  & 2.66$\pm$0.01 & $-$17.4$\pm$0.6  & 0.53$\pm$0.03 & 2.8$\pm$0.3\\
... & ... & N & 24.0$\pm$1.4  & 34.3$\pm$2.0  & 8.28$\pm$0.08 & $-$16.3$\pm$0.4  & 0.75$\pm$0.06 & ...\\
... & 20170918 & S & 328.3$\pm$16.7  & 326.2$\pm$16.8  & ... & ... & 0.53$\pm$0.01 & 10.4$\pm$0.6 \\
... & ... & C & 63.4$\pm$3.7  & 95.4$\pm$5.9  & 2.49$\pm$0.02 & $-$15.7$\pm$0.8  & 0.61$\pm$0.04 & 2.3$\pm$0.4\\
... & ... & N & 29.7$\pm$2.0  & 38.7$\pm$2.9  & 8.14$\pm$0.02 & $-$16.6$\pm$0.3  & 1.16$\pm$0.04 & ...\\
... & 20180326 & S & 275.1$\pm$13.9  & 296.5$\pm$15.1  & ... & ... & 0.47$\pm$0.01 & 11.9$\pm$0.7 \\
... & ... & S0 & 10.0$\pm$1.7  & 24.4$\pm$4.5  & 1.11$\pm$0.02 & $-$3.0$\pm$1.3  & 0.14$^{\star}$ & ...\\
... & ... & C & 71.1$\pm$3.9  & 82.1$\pm$4.7  & 2.63$\pm$0.01 & $-$16.6$\pm$0.6  & 0.57$\pm$0.03 & 2.3$\pm$0.3\\
... & ... & N & 26.0$\pm$1.4  & 36.0$\pm$2.1  & 8.24$\pm$0.02 & $-$16.1$\pm$0.3  & 0.76$\pm$0.04 & ...\\
J1939$-$1002 & 19970110 & C(core) & 238.6$\pm$11.9  & 207.2$\pm$10.4  & ... & ... & 0.09$\pm$0.01 & 236.6$\pm$27.0 \\
... & ... & J3 & 174.0$\pm$8.7  & 86.7$\pm$4.4  & 1.06$\pm$0.01 & 15.9$\pm$0.3  & 0.08$^{\star}$ & ...\\
... & ... & JX$^a$ & 82.4$\pm$4.4  & 12.3$\pm$1.6  & 1.14$\pm$0.02 & $-$160.4$\pm$1.2  & 0.17$^{\star}$ & ...\\
... & ... & J2 & 32.7$\pm$2.0  & 37.9$\pm$2.6  & 3.63$\pm$0.05 & 9.2$\pm$0.8  & 0.53$\pm$0.10 & ...\\
... & ... & J1 & 3.9$\pm$0.5  & 5.5$\pm$0.9  & 23.63$\pm$0.18 & 31.5$\pm$0.4  & 0.72$\pm$0.36 & ...\\
... & 19970507 & C(core) & 314.6$\pm$15.8  & 302.6$\pm$15.2  & ... & ... & 0.42$\pm$0.01 & 14.3$\pm$0.9 \\
... & ... & J3 & 204.3$\pm$10.3  & 93.1$\pm$4.9  & 1.21$\pm$0.01 & 9.4$\pm$0.4  & 0.11$^{\star}$ & ...\\
... & ... & JX$^a$ & 18.2$\pm$1.6  & 11.2$\pm$1.6  & 1.16$\pm$0.11 & $-$97.3$\pm$5.3  & 0.38$^{\star}$ & ...\\
... & ... & J2 & 47.3$\pm$2.6  & 55.1$\pm$3.2  & 3.65$\pm$0.03 & 7.4$\pm$0.5  & 0.65$\pm$0.07 & ...\\
... & ... & J1 & 5.1$\pm$0.7  & 5.2$\pm$1.0  & 24.05$\pm$0.19 & 26.4$\pm$0.5  & 0.52$^{\star}$ & ...\\
... & 19981001 & C(core) & 308.8$\pm$15.5  & 286.9$\pm$14.4  & ... & ... & 0.27$\pm$0.01 & 31.9$\pm$2.1 \\
... & ... & J3 & 205.3$\pm$10.3  & 84.0$\pm$4.3  & 1.14$\pm$0.01 & 13.1$\pm$0.3  & 0.09$^{\star}$ & ...\\
... & ... & J2 & 44.6$\pm$2.4  & 52.2$\pm$2.9  & 3.56$\pm$0.03 & 8.1$\pm$0.5  & 0.74$\pm$0.06 & ...\\
... & ... & J1 & 1.5$\pm$0.6  & 7.5$\pm$3.1  & 21.28$\pm$0.61 & 28.2$\pm$1.7  & 2.19$\pm$1.23 & ...\\
... & 20060711 & C(core) & 282.4$\pm$14.8  & 327.2$\pm$17.7  & ... & ... & 0.38$\pm$0.04 & 17.7$\pm$2.7 \\
... & ... & J2 & 21.9$\pm$4.3  & 49.5$\pm$10.6  & 3.08$\pm$0.24 & 8.9$\pm$4.4  & 1.12$\pm$0.48 & ...\\
... & 20170205 & C(core) & 352.5$\pm$17.6  & 365.0$\pm$18.3  & ... & ... & 0.38$\pm$0.01 & 20.9$\pm$1.1 \\
... & ... & J3 & 210.4$\pm$10.5  & 66.6$\pm$3.4  & 1.14$\pm$0.01 & 0.9$\pm$0.2  & 0.07$^{\star}$ & ...\\
... & ... & JX$^a$ & 76.6$\pm$4.0  & 36.5$\pm$2.2  & 1.46$\pm$0.02 & $-$159.1$\pm$0.7  & 0.14$^{\star}$ & ...\\
... & ... & J2 & 31.5$\pm$1.8  & 30.0$\pm$1.9  & 3.34$\pm$0.03 & 6.9$\pm$0.6  & 0.19$^{\star}$ & ...\\
... & 20190204 & C(core) & 347.7$\pm$17.4  & 337.5$\pm$16.9  & ... & ... & 0.39$\pm$0.01 & 22.3$\pm$1.2 \\
... & ... & J3 & 193.2$\pm$9.7  & 49.1$\pm$2.6  & 1.32$\pm$0.01 & $-$0.8$\pm$0.3  & 0.10$^{\star}$ & ...\\
... & ... & JX$^a$ & 153.4$\pm$7.7  & 35.9$\pm$1.9  & 1.37$\pm$0.01 & $-$163.7$\pm$0.3  & 0.10$^{\star}$ & ...\\
... & ... & J2 & 38.5$\pm$2.0  & 36.1$\pm$2.0  & 3.31$\pm$0.03 & 5.9$\pm$0.5  & 0.64$\pm$0.06 & ...\\
... & ... & J1 & 2.7$\pm$0.5  & 8.8$\pm$1.6  & 21.93$\pm$0.27 & 29.1$\pm$0.7  & 2.75$\pm$0.53 & ...\\
J2102$+$6015 & 19940812 & E0 & 138.4$\pm$7.7  & 145.1$\pm$8.8  & ... & ... & 0.25$\pm$0.07 & 22.2$\pm$4.6 \\
... & 20061218 & E0 & 91.2$\pm$4.6  & 48.6$\pm$2.4  & ... & ... & 0.31$\pm$0.01 & 4.6$\pm$0.3 \\
... & ... & E2 & 64.6$\pm$3.3  & 36.9$\pm$1.9  & 0.54$\pm$0.01 & $-$95.7$\pm$0.90  & 0.83$\pm$0.02 & ...\\
... & ... & E1 & 74.7$\pm$3.8  & 54.1$\pm$2.8  & 0.58$\pm$0.01 & 90.2$\pm$0.92  & 0.65$\pm$0.02 & ...\\
... & ... & W0 & 5.4$\pm$0.4  & 5.3$\pm$0.5  & 9.84$\pm$0.07 & $-$89.2$\pm$0.43  & 0.29$^{\star}$ & ...\\
... & 20170205 & E0 & 101.3$\pm$5.1  & 121.6$\pm$6.1  & ... & ... & 0.60$\pm$0.00 & 3.3$\pm$0.2 \\
... & ... & E1 & 26.8$\pm$1.4  & 9.8$\pm$0.6  & 0.79$\pm$0.01 & 79.7$\pm$0.67  & 0.08$^{\star}$ & ...\\
... & ... & E2 & 34.7$\pm$1.8  & 25.6$\pm$1.3  & 0.79$\pm$0.01 & $-$98.2$\pm$0.52  & 0.78$\pm$0.01 & ...\\
... & ... & C & 1.3$\pm$0.2  & 1.3$\pm$0.3  & 4.68$\pm$0.15 & $-$88.2$\pm$1.86  & 0.36$^{\star}$ & ...\\
... &  & W0 & 2.4$\pm$0.3  & 5.5$\pm$0.7  & 9.94$\pm$0.10 & $-$88.8$\pm$0.59  & 1.36$\pm$0.21 & ...\\
... & 20170501 & E0 & 102.5$\pm$5.1  & 115.3$\pm$5.8  & ... & ... & 0.60$\pm$0.01 & 2.9$\pm$0.2 \\
... & ... & E2 & 51.0$\pm$2.6  & 25.5$\pm$1.3  & 0.82$\pm$0.01 & $-$102.5$\pm$0.44  & 0.77$\pm$0.01 & ...\\
... & ... & E1 & 31.0$\pm$1.6  & 6.2$\pm$0.5  & 0.92$\pm$0.02 & 80.3$\pm$1.03  & 0.13$^{\star}$ & ...\\
... & ... & W0 & 3.0$\pm$0.3  & 5.2$\pm$0.5  & 10.12$\pm$0.09 & $-$89.5$\pm$0.49  & 1.06$\pm$0.17 & ...\\
... & 20180609 & E0 & 94.2$\pm$4.7  & 103.2$\pm$5.2  & ... & ... & 0.58$\pm$0.00 & 2.8$\pm$0.1 \\
... & ... & E1 & 36.3$\pm$1.8  & 5.9$\pm$0.3  & 0.81$\pm$0.00 & 111.1$\pm$0.27  & 0.07$^{\star}$ & ...\\
... & ... & E2 & 43.4$\pm$2.2  & 22.0$\pm$1.1  & 0.82$\pm$0.00 & $-$103.2$\pm$0.27  & 0.78$\pm$0.01 & ...\\
... & ... & P1$^a$ & 18.5$\pm$1.0  & 5.1$\pm$0.3  & 1.09$\pm$0.02 & 44.5$\pm$0.80  & 0.13$^{\star}$ & ...\\
... & ... & W2$^b$ & 2.3$\pm$0.2  & 2.5$\pm$0.3  & 10.70$\pm$0.10 & $-$89.6$\pm$0.56  & 0.41$\pm$0.21 & ...\\
\enddata
\tablenotetext{^{a}}{These components could be part of a diffuse emission or newly ejected components that cannot be fully fitted with the nearby Gaussian components. For such components, a point-source model was used to account for the extra flux density.}
\tablenotetext{^{b}}{The W feature of J2102+6015 can be fitted with 3 components (W0, W1, and W2), W2 represents the westernmost component. See details in \citealt{2021MNRAS.507.3736Z}.}
\tablenotetext{^{*}}{Upper limit of the component size.}
\tablecomments{Columns: (1) J2000 source name; (2) observing date; (3) component identifier; (4) peak intensity of the component; (5) integrated flux density of the component through Gaussian model fitting; (6) separation of the component from core; (7) position angle of the component with respect to the core, measured from north to east; (8) FWHM size of the fitted Gaussian model component.}
\end{deluxetable}

\begin{figure*}
\gridline{
	\includegraphics[width=0.24\textwidth]{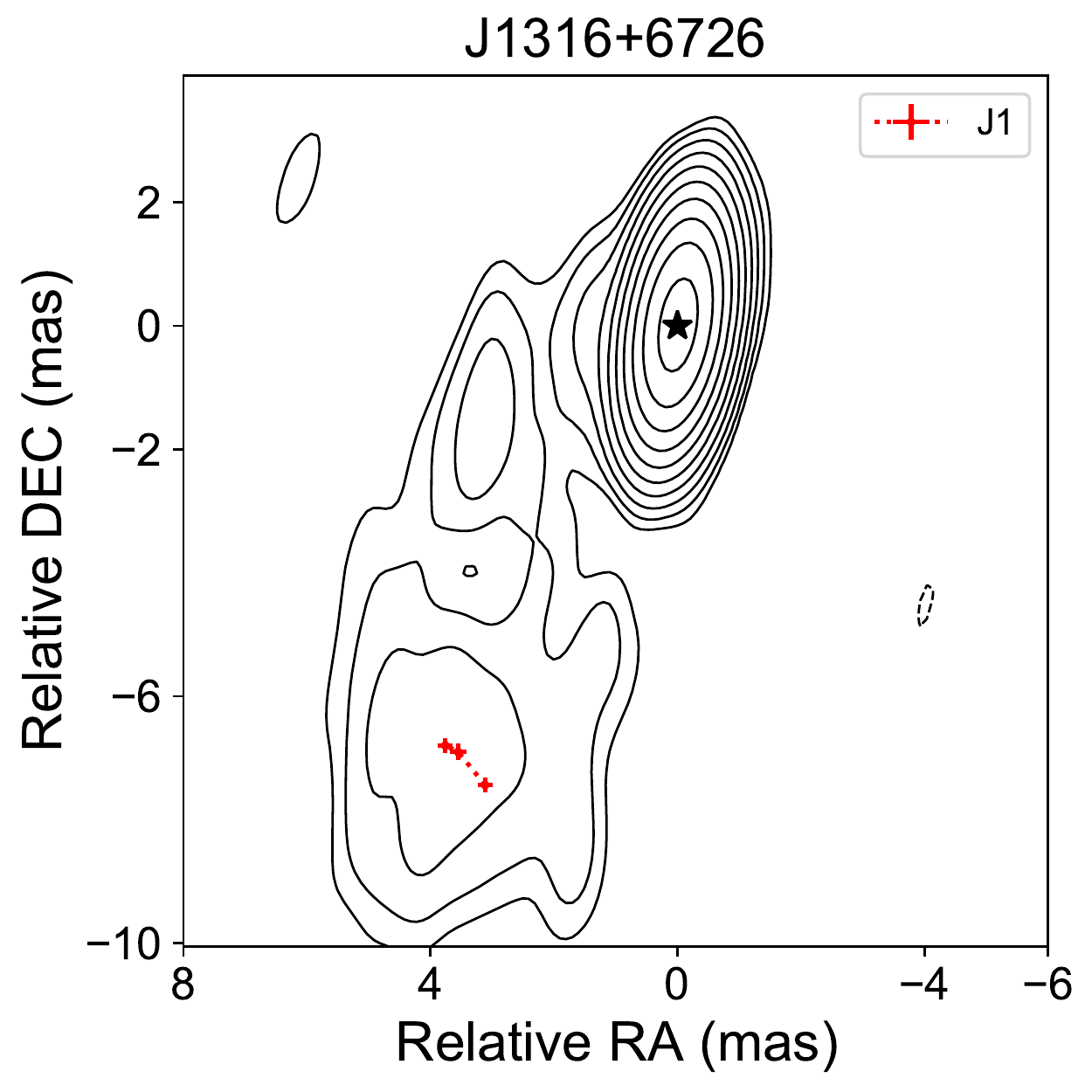}
	\includegraphics[width=0.3\textwidth]{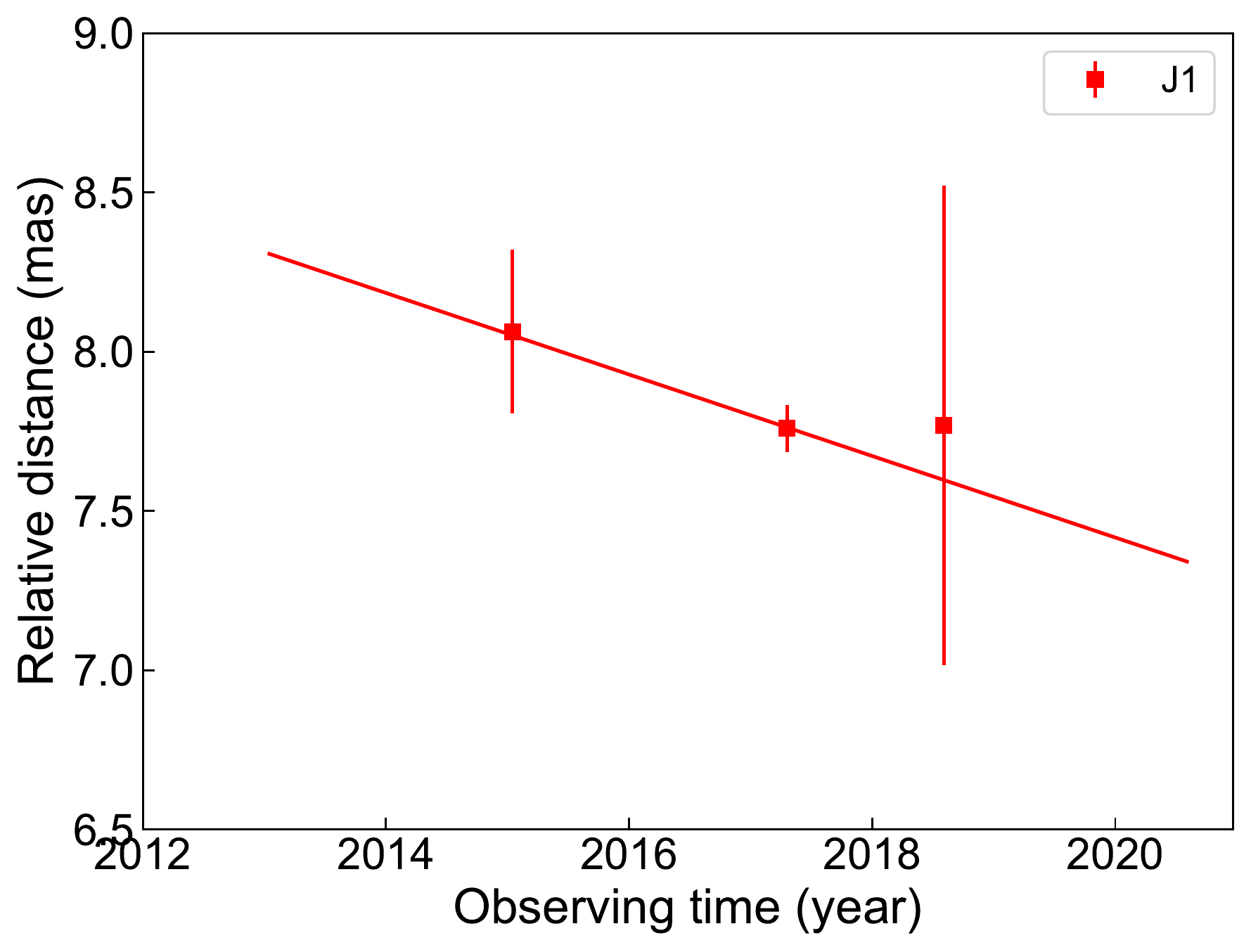}
	\includegraphics[width=0.3\textwidth]{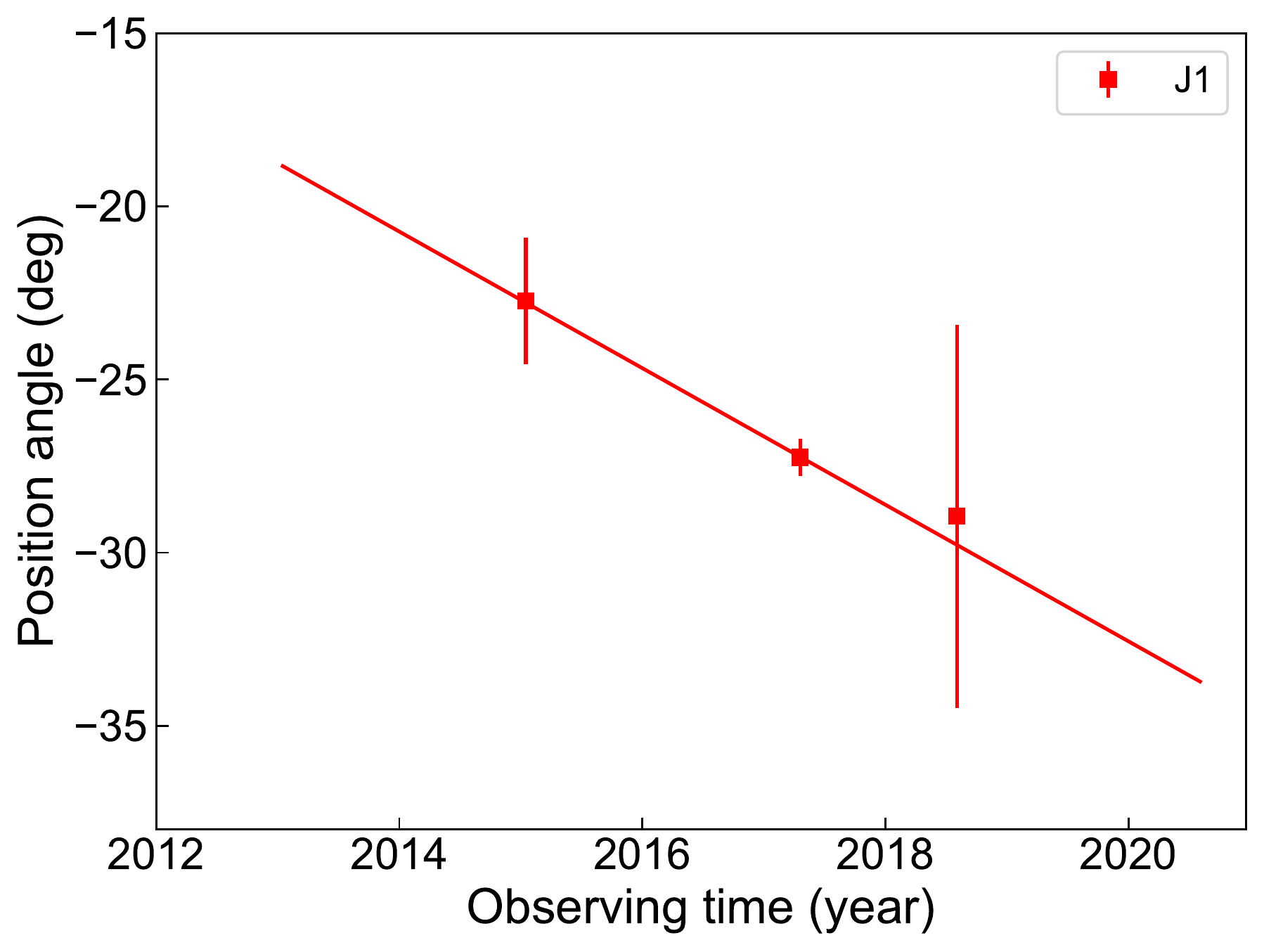}
}
\gridline{
	\includegraphics[width=0.24\textwidth]{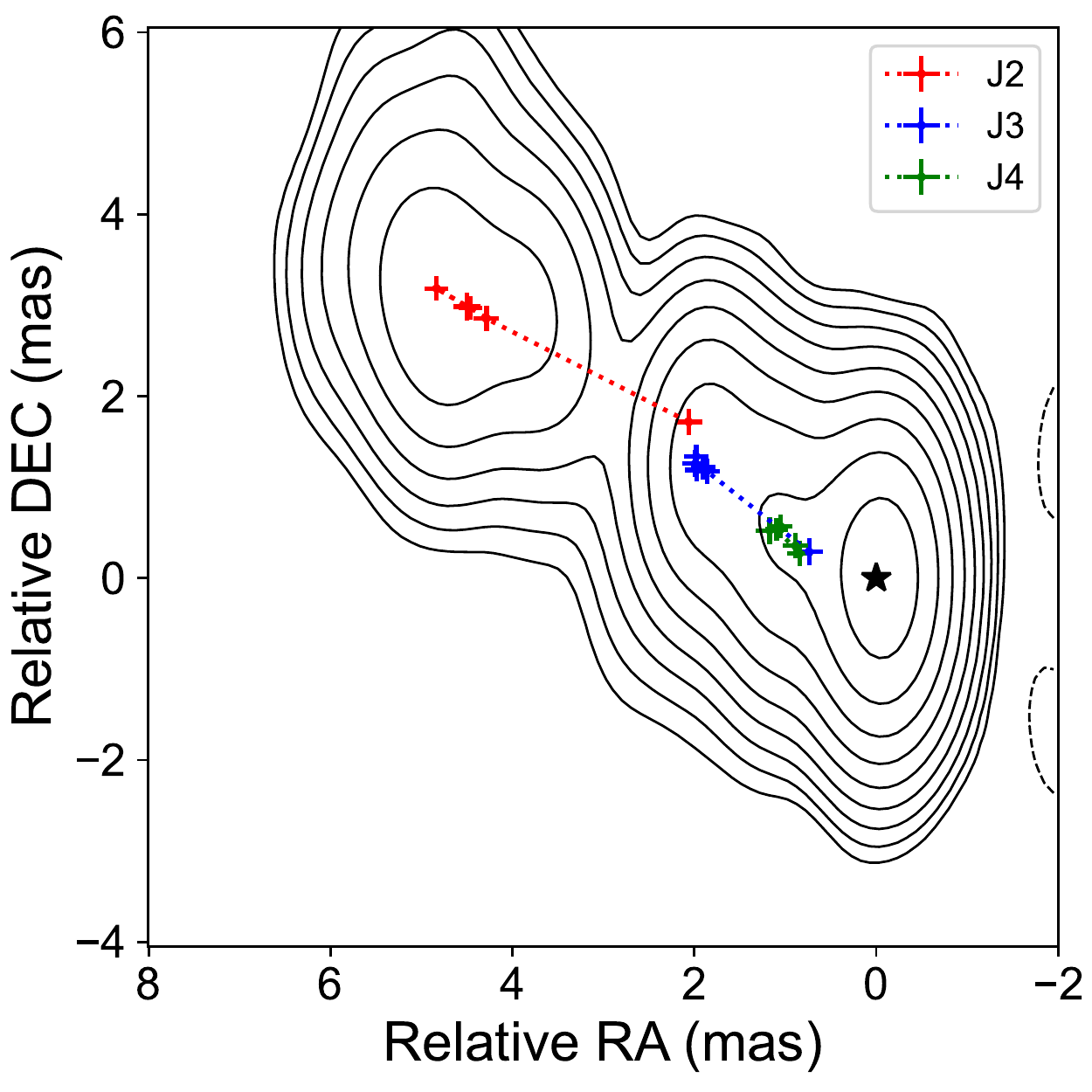}
	\includegraphics[width=0.3\textwidth]{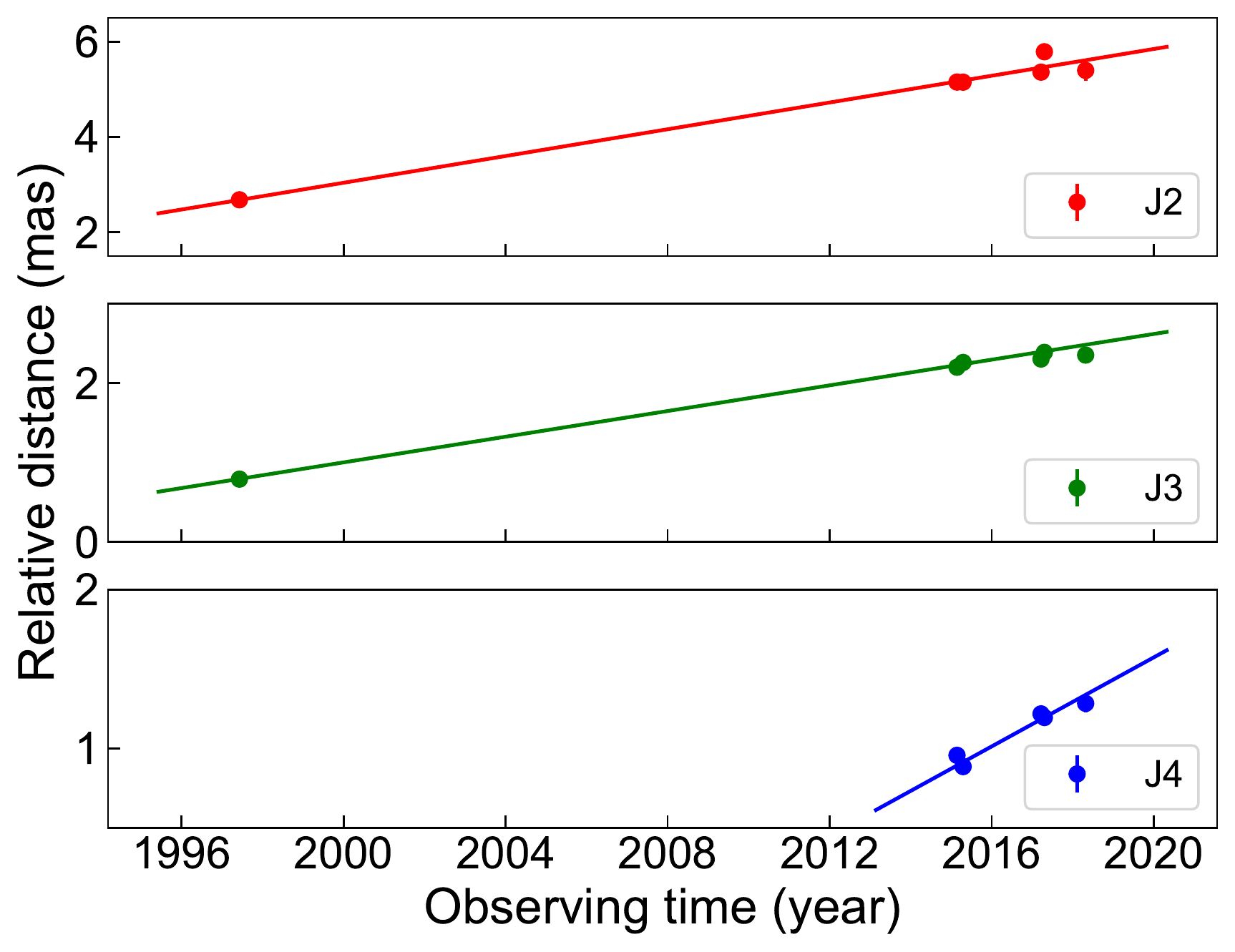}
	\includegraphics[width=0.3\textwidth]{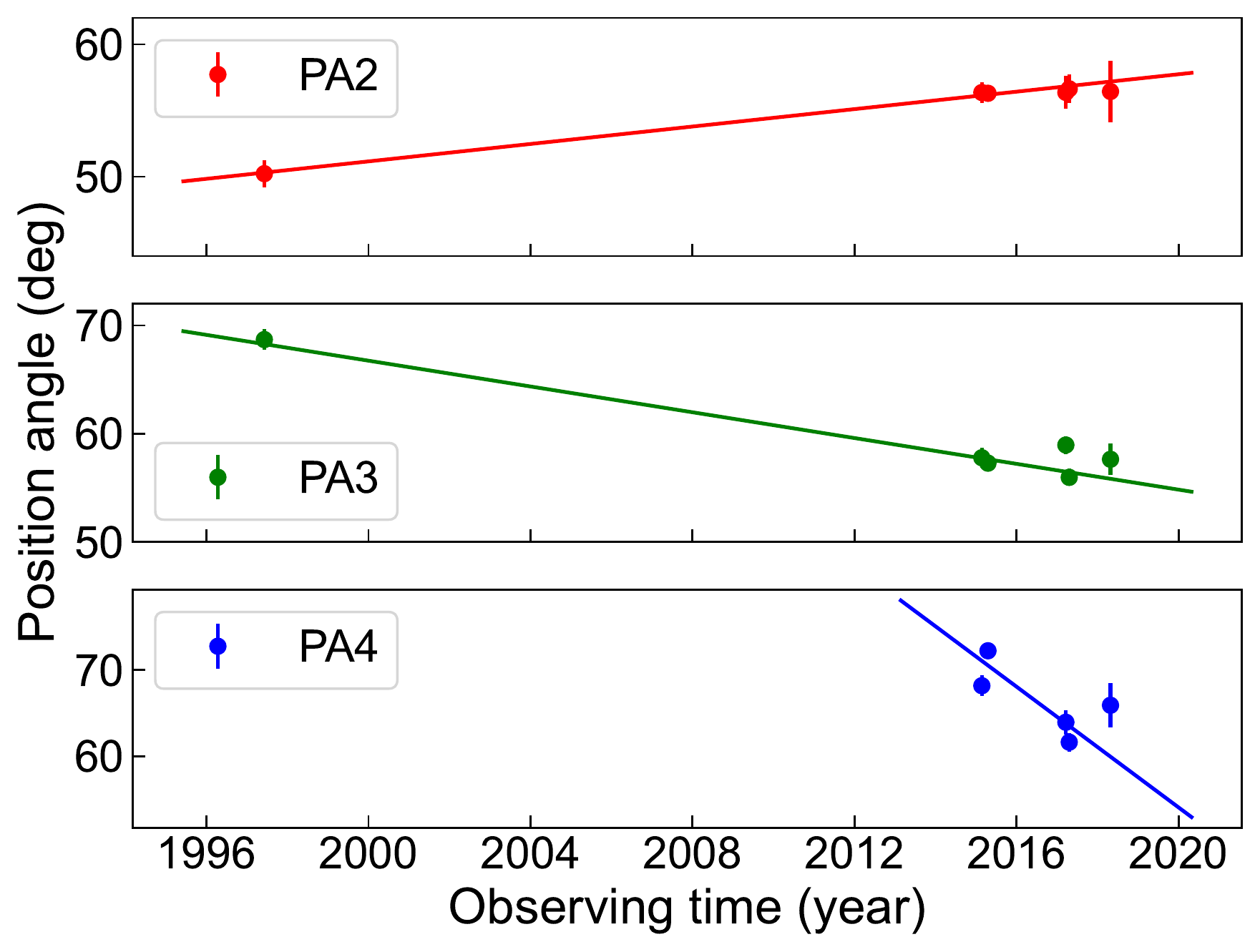}
}
\gridline{
	\includegraphics[width=0.24\textwidth]{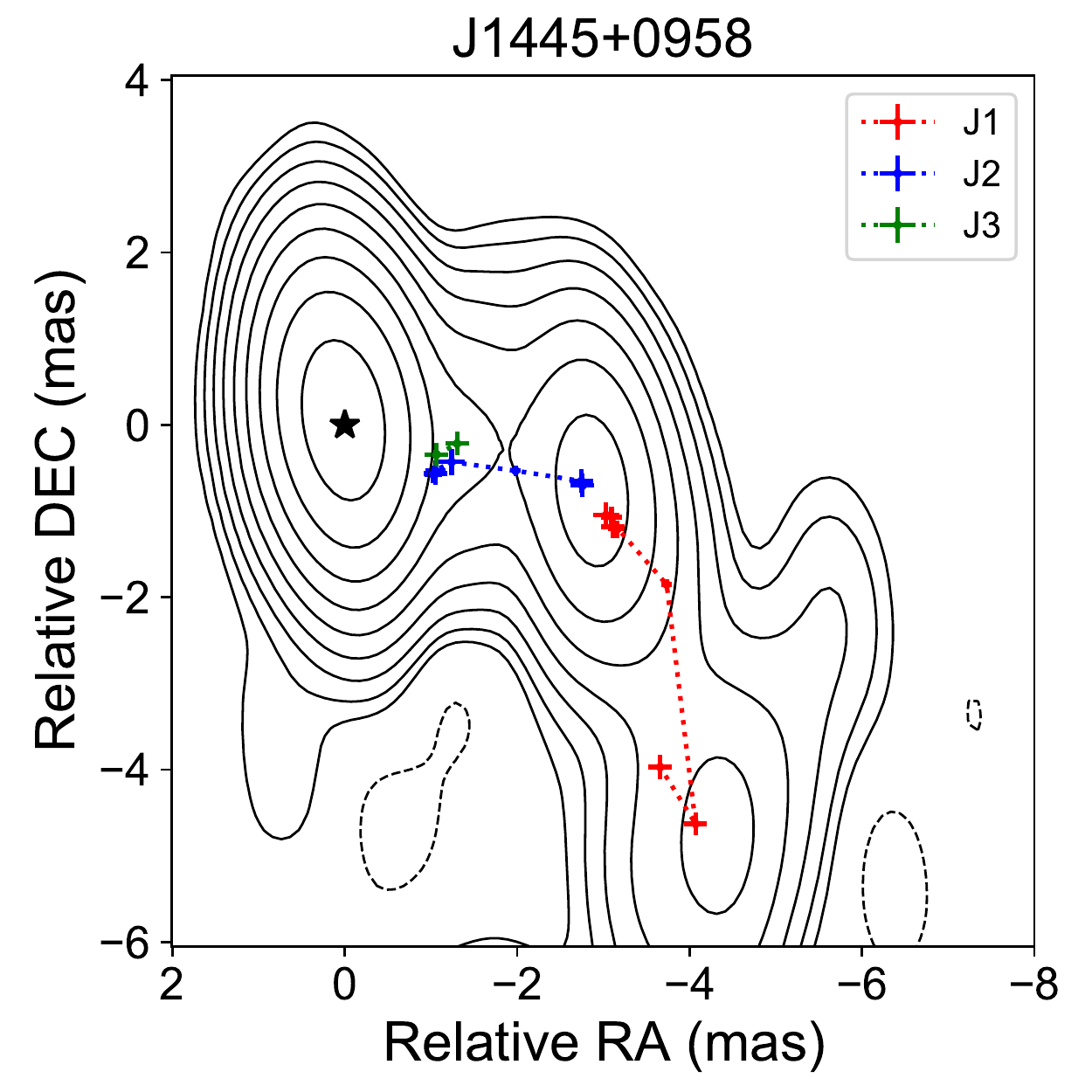}
	\includegraphics[width=0.3\textwidth]{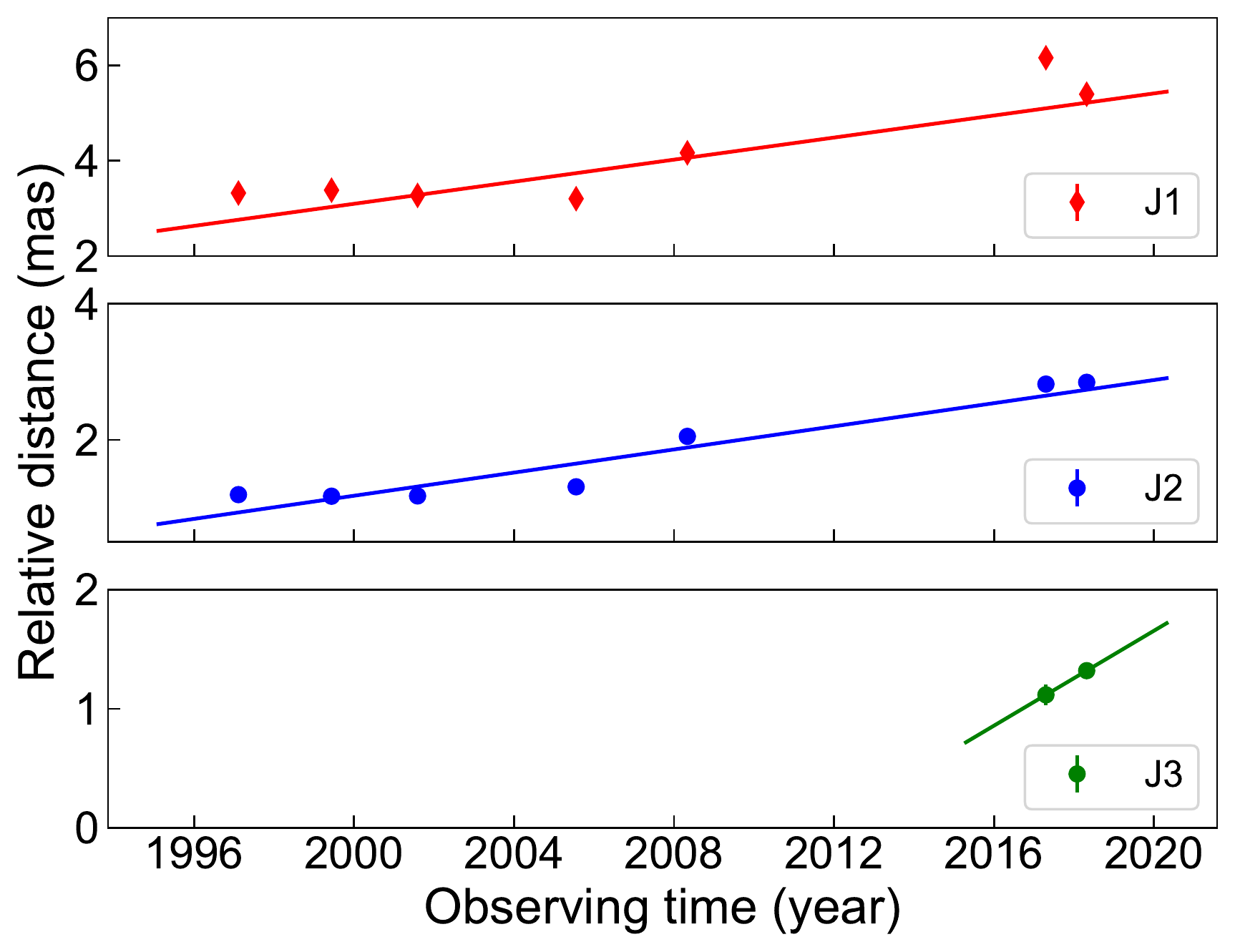}
	\includegraphics[width=0.3\textwidth]{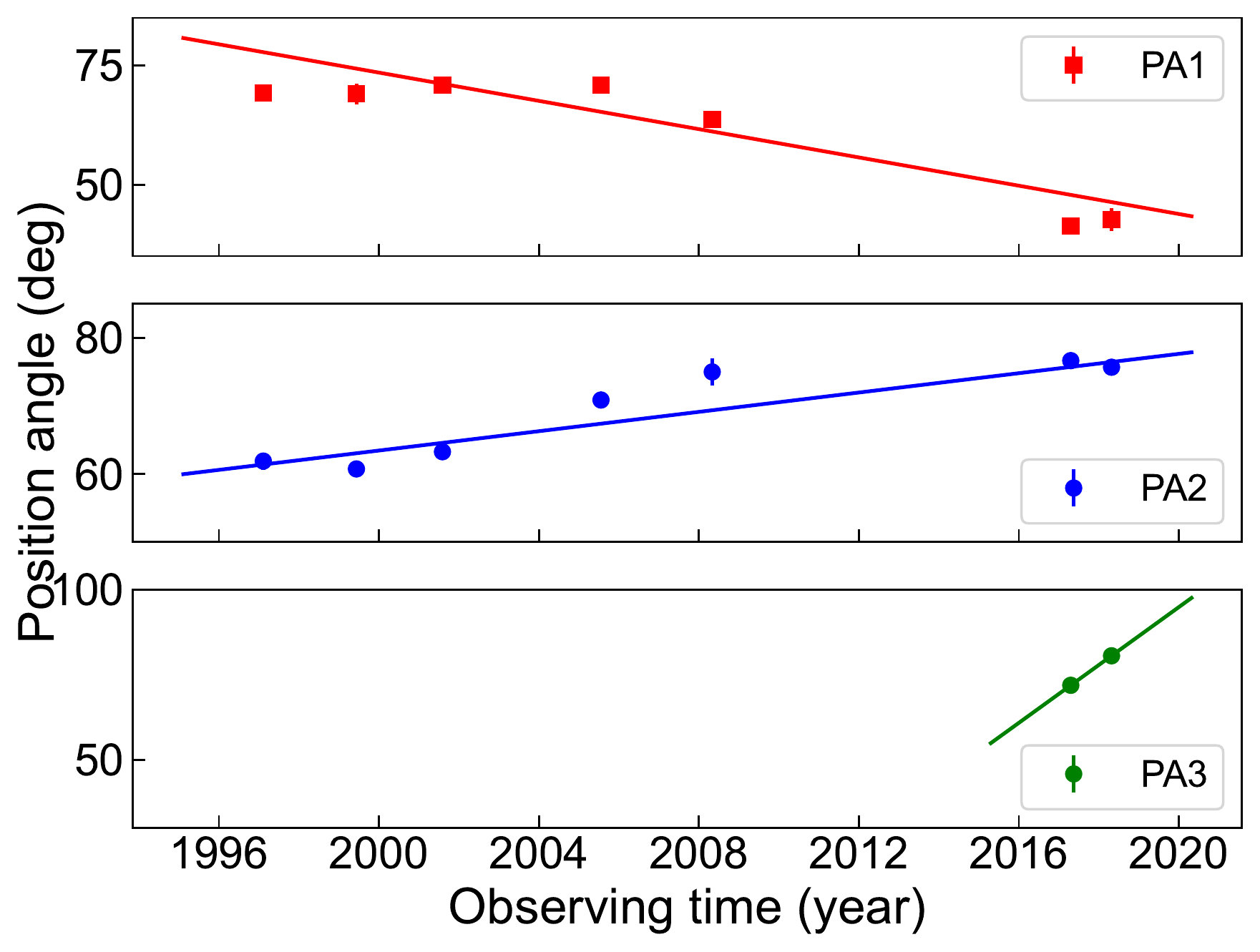}
}
\caption{Jet proper motions (continued)}
\figurenum{3}
\end{figure*}
\begin{figure*}
\gridline{
	\includegraphics[width=0.24\textwidth]{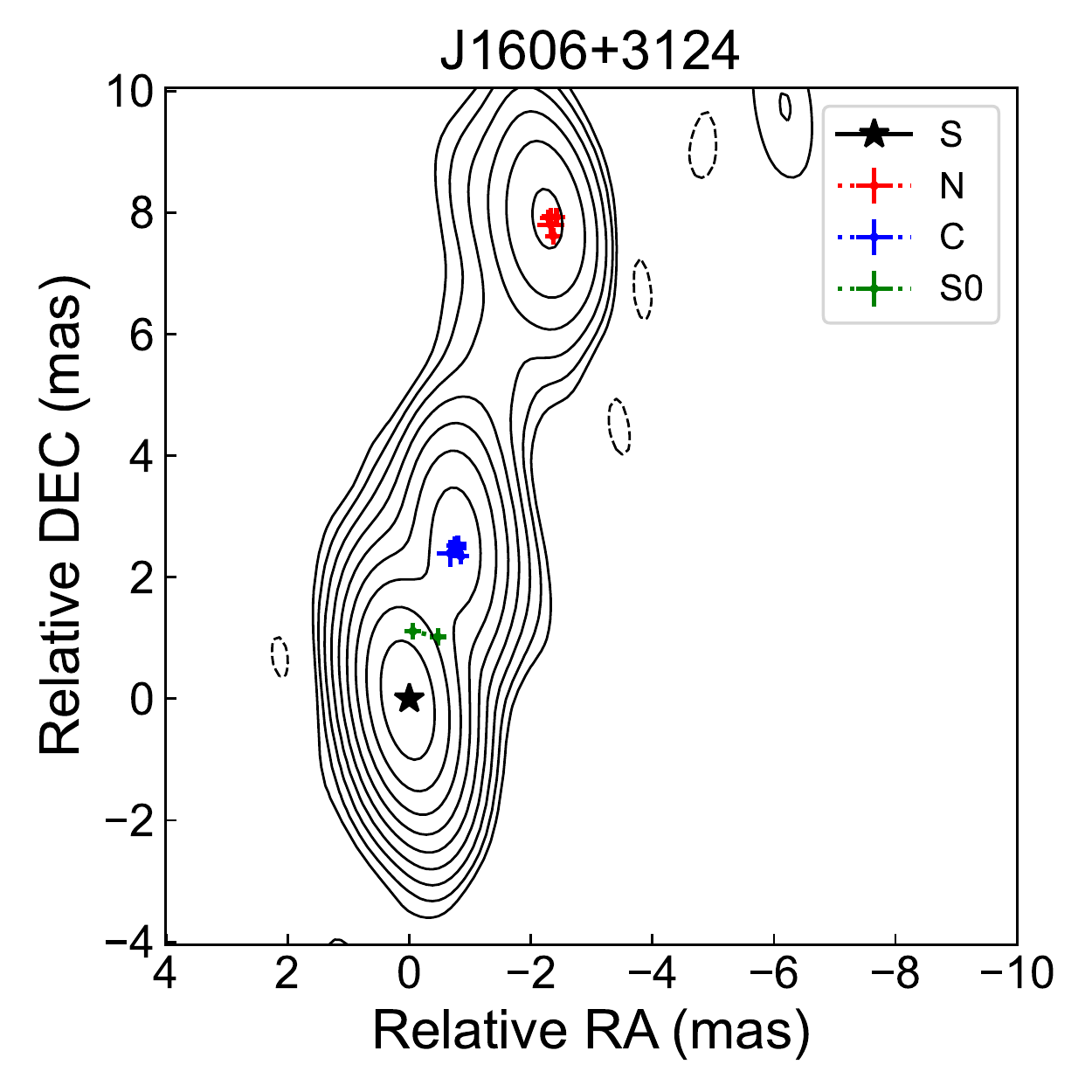}
	\includegraphics[width=0.3\textwidth]{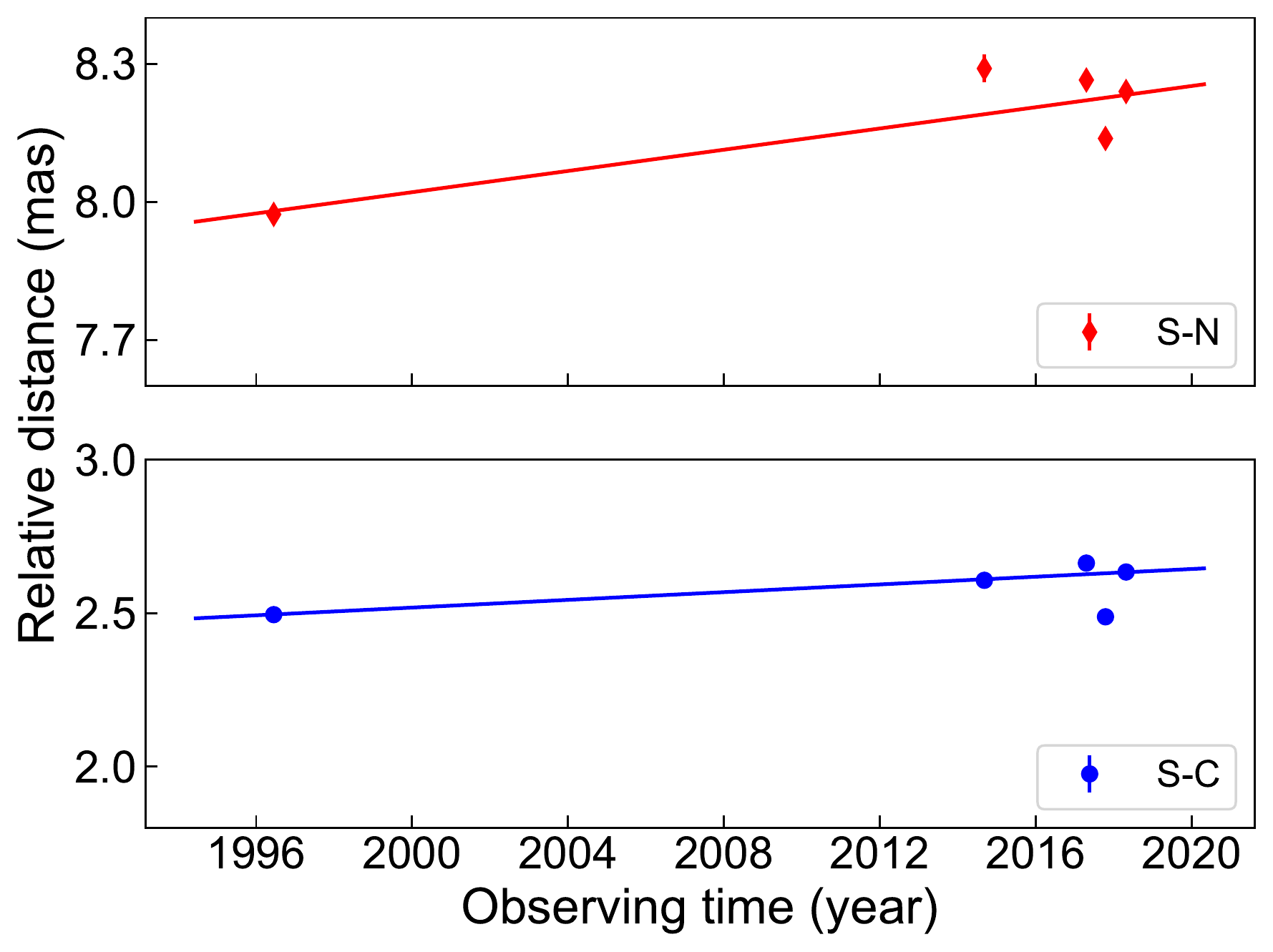}
	\includegraphics[width=0.3\textwidth]{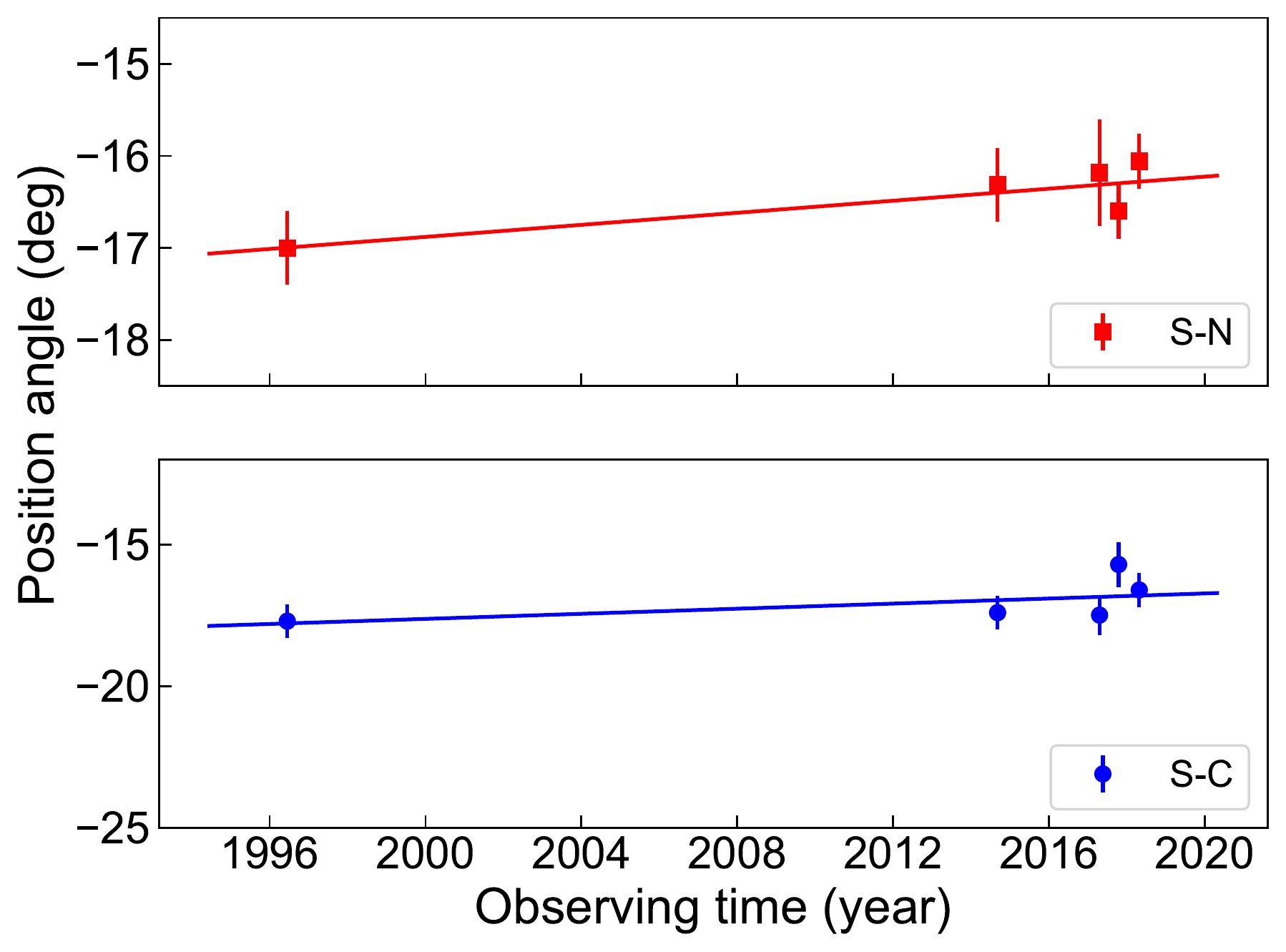}
}
\gridline{
	\includegraphics[width=0.24\textwidth]{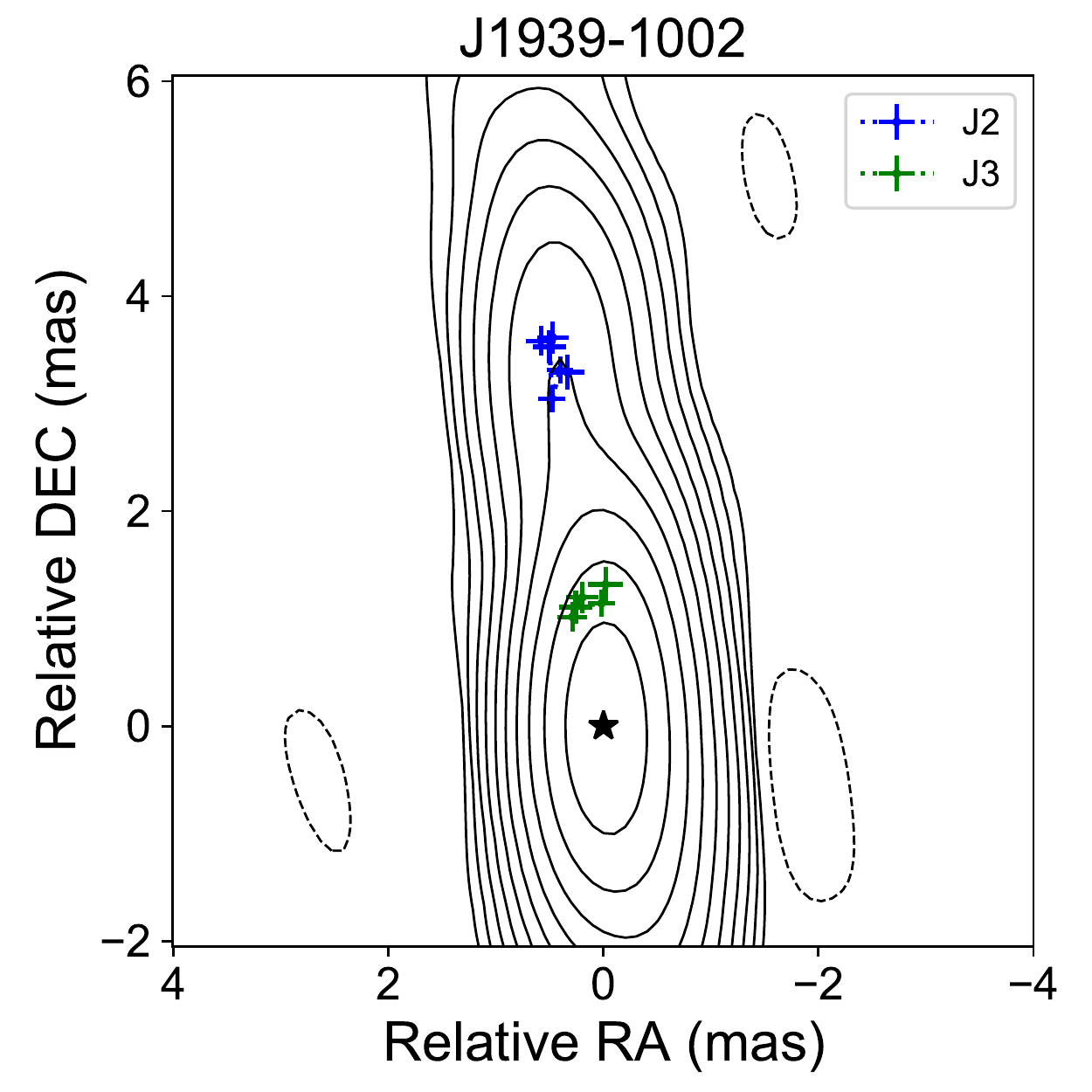}
	\includegraphics[width=0.3\textwidth]{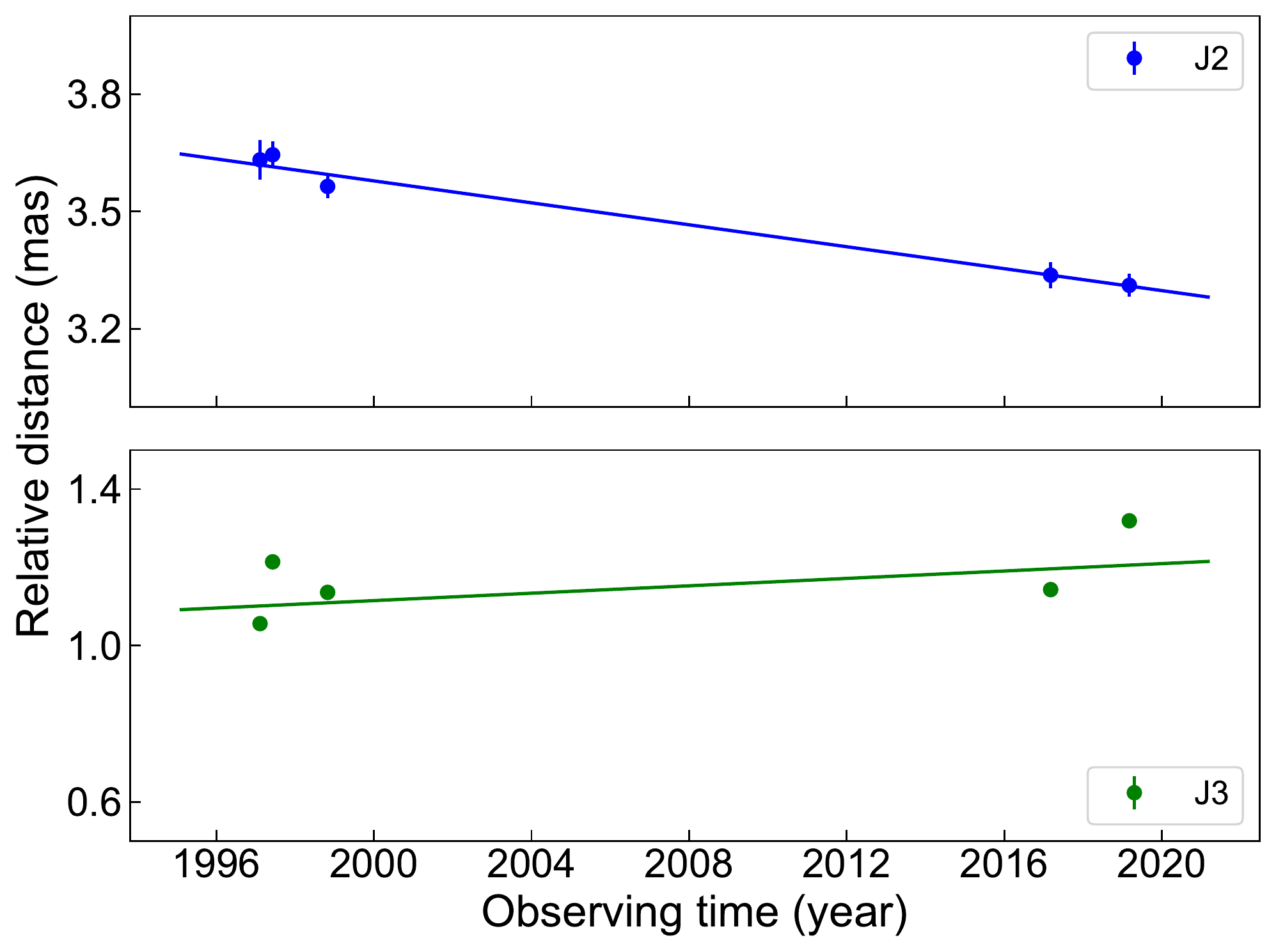}
	\includegraphics[width=0.3\textwidth]{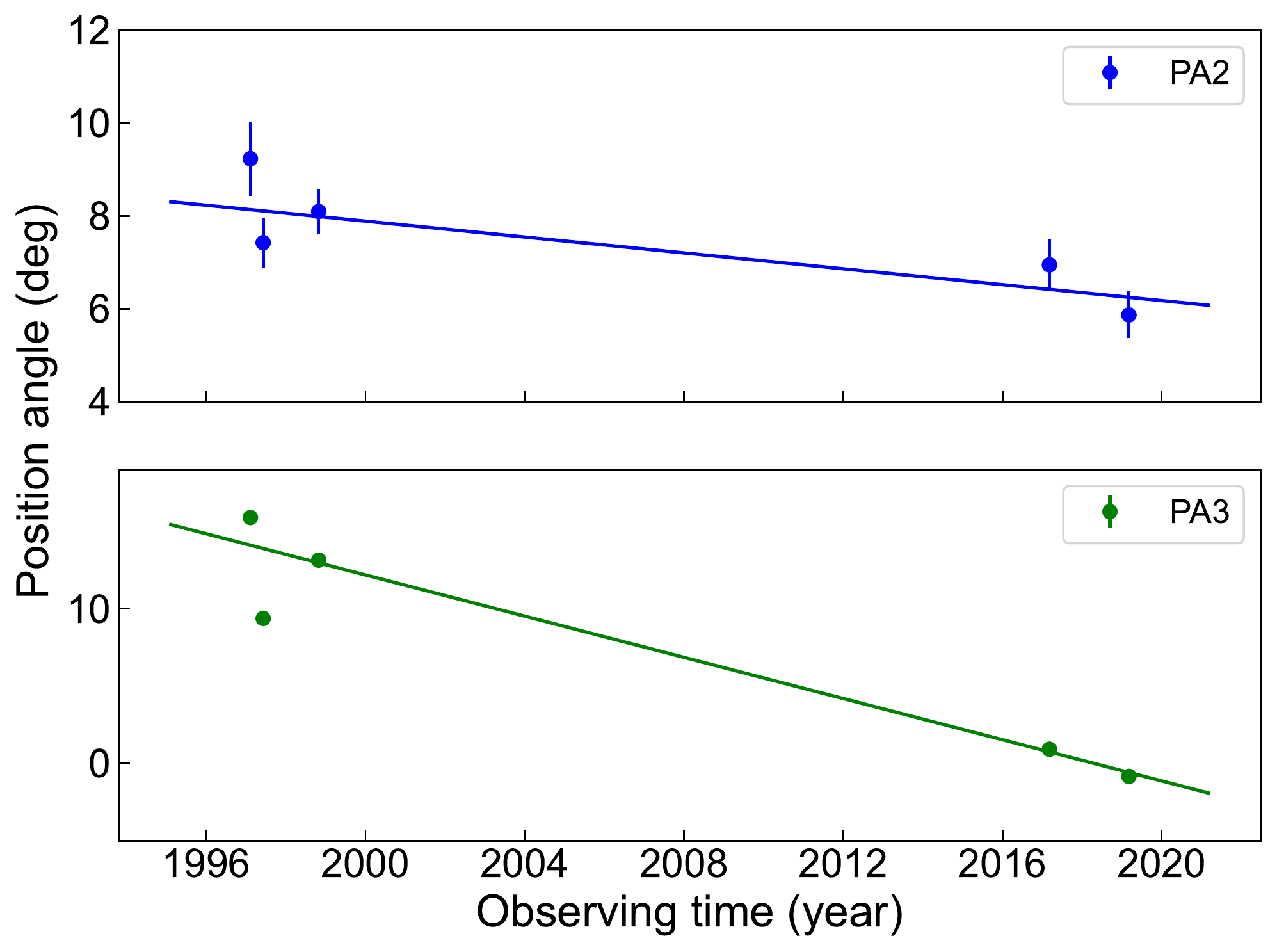} 
}
\gridline{
	\includegraphics[width=0.24\textwidth]{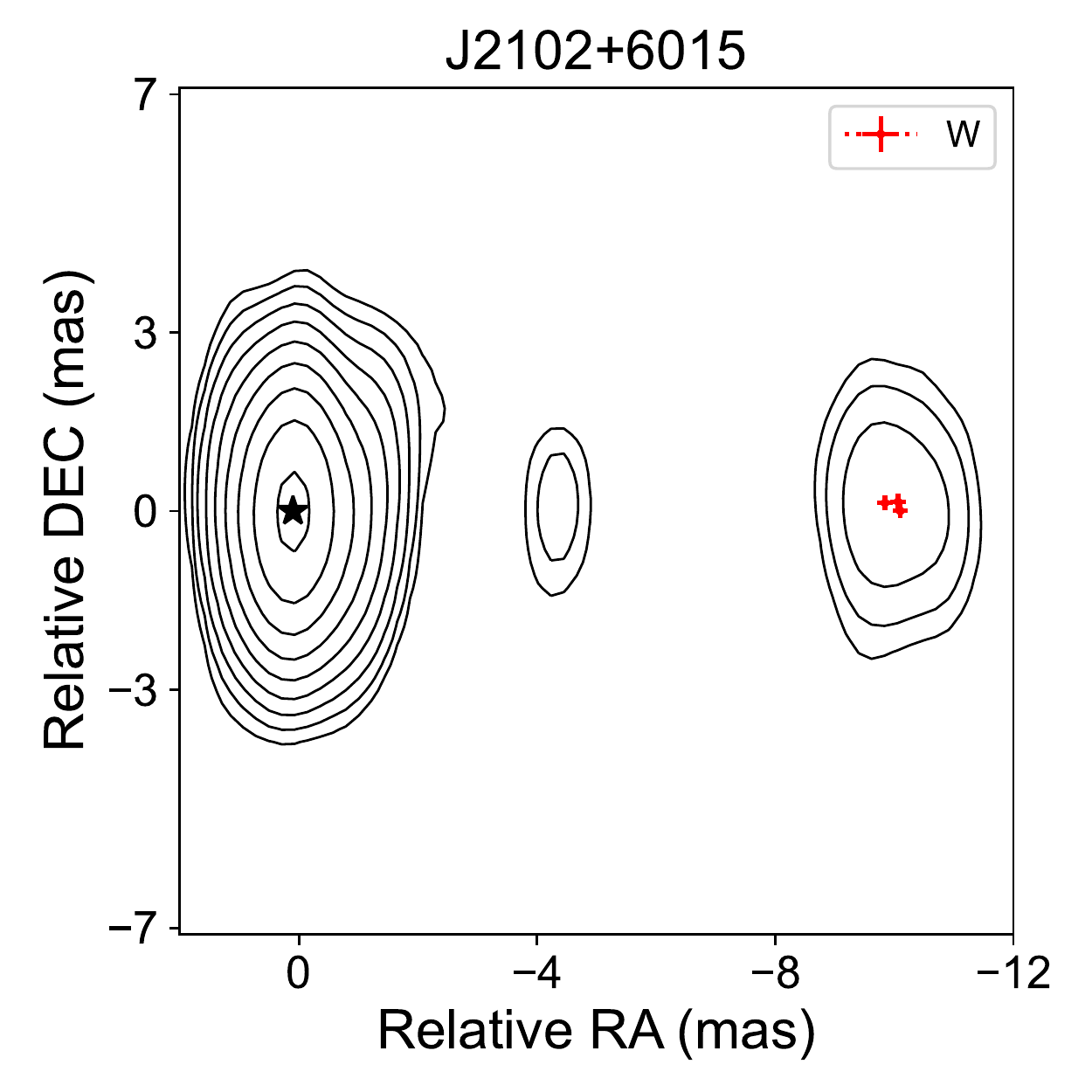}
	\includegraphics[width=0.3\textwidth]{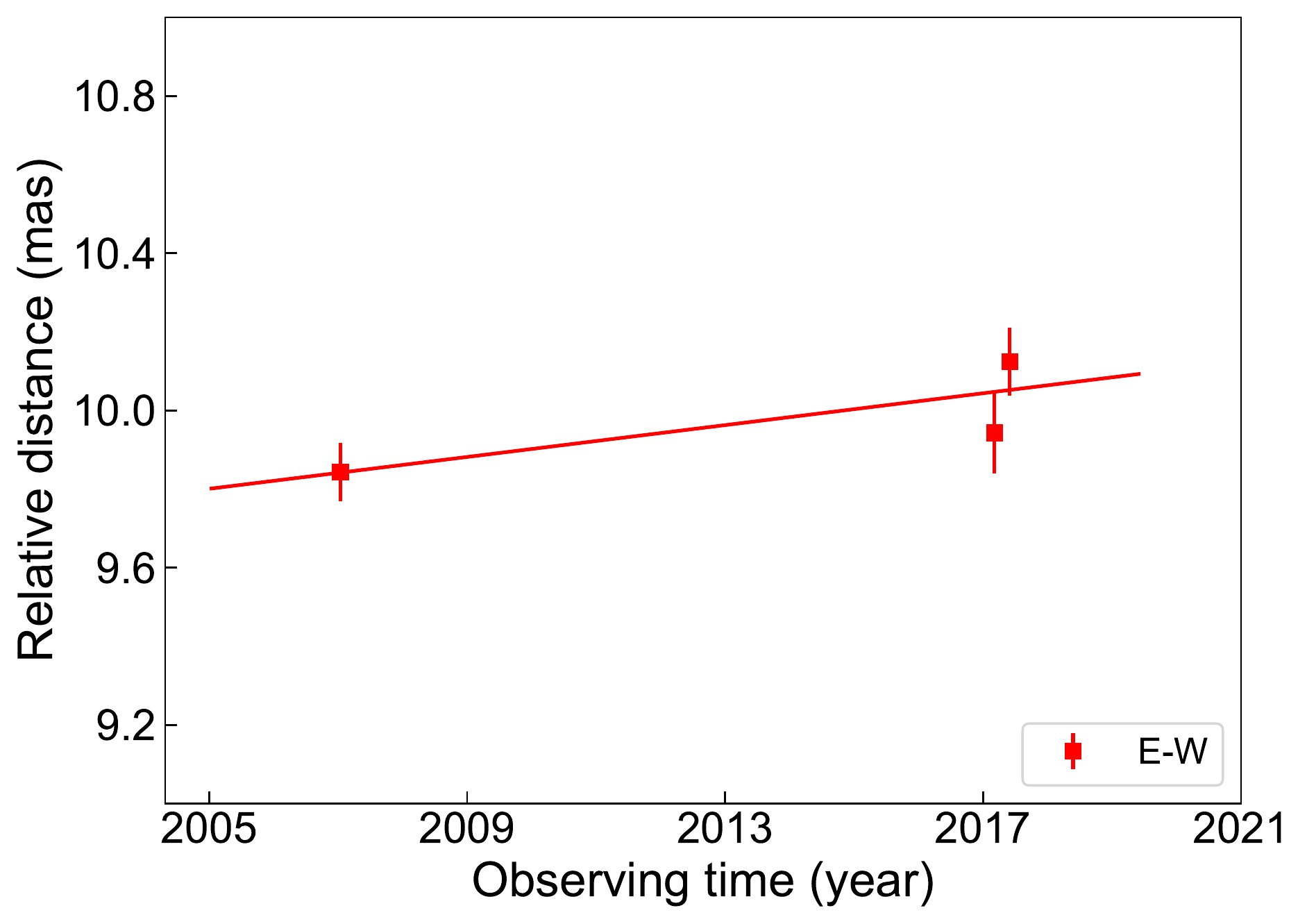}
	\includegraphics[width=0.3\textwidth]{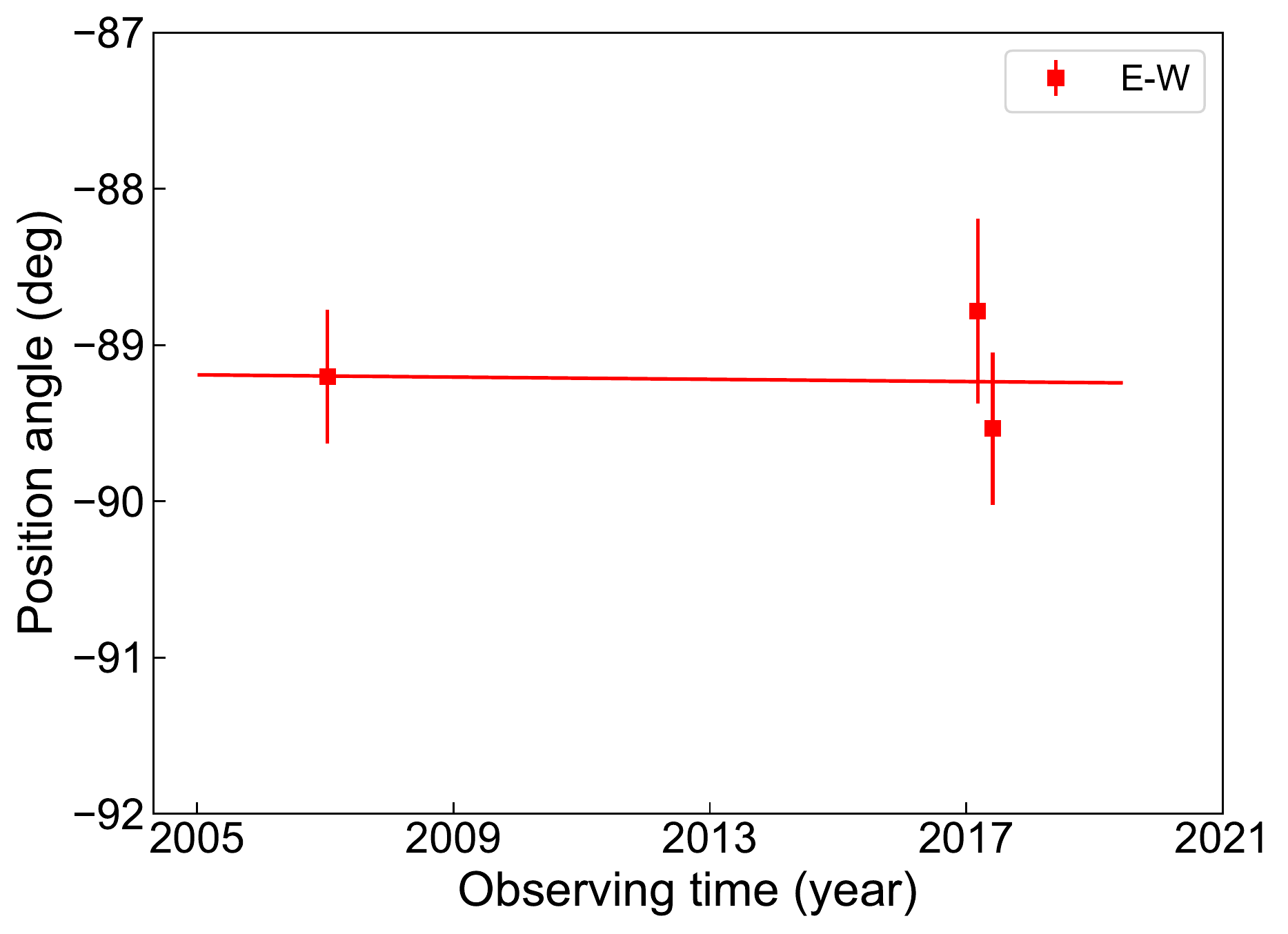}
}
	\caption{Jet proper motions (continued)}
	\figurenum{3}
\end{figure*}

\bibliography{references.bib}
\bibliographystyle{aasjournal}

\end{document}